\pdfoutput=1
\PassOptionsToPackage{table}{xcolor}
\documentclass{article}

\usepackage{iclr2027_conference,times}
\usepackage{etoolbox}
\iclrfinalcopy
\makeatletter
\patchcmd{\@maketitle}{Published as a conference paper at ICLR 2027}{Preprint}{}{\PackageError{ui2app}{Preprint header patch failed}{Check the conference style.}}
\patchcmd{\@maketitle}{\LARGE\sc}{\LARGE\normalfont\bfseries\centering}{}{\PackageError{ui2app}{Title style patch failed}{Check the conference style.}}
\patchcmd{\@maketitle}{\@title\par}{\@title\par\vskip 8pt}{}{\PackageError{ui2app}{Title spacing patch failed}{Check the conference style.}}
\patchcmd{\@maketitle}{\vskip 0.3in minus 0.1in}{\vskip 0.04in}{}{\PackageError{ui2app}{Author spacing patch failed}{Check the conference style.}}
\makeatother

\usepackage[utf8]{inputenc}
\usepackage[T1]{fontenc}
\usepackage{newunicodechar}
\newunicodechar{’}{\textquoteright}
\usepackage{hyperref}
\usepackage{url}
\usepackage{booktabs}
\usepackage{amsmath,amssymb,amsfonts}
\usepackage{nicefrac}
\usepackage{microtype}
\usepackage[table]{xcolor}
\definecolor{bestcell}{RGB}{225,242,225}
\definecolor{worstcell}{RGB}{250,228,228}
\newcommand{\B}[1]{\cellcolor{bestcell}\textbf{#1}}
\newcommand{\W}[1]{\cellcolor{worstcell}#1}
\definecolor{cmarkgreen}{RGB}{46,139,87}
\definecolor{xmarkred}{RGB}{192,57,43}
\newcommand{\cmark}{\textcolor{cmarkgreen}{\checkmark}}
\newcommand{\xmark}{\textcolor{xmarkred}{\ding{55}}}
\usepackage{graphicx}
\usepackage{subcaption}
\usepackage{tikz}
\usetikzlibrary{arrows.meta, positioning, calc}
\usepackage{fontawesome5}
\usepackage{makecell}
\usepackage{longtable}
\usepackage{array}
\usepackage{multirow}
\usepackage{enumitem}
\usepackage{placeins}
\usepackage{float}
\usepackage{wrapfig}
\usepackage{xspace}
\usepackage{pifont}
\usepackage{algorithm}
\usepackage{algpseudocode}
\usepackage{listings}
\definecolor{codebg}{RGB}{246,248,250}
\definecolor{codeframe}{RGB}{213,220,229}
\definecolor{codetext}{RGB}{46,52,64}
\definecolor{codekw}{RGB}{47,111,179}
\definecolor{codecmt}{RGB}{110,139,110}
\definecolor{codestr}{RGB}{161,90,50}
\lstdefinelanguage{tsx}{
  morekeywords={const,let,var,function,return,if,else,for,while,switch,case,break,
    import,export,from,default,new,class,extends,type,interface,async,await,void,
    useCallback,useState,useEffect,useRef,useMemo},
  morekeywords=[2]{true,false,null,undefined,this},
  sensitive=true,
  morecomment=[l]{//},
  morecomment=[s]{/*}{*/},
  morestring=[b]",
  morestring=[b]',
}
\usepackage[most]{tcolorbox}
\newtcolorbox{promptbox}[1][]{
  enhanced, breakable,
  colback=gray!8, colframe=black!65,
  boxrule=0.7pt, arc=2pt,
  left=6pt, right=6pt, top=5pt, bottom=5pt,
  fontupper=\small\ttfamily,
  parskip=0.4em,
  #1
}
\tcbuselibrary{listings,skins}
\definecolor{codetitlebg}{RGB}{59,66,82}
\newtcblisting{codecard}[1]{%
  listing engine=listings, listing only,
  colback=codebg, colframe=codeframe, boxrule=0.6pt, arc=3pt,
  title={\ttfamily\scriptsize #1}, colbacktitle=codetitlebg, coltitle=white,
  fonttitle=\bfseries, toptitle=1pt, bottomtitle=1pt,
  left=6pt, right=4pt, top=3pt, bottom=3pt,
  listing options={frame=none, backgroundcolor=\color{codebg},
    xleftmargin=0pt, framexleftmargin=0pt, aboveskip=0pt, belowskip=0pt},
}

\usepackage[disable, textwidth=2.4cm,textsize=footnotesize,backgroundcolor=yellow!20,linecolor=orange]{todonotes}

\DeclareMathOperator*{\argmin}{arg\,min}

\newcommand{\exect}[1][]{EXEC\ifx\\#1\\\else @#1\fi\xspace}
\newcommand{\vfs}{VFS\xspace}
\newcommand{\vfsstar}{VFS$^{\star}$\xspace}
\newcommand{\iis}{IIS\xspace}
\newcommand{\iisstar}{IIS$^{\star}$\xspace}

\newcommand{\sone}{S1\xspace}
\newcommand{\stwo}{S2\xspace}
\newcommand{\sthree}{S3\xspace}

\newcommand{\bench}{UI2App\xspace}

\hypersetup{
  colorlinks=true,
  linkcolor=[rgb]{0.0,0.35,0.7},
  citecolor=[rgb]{0.0,0.4,0.0},
  urlcolor=[rgb]{0.0,0.35,0.7},
  pdftitle={UI2App: Benchmarking Visual Interaction Inference in Executable Web Application Generation},
  pdfauthor={Grace Man Chen, Litao Guo, Yifan Wu, Yiyu Chen, Yenchi Tseng, Sicheng Liu, Yuyu Luo, Ying-Cong Chen}
}

\graphicspath{{./figures/}{./}}

\title{UI2App: Benchmarking Visual Interaction Inference\texorpdfstring{\\}{ }in Executable Web Application Generation}

\author{%
  \begin{minipage}{0.92\textwidth}
  \centering\normalfont
  \bfseries Grace Man Chen$^{1}$, Litao Guo$^{1}$, Yifan Wu$^{1}$, Yiyu Chen$^{1}$, Yenchi Tseng$^{1}$, Sicheng Liu$^{1}$, Yuyu Luo$^{1,2,\dagger}$, Ying-Cong Chen$^{1,2,\dagger}$%
  \\[5pt]
  \mdseries $^{1}$The Hong Kong University of Science and Technology (Guangzhou)\\
  $^{2}$The Hong Kong University of Science and Technology%
  \\[3pt]
  {\small $^{\dagger}$Corresponding author}%
  \end{minipage}%
}

\begin{document}
\raggedbottom 

\maketitle

\vspace{0.04in}
{\centering\small
\faGlobe~\href{https://chenmancm169.github.io/UI2App-ProjectPage/}{\textbf{Project}}\hspace{1.8em}
\faGithub~\href{https://github.com/chenmancm169/UI2App}{\textbf{Code}}\hspace{1.8em}
\raisebox{-2.2pt}{\includegraphics[height=1em]{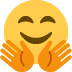}}~\href{https://huggingface.co/datasets/UI2App/UI2App}{\textbf{Dataset}}\par}
\vspace{0.16in}

\begin{abstract}
Large language models (LLMs) have demonstrated growing competence in generating web pages from UI screenshots, which convey both visual structure and cues to application behavior. Yet most screenshot-to-code benchmarks emphasize visual fidelity, while interactive generation benchmarks often supply behavioral specifications or demonstrated transitions. Whether models can infer and realize interactions from static screenshots alone remains insufficiently evaluated. We introduce \bench{} to evaluate \emph{interaction inference}: inferring and realizing application behavior from static visual cues without added behavioral guidance. \bench{} comprises $600$ screenshots organized into $95$ state-coherent sets for runnable multi-route web applications. Our end-to-end pipeline evaluates each artifact along three dimensions: executability, visual fidelity, and interaction inference. The interaction metric (\iis{}) assesses functional correctness and state-management complexity, crediting valid implementations rather than requiring a match to a single reference. Experiments on six frontier vision-language models reveal a marked mismatch between visual fidelity and interaction realization: the visual-fidelity leader scores only $8.1$ on \iis{}, ranking fourth, while the \iis{} leader achieves $4.5\times$ that score. High-complexity interactions such as cross-route state persistence remain a major bottleneck, with five of the six models scoring at most $3.5$ on this dimension. Overall, these results highlight interaction inference as a key challenge in generating functional web applications from static screenshots.
\end{abstract}

\section{Introduction}
\label{sec:introduction}

\begin{figure}[t]
  \centering
  \includegraphics[width=\linewidth]{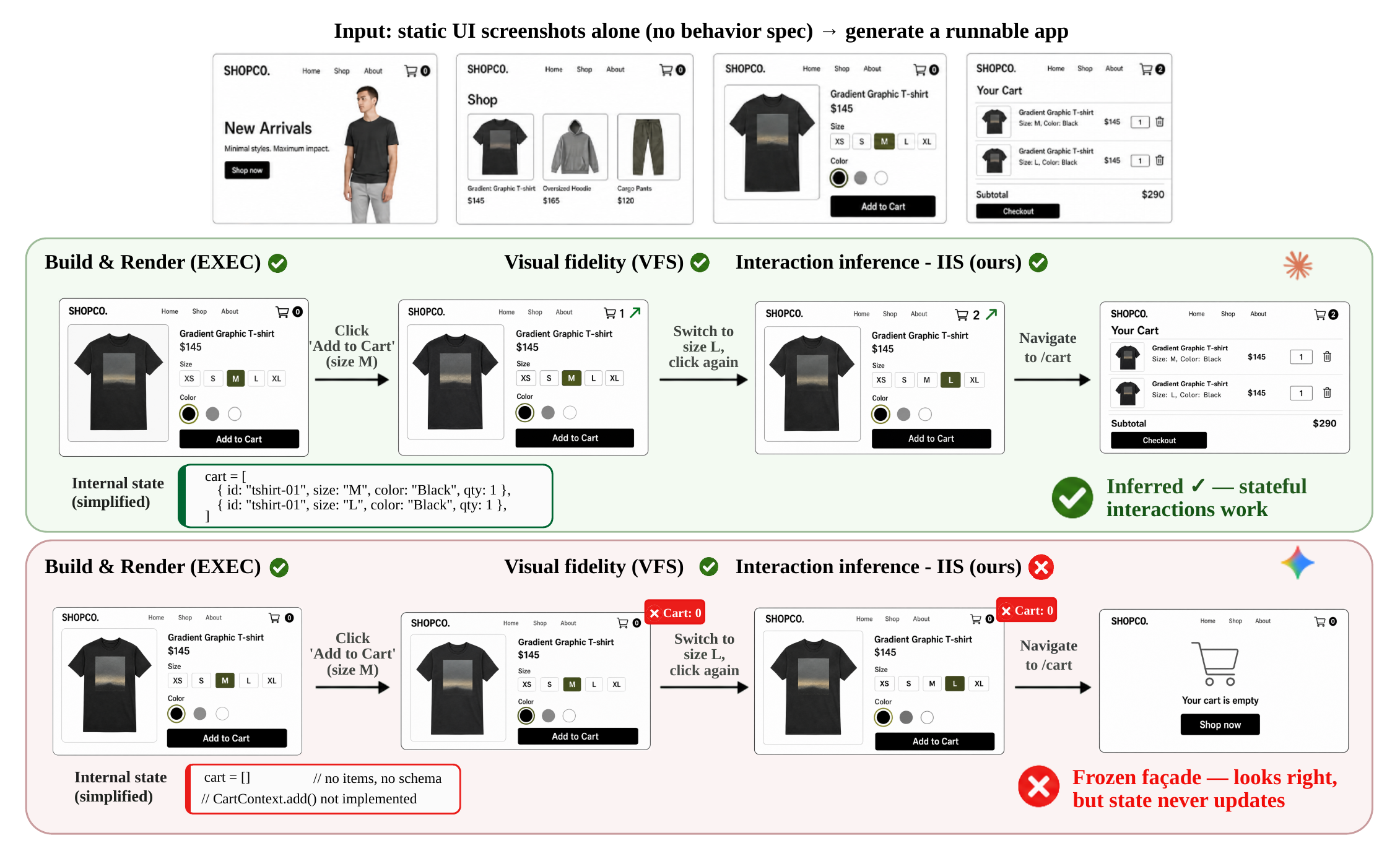}
  \caption{\textbf{\bench{} task overview: visual fidelity does not
  guarantee interaction-inference capability.} Two applications generated
  from the same e-commerce screenshots are tested using the same action
  sequence (Add to Cart twice, then navigate to \texttt{/cart}). Claude
  Sonnet~4.6 updates the cart counter and propagates state across routes,
  whereas Gemini~3.1~Pro~Preview achieves higher visual fidelity but fails
  to realize these cart-state updates. \iis{} captures this behavioral
  difference beyond visual fidelity.}
  \label{fig:task_overview}
\end{figure}

Large language models (LLMs) have made notable progress in code generation,
task planning, and reasoning~\citep{roziere2023code, guo2024deepseek,
guo2025deepseek}. Building on these advances, recent work has demonstrated
the feasibility of end-to-end software development from natural-language
instructions~\citep{hong2023metagpt, qian2024chatdev}. However, existing
text-driven paradigms largely rely on carefully designed or highly detailed
prompts as input~\citep{lu2025webgenbench, zhang2025artifactsbench}, which
introduces a practical limitation. Textual descriptions are limited in
precisely specifying detailed visual requirements, making it difficult to
accurately define page layouts and maintain visual coherence across multiple
pages.

To reduce the burden of specifying visual requirements, image-to-webpage
generation offers an alternative paradigm. Instead of relying on textual
specifications, this approach takes UI designs as input and reconstructs
webpage structure and appearance directly from visual signals. This paradigm
better aligns with the common designer workflow of turning mockups into
functional prototypes. It also addresses the demand from non-technical users
who want functional apps directly from screenshots. Several benchmarks have
been developed along this line~\citep{beltramelli2018pix2code,
si2025design2code, laurenccon2024unlocking, li2024sketch2code}.

However, these visual-reconstruction benchmarks largely focus on visual
fidelity. They measure only how closely the rendered output matches a
reference image, not whether the artifact \emph{works} as an application. A
generated interface may appear correct yet remain a behaviorally inert
fa\c{c}ade. Figure~\ref{fig:task_overview} illustrates this distinction:
both models produce runnable applications, but the model with higher visual
fidelity fails to update and propagate shopping-cart state across routes,
whereas the other realizes these interactions from the same static
screenshots.

This gap reflects a deeper distinction in capability. When interaction
behavior is provided through textual instructions or demonstrated
transitions~\citep{he2026vision2web,chen2025iwr,webvr2026}, models implement
behavior that has already been specified or shown. Without such guidance,
models must infer and realize screenshot-implied interactions from static
visual cues. We call this capability \emph{interaction inference}. Existing
benchmarks evaluate interactive generation, but do not explicitly isolate
this inference setting.

We introduce \bench{}, a benchmark that makes interaction inference
measurable. Each task supplies a set of screenshots and asks the model to
emit a runnable multi-page artifact. Screenshots arrive without added
behavioral guidance. The model must therefore interpret UI structure and
semantics, infer the implied interactions, and implement a runnable
framework-based application with shared components, routing, and coordinated
state management across pages. The benchmark comprises $600$ screenshots
organized into $95$ state-coherent screenshot sets, with $665$
application-category evaluation items spanning seven interaction
categories. Rather than isolated page images, each set spans multiple routes
and captures consistent cross-page state, providing visual context for
inferring and realizing interactions within and across routes. To evaluate
the resulting artifacts, we design an end-to-end pipeline that scores each
along three dimensions: whether the application builds and renders
(\exect[1] / \exect[3]), how faithfully the rendered UI matches the reference
(\vfs{}), and whether screenshot-implied interactions are realized (\iis{}).

A central challenge in designing \iis{} is that static screenshots
underdetermine behavior and admit multiple valid realizations, since the
same visual state may correspond to different correct implementations of
interaction logic (e.g., a search box that can be triggered on Enter,
button click, or real-time query). To address this challenge, we organize
interactions commonly found in web applications into a taxonomy of seven
categories and evaluate each application-category pair along three
complementary axes: interaction coverage, implementation result, and
state-management scope. Category-specific rubrics assess whether
screenshot-implied functionality is correctly realized while allowing
alternative implementations consistent with the visual evidence. Each
category covers multiple sub-functions and is rated fully functional only
when all sub-functions required by its rubric work correctly. This
taxonomy-driven design enables structured assessment of inferred behaviors
without requiring a unique ground-truth execution trace, while providing
fine-grained diagnostic signals beyond a single aggregate score.

We evaluate six frontier vision-language models (VLMs) on \bench{}. Under
our evaluation protocol, even the strongest reaches an overall \iis{} of
only $36.8$, highlighting the difficulty of inferring and realizing
screenshot-implied interactions. Moreover, visual fidelity does not
guarantee interaction realization: the visual-fidelity leader scores only
$8.1$ on \iis{} and ranks fourth, while the \iis{} leader achieves
$4.5\times$ that score. Cross-route state persistence ($\sthree$-scope
interactions) is particularly challenging: five of six models score at most
$3.5$, and even the best reaches only $15.8$ out of $100$.

Our contributions are as follows:
\begin{itemize}[leftmargin=1.4em,itemsep=1pt,topsep=1pt]

\item \textbf{Task formulation and benchmark.}
We formulate interaction inference: inferring and realizing
screenshot-implied interactions without added behavioral guidance. We
introduce \bench{}, comprising $600$ screenshots in $95$ state-coherent sets
for runnable multi-route web applications. To our knowledge, \bench{} is the
first benchmark dedicated to evaluating this setting.

\item \textbf{End-to-end evaluation protocol and \iis{} taxonomy.}
We introduce a three-dimensional evaluation protocol spanning executability
(\exect[1] / \exect[3]), visual fidelity (\vfs{}), and interaction inference
(\iis{}), with a taxonomy-based \iis{} that assesses functional correctness
and state-management scope while allowing valid alternative implementations.

\item \textbf{Extensive evaluation and analysis.}
We benchmark six frontier VLMs and a Qwen2.5-VL scaling ladder, establishing
baselines for screenshot-only interaction inference. Our results show that
high visual fidelity does not guarantee successful interaction realization,
with cross-route state persistence a common bottleneck across the evaluated
models.

\end{itemize}

\section{Related Work}
\label{sec:related_work}

\paragraph{Static Visual-to-Code.}
Building on advances in screenshot and UI understanding~\citep{lee2023pix2struct, baechler2024screenai, liuvisualwebbench, wu-etal-2024-chartinsights}, vision-language models have made rapid progress on translating UI visuals into frontend code. Early benchmarks established the screenshot-to-code task with real-world webpages and visual similarity scoring~\citep{si2025design2code, xiao2026designbench}, then scaled the paradigm with synthetic and large-scale corpora~\citep{yun2024web2code, gui2025webcode2m}. MRWeb extends generation to multi-page sites by pairing screenshots with a structured list of navigation links and resources~\citep{wan2024mrweb}. The underlying paradigm treats UI generation as a static rendering problem: tasks specify what an interface looks like, not how it behaves, and runtime interaction is left outside evaluation. In contrast, \bench{} inherits the screenshot-based input but reframes the task around behavior, asking models to produce a runnable application whose interaction implementations are assessed alongside rendered appearance.

\paragraph{Generation with Behavioral Guidance.}
WebGen-Bench~\citep{lu2025webgenbench} and ArtifactsBench~\citep{zhang2025artifactsbench} specify functionality in text, while Vision2Web~\citep{he2026vision2web} pairs multi-screen prototypes with interaction descriptions. Interaction2Code~\citep{xiao2025interaction2code}, IWR-Bench~\citep{chen2025iwr}, and WebCompass~\citep{lei2026webcompass} provide pre-/post-interaction screenshots, interaction videos, and navigation annotations with state-change keyframes, respectively. These inputs provide behavioral guidance; \bench{} instead isolates interaction inference from static screenshots without behavioral annotations or demonstrated transitions (Table~\ref{tab:benchmark-comparison}).

\paragraph{Framework-Based Application Generation.}
Recent benchmarks extend evaluation beyond standalone HTML to framework-based code and applications, including DesignBench~\citep{xiao2026designbench}, WebGen-Bench~\citep{lu2025webgenbench}, and Vision2Web~\citep{he2026vision2web}. This setting better reflects component-based development, but requires consistent cross-file interfaces, resolvable dependencies, and successful compilation and bundling. \bench{} uses a fixed React--TypeScript scaffold and evaluates executability alongside visual fidelity and screenshot-implied interactions.

\begin{table}[t!]
\centering
\caption{Comparison of benchmarks for web interface and application generation.}
\label{tab:benchmark-comparison}
\scriptsize
\setlength{\tabcolsep}{2pt}
\renewcommand{\arraystretch}{1.15}
\resizebox{\textwidth}{!}{%
\begin{tabular}{@{}l l l l c c c l@{}}
\toprule
Benchmark & \shortstack[l]{Input\\modality} & Dataset size & \shortstack[l]{Output\\stack} & \shortstack{Multi-\\page} & \shortstack{No behavior\\hints} & \shortstack{Interaction\\eval} & \shortstack[l]{Interaction\\eval size} \\
\midrule
\multicolumn{8}{@{}l}{\textbf{Visual reconstruction}} \\
Design2Code~\citep{si2025design2code} & Single image & 484 pages & HTML & \xmark & \cmark & \xmark & -- \\
DesignBench--Gen~\citep{xiao2026designbench} & Single image & 430 pages & HTML / Framework & \xmark & \cmark & \xmark & -- \\
MRWeb~\citep{wan2024mrweb} & Image + Text & 500 samples & HTML & \cmark & \xmark & \xmark & -- \\
ComUIBench~\citep{xiao2026comuicoder} & Images & 150 pages / 40 sites & HTML & \cmark & \cmark & \xmark & -- \\
\midrule
\multicolumn{8}{@{}l}{\textbf{Interactive HTML generation}} \\
Interaction2Code~\citep{xiao2025interaction2code} & Paired images & 127 pages & HTML & \xmark & \xmark & \cmark & 374 interactions \\
IWR-Bench~\citep{chen2025iwr} & Video & 113 tasks & HTML & \xmark & \xmark & \cmark & 1,001 actions \\
\midrule
\multicolumn{8}{@{}l}{\textbf{Interactive application generation}} \\
WebGen-Bench~\citep{lu2025webgenbench} & Text & 101 tasks & Framework & \cmark & \xmark & \cmark & 647 test cases \\
Vision2Web--L2~\citep{he2026vision2web} & Images + Text & 388 images / 66 apps & Framework & \cmark & \xmark & \cmark & 495 test cases \\
\midrule
\textbf{\bench{} (ours)} & \textbf{Images} & \textbf{600 images / 95 apps} & \textbf{Framework} & \textbf{\cmark} & \textbf{\cmark} & \textbf{\cmark} & \textbf{665 evaluation items} \\
\bottomrule
\end{tabular}%
}
\end{table}

\section{UI2App}
\label{sec:ui2app}

\subsection{Task Definition}
\label{sec:task_def}

A \bench{} task supplies only a set of $M$ screenshots ($M = 3$--$14$, mean $6.3$), each capturing a distinct route or visual state of the target application. \emph{No action trace or interaction description is provided}. Given this input, the model must produce the source code of a runnable application that reproduces the screenshots. All tasks and models share a fixed React + TypeScript scaffold, chosen as the most widely deployed front-end stack so that the benchmark reflects current production practice.

\subsection{Dataset Construction}
\label{sec:benchmark_dataset}

\paragraph{Source pool and filtering.}
\begin{wrapfigure}{r}{0.55\linewidth}
  \centering
  \vspace{-0.8em}
  \includegraphics[width=\linewidth]{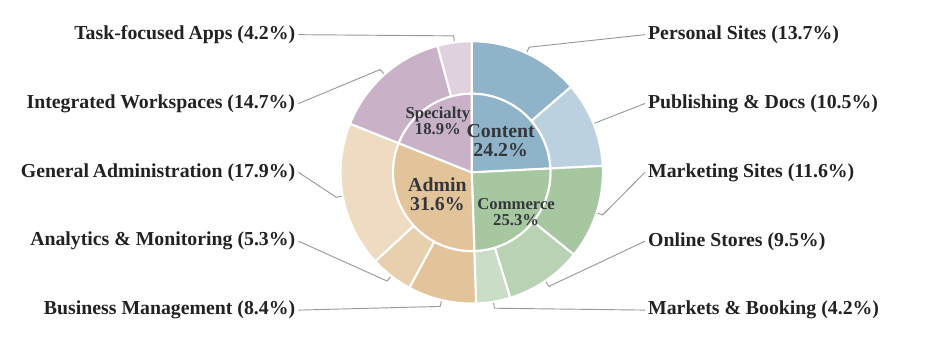}
  \caption{\textbf{\bench{} dataset diversity.}}
  \label{fig:dataset-distribution}
  \vspace{-0.8em}
\end{wrapfigure}
We curate \bench{}'s reference screenshots from open-source GitHub projects. To obtain a broad initial pool, we use $24$ archetype-aware GitHub search queries spanning $12$ application categories, yielding $4{,}254$ candidate repositories. We then apply a four-stage automated pipeline that checks permissive licensing, structural validity, buildability, and authentication-wall detection. This leaves $423$ repositories for expert review.

\paragraph{Expert selection.} Experts select $95$ of the $423$ candidates using three criteria: \textbf{Page-level discriminability.} Pages should combine non-trivial layouts, diverse UI components, and real text or images. \textbf{Application-level complexity.} Apps must offer distinct routes and screenshot-implied interactions across multiple \iis{} categories, and render stably at the capture resolution. \textbf{Corpus-level diversity.} Reviewers exclude repeated visual themes, duplicated upstream starters, and near-identical templates.
\paragraph{Capture pipeline.}
Each application is captured by a headless-browser script at $1440{\times}900$, with timing and content-validation rules producing post-hydration screenshots free of dev-server overlays (full pipeline in Appendix~\ref{app:dataset_details}). To assess generation from visual evidence, we exclude repository names, URLs, and reference source code from model inputs and evaluate visual fidelity and screenshot-implied functionality rather than source-code similarity.

\paragraph{Stratification and statistics.}
Applications are organised into four \emph{application categories} by user-facing function: \textbf{Content}, \textbf{Commerce}, \textbf{Admin}, and \textbf{Specialty}. Subcategory listings appear in Table~\ref{tab:category_summary} (Appendix).

\subsection{Evaluation Protocol}
\label{sec:pipeline}

\begin{figure}[!ht]
  \centering
  \includegraphics[width=\linewidth]{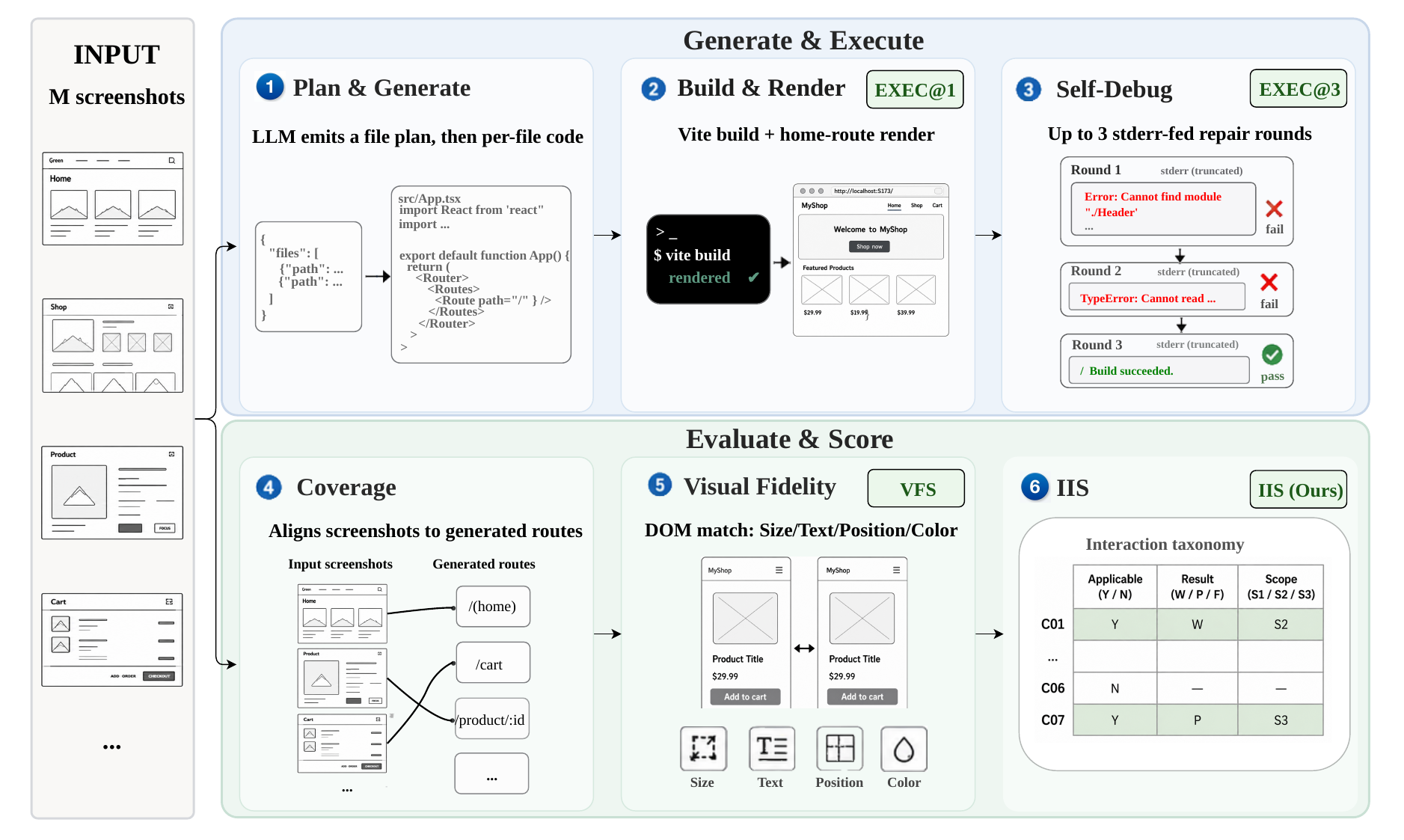}
  \caption{\textbf{The evaluation protocol of \bench{}.}}
  \label{fig:pipeline}
\end{figure}

\FloatBarrier

The protocol decomposes evaluation into a sequence of stages. Figure~\ref{fig:pipeline} illustrates the full chain. (1) \textbf{Plan \& Generate} splits generation into a file-list pass followed by per-file content. (2) \textbf{Build \& Render} attempts a production build and a home-route render, producing the per-attempt EXEC pass/fail signal. (3) \textbf{Self-Debug} on EXEC failure feeds the build's error messages back to the same model for up to three repair attempts. (4) \textbf{Coverage} maps each input screenshot to its generated-app route and emits the page pairs consumed by \vfs{}. (5) \textbf{Visual Fidelity} computes DOM-to-DOM similarity on the matched pairs, defining \vfs{}. (6) \textbf{Automated interaction evaluation} uses GPT-6 Luna (medium) to label each interaction from generated source code and rendered screenshots along the interaction taxonomy.

\subsection{Standard Metrics: \exect, \vfs}
\label{sec:metrics}

\paragraph{\exect[k] (Executability).}
\exect measures whether the generated source code \emph{actually runs}, the protocol's entry gate. An app passes when the production build completes without error and the home route renders meaningful content. \exect[1] is the one-shot pass rate. \exect[3], the pass rate after up to three rounds of error-feedback retry, separately measures iterate-fix capability.

\paragraph{\vfs{} (Visual Fidelity Score).}
\vfs{} scores \emph{what is rendered where} via judge-free block-level matching: visible content blocks from each page are paired by optimal bipartite matching, and four sub-metrics (Size, Text, Position, Color) on the matched pairs are averaged unweighted. This design is robust to implementation differences in the underlying DOM. Algorithm and sub-metric formulas in Appendix~\ref{app:vfs}.

\paragraph{Conventions.}
\textit{URL-direct loading}: Rendered screenshots for \vfs{} and \iis{} are collected by loading each route directly via URL, isolating page implementation from routing. \textit{Zero-imputation}: apps failing \exect[3] contribute zero to \vfs{}/\iis{}, keeping models with different \exect[3] rates directly comparable.

\subsection{Interaction Inference Score}
\label{sec:iis}

\paragraph{Definition.}
\iis{} quantifies how well a model infers and realizes screenshot-implied interactions, jointly considering interaction coverage, implementation result, and state-logic complexity. Rather than comparing against a single reference behavior, \iis{} adopts a rubric-based evaluation in which each interaction is judged against the functional criteria. This implementation-agnostic formulation accommodates diverse yet semantically equivalent realizations of the same screenshot-implied interaction while enabling consistent evaluation across different applications.

\paragraph{Interaction taxonomy and axes.}
To make interaction inference measurable, we organize interactions commonly found in web applications into seven categories: toggle, expand/collapse, list operations, data CRUD, form validation, notification, and cross-route state (C01--C07, cataloged in Appendix~\ref{app:iis_categories}). Each category is evaluated along three axes: coverage, indicating whether the interaction category is implied by the screenshots; result, indicating whether the generated application realizes the interaction category (fully realized, partially realized, or failed); and scope ($\sone$ UI-state, $\stwo$ data-state, $\sthree$ cross-route persistence), reflecting increasing state-management complexity in real front-end engineering. Separating these axes enables failures to be attributed to specific interaction types and aspects, rather than being collapsed into a single aggregate score.

\paragraph{Scoring and aggregation.}
For each application, we first construct a Reference Interaction Inventory (RII) from the input screenshots by labeling each interaction category with its coverage (whether the interaction is implied by the screenshots) and, when covered, its reference scope $s^{\text{ref}}_i$. The covered categories form the reference set $G_a$, ensuring that only interactions inferable from the screenshots are evaluated. For each generated application, every category is annotated along the same three axes: coverage $y_i \in \{0,1\}$, implementation scope $s^{\text{gen}}_i \in \{1,2,3\}$, and result $r_i \in \{1, 0.5, 0\}$ (working, partial, or failed). Each scope level carries a linear weight, $w_{\sone}=1$, $w_{\stwo}=2$, $w_{\sthree}=3$. Combining the three axes yields the per-application quality score and the model-level \iis{}:
\begin{equation}
\mathrm{QS}_{a,m} = \frac{\sum_{i \in G_a} y_i \cdot \min\!\bigl(w_{s^{\text{gen}}_i},\;w_{s^{\text{ref}}_i}\bigr)\,r_i}{\sum_{i \in G_a} w_{s^{\text{ref}}_i}},\qquad
\iis_m = \frac{100}{N} \sum_{a=1}^{N} \mathrm{QS}_{a,m},
\label{eq:iis}
\end{equation}
with $N$ the number of applications. The denominator is the sum of reference scope weights, which makes \iis{} recall-oriented.

\paragraph{Human IIS evaluation.}
Two front-end engineering experts construct the RII from input screenshots alone. On a 45-application validation subset, two annotators independently assess each generated application's runtime behavior across all seven categories. Adjacent outcome disagreements are averaged; larger disagreements are adjudicated by a third annotator. Inter-annotator agreement is substantial (Krippendorff $\alpha$ in $0.72$--$0.84$; Appendix~\ref{app:iis_handbook}).

\paragraph{Automated IIS evaluation.}
To improve evaluation efficiency and support scalable, repeatable assessment, we implement Auto IIS with GPT-6 Luna (medium), which assigns category-level labels from the generated application's source code and rendered screenshots without access to the generator identity.
We validate these static assessments against human annotations obtained by executing interactions and observing runtime behavior on the 45-application validation subset.
Coverage agreement is $86.4\%$; on items judged applicable by both evaluators, outcome agreement is $81.8\%$ and exact scope agreement is $91.1\%$.
The resulting model-level IIS scores preserve the human-derived ranking of all six models, with a Pearson correlation of $0.991$ and a mean absolute error of $1.57$ IIS points (Appendix~\ref{app:auto_iis}).

\section{Experiments}
\label{sec:experiments}

\subsection{Experimental Setup}
\label{sec:exp_setup}

We evaluate six frontier VLMs: Claude Sonnet 4.6~\citep{claude-4-5-card}, Kimi K2.5~\citep{kimiteam2026kimik25visualagentic} (thinking mode enabled), GPT-5.4~\citep{gpt-5-4-card}, Gemini 3.1 Pro Preview~\citep{gemini-3-1-pro-card}, Qwen3.5-397B-A17B~\citep{qwen-3-5-card}, and GLM-4.6V~\citep{glm-4-6v-card}. To make performance differences more directly attributable to the underlying models, we query each model through its vendor’s official API under a unified generation protocol. The main table reports one generation per application--model pair on all 95 applications, with IIS evaluated by GPT-6 Luna (medium). To validate automated scoring and assess generation variability, we use human annotations and three independent generations on a 45-application subset; evaluator agreement and repeated-generation results are detailed in Appendices~\ref{app:auto_iis} and~\ref{app:repeated_evaluation}. We additionally evaluate four Qwen2.5-VL checkpoints on this subset to study within-family scaling (Section~\ref{sec:exp_scaling}).

\subsection{Main Results}
\label{sec:exp_main}

\paragraph{Overall ranking across the three metrics.}
Table~\ref{tab:main_results} shows the three-metric panorama. Claude~Sonnet~4.6 leads \exect[1] ($95.8\%$) and \exect[3] ($100\%$). Gemini leads \vfs{} ($73.2$) and Claude leads \iis{} ($36.8$), while GLM-4.6V is lowest on every metric. Visual and interaction quality yield different model orderings.

\begin{table}[!ht]
\centering
\caption{\textbf{Per-category main results on \bench{}.} All scores except EXEC (\%) are on a $0$--$100$ scale. Best per (category, column) shown with \B{sage cell + bold}; worst with blush cell.}
\label{tab:main_results}
\normalsize
\renewcommand{\arraystretch}{1.09}
\setlength{\tabcolsep}{3pt}
\begin{tabular}{ll rrr rrrr r r}
\toprule
\multirow{2}{*}{Category} & \multirow{2}{*}{Model}
  & \multicolumn{3}{c}{EXEC (\%)}
  & \multicolumn{4}{c}{\vfs{} components} & \multirow{2}{*}{\vfs{}} & \multirow{2}{*}{\iis{}} \\
\cmidrule(lr){3-5} \cmidrule(lr){6-9}
 &  & @1 & @3 & $\Delta$ & Size & Text & Pos. & Color & & \\
\midrule
\multirow{6}{*}{Content} & Claude Sonnet 4.6 & \B{100.0} & \B{100.0} & 0.0 & 64.9 & 78.7 & 75.9 & 66.6 & 71.6 & \B{57.3} \\
 & Gemini 3.1 Pro Preview & \B{100.0} & \B{100.0} & 0.0 & \B{71.1} & \B{82.2} & \B{79.8} & \B{69.8} & \B{75.7} & 13.7 \\
 & Kimi K2.5 & 82.6 & 87.0 & $+4.3$ & 57.4 & 70.1 & 66.1 & 62.9 & 64.1 & 37.2 \\
 & GPT-5.4 & 69.6 & 78.3 & $+8.7$ & 49.2 & 59.0 & 58.3 & 52.9 & 54.9 & 16.3 \\
 & Qwen3.5-397B-A17B & 60.9 & 78.3 & $+17.4$ & 47.4 & 57.7 & 55.5 & 53.9 & 53.6 & 29.4 \\
 & GLM-4.6V & \W{43.5} & \W{69.6} & \B{$+26.1$} & \W{43.8} & \W{52.1} & \W{46.1} & \W{47.8} & \W{47.4} & \W{2.6} \\
\midrule
\multirow{6}{*}{Commerce} & Claude Sonnet 4.6 & \B{95.8} & \B{100.0} & $+4.2$ & \B{69.1} & 75.0 & \B{74.6} & 68.8 & 71.9 & \B{36.7} \\
 & Gemini 3.1 Pro Preview & \B{95.8} & \B{100.0} & $+4.2$ & 68.9 & \B{75.1} & 73.7 & \B{71.2} & \B{72.2} & 10.2 \\
 & Kimi K2.5 & 87.5 & 87.5 & 0.0 & 61.8 & 67.3 & 65.4 & 65.0 & 64.8 & 12.5 \\
 & GPT-5.4 & 83.3 & 91.7 & $+8.3$ & 59.8 & 67.5 & 66.5 & 65.7 & 64.9 & \W{7.8} \\
 & Qwen3.5-397B-A17B & 83.3 & 87.5 & $+4.2$ & 58.5 & 63.7 & 62.7 & 60.6 & 61.4 & 15.0 \\
 & GLM-4.6V & \W{20.8} & \W{62.5} & \B{$+41.7$} & \W{39.9} & \W{40.0} & \W{40.6} & \W{38.3} & \W{39.7} & 10.3 \\
\midrule
\multirow{6}{*}{Admin} & Claude Sonnet 4.6 & \B{90.0} & \B{100.0} & $+10.0$ & 68.4 & 77.1 & 78.1 & 62.1 & 71.4 & \B{21.3} \\
 & Gemini 3.1 Pro Preview & 86.7 & 96.7 & $+10.0$ & \B{71.0} & \B{78.8} & \B{78.5} & \B{63.8} & \B{73.0} & 4.9 \\
 & Kimi K2.5 & 60.0 & 66.7 & $+6.7$ & 43.9 & 49.1 & 49.9 & 39.5 & 45.6 & 13.9 \\
 & GPT-5.4 & 56.7 & 73.3 & $+16.7$ & 47.7 & 54.5 & 54.7 & 45.5 & 50.6 & \W{1.1} \\
 & Qwen3.5-397B-A17B & 70.0 & 76.7 & $+6.7$ & 54.6 & 59.5 & 59.0 & 49.6 & 55.7 & 6.9 \\
 & GLM-4.6V & \W{20.0} & \W{46.7} & \B{$+26.7$} & \W{25.9} & \W{29.4} & \W{29.2} & \W{25.7} & \W{27.6} & 3.5 \\
\midrule
\multirow{6}{*}{Specialty} & Claude Sonnet 4.6 & \B{100.0} & \B{100.0} & 0.0 & \B{71.4} & 79.5 & 74.7 & 55.3 & 70.3 & \B{36.7} \\
 & Gemini 3.1 Pro Preview & 72.2 & 94.4 & $+22.2$ & 65.6 & \B{81.8} & \B{80.1} & 59.1 & \B{71.7} & 3.7 \\
 & Kimi K2.5 & 66.7 & 66.7 & 0.0 & 47.4 & 56.5 & 53.1 & 37.4 & 48.6 & 16.5 \\
 & GPT-5.4 & 66.7 & 94.4 & \B{$+27.8$} & 66.3 & 77.1 & 74.6 & \B{59.6} & 69.4 & 3.5 \\
 & Qwen3.5-397B-A17B & 72.2 & 88.9 & $+16.7$ & 64.2 & 72.4 & 69.6 & 56.5 & 65.7 & 11.9 \\
 & GLM-4.6V & \W{11.1} & \W{27.8} & $+16.7$ & \W{13.5} & \W{17.0} & \W{16.8} & \W{15.0} & \W{15.6} & \W{1.4} \\
\specialrule{\heavyrulewidth}{0pt}{0pt}
\multirow{6}{*}{\textbf{Overall}} & Claude Sonnet 4.6 & \B{95.8} & \B{100.0} & $+4.2$ & 68.3 & 77.4 & 76.1 & 63.6 & 71.3 & \B{36.8} \\
 & Gemini 3.1 Pro Preview & 89.5 & 97.9 & $+8.4$ & \B{69.5} & \B{79.3} & \B{77.9} & \B{66.3} & \B{73.2} & 8.1 \\
 & Kimi K2.5 & 73.7 & 76.8 & $+3.2$ & 52.3 & 60.2 & 58.3 & 51.2 & 55.5 & 19.7 \\
 & GPT-5.4 & 68.4 & 83.2 & $+14.7$ & 54.7 & 63.2 & 62.3 & 55.1 & 58.8 & 6.9 \\
 & Qwen3.5-397B-A17B & 71.6 & 82.1 & $+10.5$ & 55.7 & 62.6 & 61.1 & 54.7 & 58.5 & 15.3 \\
 & GLM-4.6V & \W{24.2} & \W{52.6} & \B{$+28.4$} & \W{31.4} & \W{35.2} & \W{33.8} & \W{32.2} & \W{33.2} & \W{4.6} \\
\bottomrule
\end{tabular}
\end{table}

\FloatBarrier

\paragraph{EXEC self-debug heterogeneity.}
Self-debug gains vary across models (Table~\ref{tab:main_results}): GLM-4.6V gains $28.4$ points, GPT-5.4 gains $14.7$, and Gemini and Qwen3.5 gain $8.4$ and $10.5$, respectively; Kimi gains $3.2$ and Claude $4.2$. GPT-5.4 still ranks $5/6$ on \iis{}, indicating interaction deficiencies after build repair.

\paragraph{VFS spread is largely build-driven.}
Conditioning on \exect[3]-passing apps narrows the \vfs{} range from $40.1$ to $11.8$ points and raises GLM-4.6V from $33.2$ to $63.0$ (Table~\ref{tab:cond_vfs}). Appendix~\ref{app:cond_submetric} breaks down the remaining visual differences.

\paragraph{Consistency across generations.}
On the 45-app subset, Gemini leads visual fidelity in all three generations, while Claude consistently leads IIS ($36.86\pm0.39$), exceeding Gemini by $27.91$ points on average (95\% hierarchical bootstrap CI: $[20.95,35.06]$; Appendix~\ref{app:repeated_evaluation}).

\subsection{Interaction Inference Deep Dive}
\label{sec:exp_iis}

\begin{figure}[!ht]
  \centering
  \begin{minipage}[c]{0.54\linewidth}
    \centering
    \includegraphics[width=\linewidth]{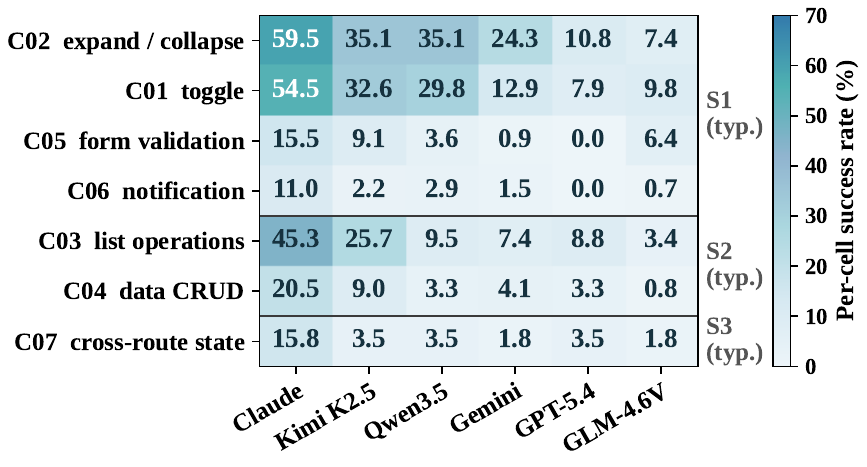}
    \captionof{figure}{\textbf{\iis{} per-cell heatmap.} Per-(category, model) success rate ($\times 100$, capped, zero-imputation) over $7$ categories $\times$ $6$ baselines; rows grouped by typical scope (\sone/\stwo/\sthree); columns ordered by overall \iis{}.}
    \label{fig:iis_heatmap}
  \end{minipage}\hfill
  \begin{minipage}[c]{0.44\linewidth}
    \centering
    \captionof{table}{$S$-restricted \iis{} (capped, zero-imputation, \S\ref{sec:iis}). Population per scope: apps with $\ge 1$ $S$-typical RII-positive item ($n=95/80/57$). Best per column \textbf{bold}, worst \underline{underlined}.}
    \label{tab:per_scope_iis}
    \scriptsize
    \setlength{\tabcolsep}{2pt}
    \begin{tabular}{@{}l@{\hskip 4pt}rrr@{}}
    \toprule
    Model & $\sone$ & $\stwo$ & $\sthree$ \\
    \midrule
    Claude Sonnet 4.6 & \textbf{39.9} & \textbf{37.8} & \textbf{15.8} \\
    Kimi K2.5 & 24.1 & 19.7 & 3.5 \\
    Qwen3.5-397B-A17B & 20.8 & 8.8 & 3.5 \\
    Gemini 3.1 Pro Prev. & 10.8 & 8.4 & \underline{1.8} \\
    GPT-5.4 & \underline{5.7} & 8.8 & 3.5 \\
    GLM-4.6V & 6.6 & \underline{2.8} & \underline{1.8} \\
    \midrule
    spread & 34.2 & 35.0 & 14.0 \\
    \bottomrule
    \end{tabular}
  \end{minipage}
\end{figure}

Through a fine-grained analysis of \iis{} across interaction categories, scope tiers, and application categories (Figure~\ref{fig:iis_heatmap}, Tables~\ref{tab:per_scope_iis} and~\ref{tab:category_iis}), we derive the four findings below.

\paragraph{Finding~1 --- Visual fidelity does not imply interaction-inference capability.}
The \vfs{} leader and the \iis{} leader are different models: Gemini~3.1~Pro~Preview ranks first on \vfs{} ($73.2$) but fourth on \iis{} ($8.1$), while Claude~Sonnet~4.6, second on \vfs{} at $71.3$, leads \iis{} at $36.8$ ($4.5\times$ ahead of Gemini). Per-cell row-means (Figure~\ref{fig:iis_heatmap}) put Gemini at only $4.1$, $1.5$, and $1.8$ on C04 (CRUD), C06 (notification), and C07 (cross-route state), showing that high VFS scores can coexist with missing core interaction logic.

\paragraph{Finding~2 --- The closed-vs-open quality hierarchy does not transfer to \iis{}.}
Two of the three open-weight entries (Kimi $19.7$, Qwen3.5 $15.3$) outscore the two closed-frontier entries below Claude (Gemini $8.1$, GPT-5.4 $6.9$) on \iis{}, although both trail Gemini and GPT-5.4 on \vfs{}. Visual-quality ordering therefore does not carry over to interaction inference.

\paragraph{Finding~3 --- Cross-route state is a shared gap among the evaluated models.}
$S$-restricted \iis{} (Table~\ref{tab:per_scope_iis}) decreases as scope becomes harder: $\sone > \stwo > \sthree$ holds for five of six models. The $\sthree$ end is a shared limitation: five models score only $1.8$--$3.5$, and Claude reaches $15.8$. Between-model spread contracts from $34.2$ on $\sone$ to $14.0$ on $\sthree$. Cross-route state remains a challenging dimension of interaction inference.

\paragraph{Finding~4 --- Admin apps have the lowest average \iis{}.}
Admin apps yield the lowest six-model mean \iis{} ($8.6$), compared with $26.1$ on Content, $15.4$ on Commerce, and $12.3$ on Specialty (Appendix Table~\ref{tab:category_iis}). Admin also has the narrowest between-model spread. Its low scores suggest a shared difficulty, although item composition and application complexity may contribute to the category gap.

\FloatBarrier

\subsection{Failure Analysis}
\label{sec:exp_failures}
Across the six frontier models, distinct failure mechanisms emerge on the executability and interaction dimensions. Of the $101$ \exect[3] build/runtime failures, $38.6\%$ violate one of four explicit plan-prompt constraints. Instruction compliance ranges from $100.0\%$ to $71.6\%$. Hallucinated icon exports and scaffold overwrites are the two largest failure categories. Cross-route state remains weak even among runnable applications. More detailed failure analysis is provided in Appendix~\ref{app:failures}.

\FloatBarrier

\subsection{Within-Family Scaling (Qwen2.5-VL)}
\label{sec:exp_scaling}

To assess how executability, visual fidelity, and interaction inference scale with model size, we evaluate Qwen2.5-VL models at 3B, 7B, 32B, and 72B on the 45-application subset. We use the vendor's official API and the same prompt templates for generation. We report \exect[1], \exect[3], \vfs{}, and \iis{}, evaluating the latter with the main experiment's Luna protocol.

\begin{figure}[!ht]
  \centering
  \begin{subfigure}[t]{0.48\linewidth}
    \includegraphics[width=\linewidth]{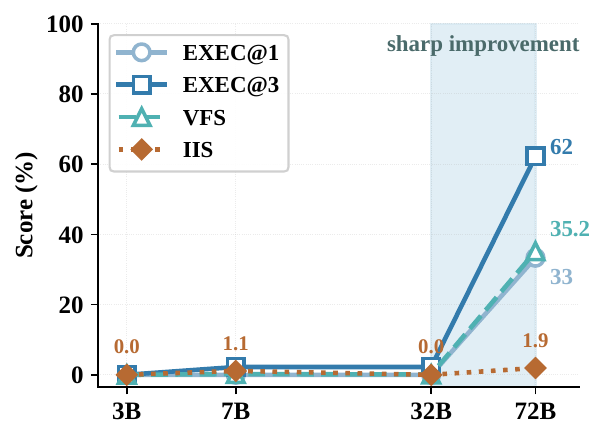}
    \caption{Scaling on four automatic metrics.}
    \label{fig:qwen_scaling_main}
  \end{subfigure}\hfill
  \begin{subfigure}[t]{0.48\linewidth}
    \includegraphics[width=\linewidth]{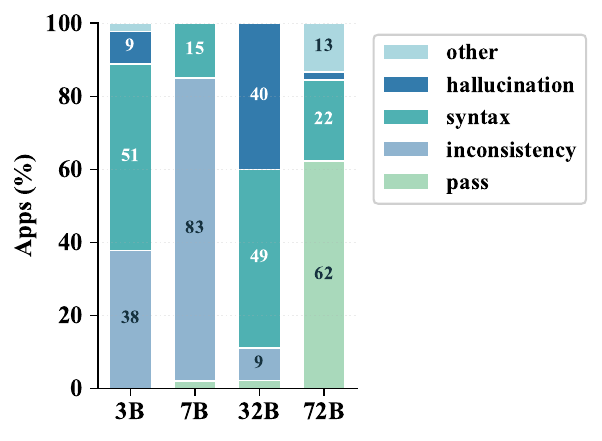}
    \caption{Failure-mode composition.}
    \label{fig:qwen_scaling_failures}
  \end{subfigure}
  \caption{\textbf{Qwen2.5-VL scaling on the 45-app subset (3B~$\to$~72B).}
  \subref{fig:qwen_scaling_main}: \exect[1] (circles), \exect[3] (squares), \vfs{} (triangles, dashed), and \iis{} (diamonds, dotted); the shaded band marks the 32B~$\to$~72B improvement region. \subref{fig:qwen_scaling_failures}: per-app build-failure category, stacked to $100\%$. Categories: \emph{syntax}, \emph{inconsistency}, \emph{hallucination}, \emph{other}, \emph{pass}.}
  \label{fig:qwen_scaling}
\end{figure}

\paragraph{Sharp improvement from 32B to 72B.}
Across the $24\times$ size range, \exect and \vfs{} are essentially flat below 32B and jump sharply at 72B (Figure~\ref{fig:qwen_scaling_main}): \exect[3] from $\le 2.2\%$ to $62.2\%$, \vfs{} from $\le 0.2$ to $35.2$. Even at 72B, $37.8\%$ of generations still fail to build, indicating that within-family parameter scaling alone does not saturate UI2App's executability requirement. Interaction inference remains weak: \iis{} remains at $0.0$, $1.1$, $0.0$, and $1.9$ for 3B, 7B, 32B, and 72B, respectively.

\paragraph{Build-failure composition shifts with size, not just rate.}
The dominant build-error category varies with model size (Figure~\ref{fig:qwen_scaling_failures}). The smaller checkpoints fail at basic source-code competence (3B on \emph{syntax}, 7B on \emph{inconsistency}), while the larger ones fail in more semantically loaded ways (32B on a mix of \emph{syntax} and \emph{hallucinated} dependencies, 72B at deeper build stages). Per-size counts and interpretation caveats are in Appendix~\ref{app:failures-scaling}.

\section{Conclusion}
\label{sec:conclusion}

In this paper, we introduced \bench{}, to our knowledge the first benchmark dedicated to
\emph{interaction inference}: inferring application behavior from static multi-page screenshots without explicit behavioral guidance. Our protocol evaluates executability, visual fidelity, and \iis{}, whose interaction taxonomy captures state-logic complexity. Across six frontier
vision-language models, visual fidelity proves a poor proxy for
interaction realization: the \vfs{} leader trails the \iis{} leader by
$4.5\times$. Cross-route state persistence is a bottleneck among the evaluated models,
with five of six models scoring at most $3.5$ and the strongest
reaching only $15.8$ on $\sthree$-scope interactions. \bench{} provides a testbed for closing this $\sthree$ headroom. We hope it will drive vision-language models toward both stronger inference and more faithful realization of screenshot-implied interactions.

\clearpage  
\bibliographystyle{iclr2027_conference}
\bibliography{references}

\clearpage
\appendix
\section{Dataset construction details}
\label{app:dataset_details}

\paragraph{Source-pool filter (\S\ref{sec:benchmark_dataset}).}
The four-stage filter referenced in the main text applies the following per-stage criteria to each of the $4{,}254$ candidate repositories:
\begin{enumerate}
\itemsep0.05em
\item \textbf{Permissive license.} The project's upstream \texttt{LICENSE} file (not package-manager metadata) must be MIT, Apache-2.0, BSD, or ISC.
\item \textbf{Structural validity.} The repository exposes a recognizable mainstream front-end framework layout with detectable component and route directories.
\item \textbf{Buildability.} Dependency installation and the production build both complete within a $180$-second timeout on a clean environment.
\item \textbf{Authentication-wall detection.} Static heuristics over README/config files combined with HTTP-redirect probing of the served root.
\end{enumerate}
Repositories surviving all four stages enter expert review. The cumulative pass rate is $423/4{,}254 \approx 9.9\%$.

\paragraph{Screenshot capture pipeline.}
Captures use Playwright with headless Chromium at viewport $1440{\times}900$ and \texttt{device\_scale\_factor}{=}1. The four-stage timing waterfall referenced in \S\ref{sec:benchmark_dataset} is:
\begin{enumerate}
\itemsep0.05em
\item \texttt{domcontentloaded} plus a $3$-second hydration settle;
\item injection of dev-overlay suppression CSS (Next.js HUD, Vite, webpack);
\item \texttt{wait\_for\_function} polling for asset readiness ($6$-second timeout);
\item animation completion via \texttt{document.getAnimations()} ($3$-second timeout), followed by a $13$-selector loading-indicator detector with a $2$-second recheck.
\end{enumerate}
A per-route login-redirect detector complements the repo-level authentication-wall gate. Surviving screenshots are deduplicated by perceptual hashing on the content area ($32{\times}32$ average-hash, Hamming distance $<6\%$).

\section{\iis{} taxonomy, and inter-annotator agreement}
\label{app:iis_handbook}

\subsection{Category catalog}
\label{app:iis_categories}
The seven interaction families spanned by \iis{} and their typical scopes are listed below.
\begin{center}\small
\begin{tabular}{lll}
\toprule
\textit{Category} & \textit{Description} & \textit{Typical scope} \\
\midrule
C01 toggle & show/hide, style switch, on/off & \sone \\
C02 expand/collapse & accordion, dropdown, drawer & \sone \\
C03 list operations & search, filter, sort, paginate & \stwo \\
C04 data CRUD & create, read, update, delete & \stwo \\
C05 form validation & required fields, format checks & \sone \\
C06 notification & toast, banner, confirmation feedback & \sone \\
C07 cross-route state & state persistence across navigation & \sthree \\
\bottomrule
\end{tabular}
\end{center}

\subsection{State management and scope design}
In production front-end engineering, interactive state is managed at increasing levels of architectural scope. The cheapest is \emph{component-local state} (e.g., React \texttt{useState}/\texttt{useReducer}): state owned by a single component, bound to its mount lifecycle and not shared externally, such as whether a dropdown is open. When a piece of state is read or mutated by several sibling components, it must be \emph{lifted to a shared owner} (a parent component, a React Context, or a store) so that updates stay consistent across the subtree; typical cases are a list with an active filter, or a selection shared between a toolbar and a table. The most demanding tier is \emph{global state that must survive client-side route changes}: a single-page app re-mounts the route subtree on navigation, so any state that must persist (a cart, an auth session, a draft) cannot live in the page that gets unmounted and must instead live in a global store \emph{whose lifecycle is independent of routing} (a Context at the app root, Redux/Zustand/Jotai, or storage-backed). Implementation cost broadly increases along this hierarchy (a local handler, then cross-component state lifting, then a routing-independent global store), and the difficulty of \emph{inferring} the requirement from the screenshots alone rises with it.

We turn this hierarchy into the three-level scope axis used by \iis{}, with an operational classification rule. An interaction is \sone{} (UI-state) when its effect is confined to the immediate rendered subtree (e.g., a toggle); \stwo{} (data-state) when it performs a collection-level transformation (filter/sort/group/CRUD) that other components must observe; and \sthree{} (cross-route persistence) when its effect must outlive a route change (e.g., a cart that stays non-empty after navigating to \texttt{/checkout}).

Because implementation cost broadly rises across \sone{}${<}$\stwo{}${<}$\sthree{}, we weight each item by its scope with the linear ordinal weights $w_{\sone}{=}1$, $w_{\stwo}{=}2$, $w_{\sthree}{=}3$ in Eq.~\ref{eq:iis}, the simplest choice that respects this ordinal ranking without imposing an arbitrary curvature. The six-model \iis{} ordering is insensitive to this choice: recomputing model-level \iis{} under equal weights ($1{:}1{:}1$), the default ($1{:}2{:}3$), and an alternative ($1{:}2{:}4$) leaves the full ranking unchanged (Table~\ref{tab:iis_weight_robustness}; Kendall $\tau{=}1.0$ between every pair of weightings).

\begin{table}[h]
\centering\small
\caption{\textbf{\iis{} is insensitive to the scope-weight choice.} Model-level \iis{} recomputed under three weightings of $(\sone,\stwo,\sthree)$; the six-model ranking is identical, and Kendall $\tau{=}1.0$ between every pair of weightings. The $1{:}2{:}3$ column reproduces the headline \iis{} of Table~\ref{tab:main_results}.}
\label{tab:iis_weight_robustness}
\begin{tabular}{lrrr}
\toprule
Model & $1{:}1{:}1$ & $1{:}2{:}3$ (default) & $1{:}2{:}4$ \\
\midrule
Claude Sonnet 4.6 & $40.1$ & $36.8$ & $36.2$ \\
Kimi K2.5 & $23.0$ & $19.7$ & $19.1$ \\
Qwen3.5-397B-A17B & $18.5$ & $15.3$ & $15.0$ \\
Gemini 3.1 Pro Preview & $9.7$ & $8.1$ & $7.9$ \\
GPT-5.4 & $6.9$ & $6.9$ & $6.9$ \\
GLM-4.6V & $5.5$ & $4.6$ & $4.5$ \\
\bottomrule
\end{tabular}
\end{table}

\subsection{Outcome rubric and paired-merge interpolation}
\label{app:iis_outcome}
For each application, each interaction category is assessed through its constituent interaction sub-features present in the generated application. Annotators evaluate these sub-features and assign a category-level outcome: \emph{W} (working), all such sub-features within the category are realized; \emph{P} (partial), some but not all such sub-features are realized, or at least one is only partially realized; and \emph{F} (failed), none of these sub-features are realized. These map to the per-item score $r_i \in \{1, 0.5, 0\}$ used in Eq.~\ref{eq:iis}. Under the paired design (\S\ref{app:iis_handbook}, two annotators per cell), the cell's $r_i$ is set as follows: agreement uses the agreed value; a one-step disagreement (W/P or P/F) is merged by interpolation to $0.75$ or $0.25$, so that a partial implementation flagged by one annotator does not collapse to an extreme; a two-step disagreement (W/F) triggers arbitration by the third annotator, after which $r_i$ takes the arbitrated label.

\subsection{Inter-annotator agreement}
The 45-app human validation subset contains 315 interaction-category items per model and 1,890 across six models.
All generated applications in this validation subset were independently double-annotated, with a third annotator resolving disagreements. Inter-annotator agreement is assessed using paired annotations from the 45-application validation subset and summarized in Table~\ref{tab:iis_agreement}. $95\%$ bootstrap CIs are computed by resampling this subset with $B{=}1{,}000$, seed~$42$.

\begin{table}[h]
\centering\small
\caption{Inter-annotator agreement on the three IIS annotation fields. \texttt{gen\_appl} (the coverage axis) is binary so reports nominal $\alpha$. \texttt{gen\_result} (result) and \texttt{gen\_scope} (scope) are ordinal so report $\alpha_{\text{ordinal}}$.}
\label{tab:iis_agreement}
\begin{tabular}{llrr}
\toprule
Axis & Statistic & Value & 95\% CI \\
\midrule
\texttt{gen\_appl}        & $\alpha_{\text{nominal}}$ & $0.77$ & $[0.71,\,0.83]$ \\
\texttt{gen\_result}           & $\alpha_{\text{ordinal}}$                  & $0.72$           & $[0.66,\,0.78]$ \\
\texttt{gen\_scope}       & $\alpha_{\text{ordinal}}$                  & $0.84$           & $[0.79,\,0.88]$ \\
\bottomrule
\end{tabular}
\end{table}

\section{Provider-rejected projects}
\label{app:provider_rejection}

For GPT-5.4, one project was rejected by the provider before generation. It contributes zero to EXEC, VFS, and IIS in the 95-app main results. Excluding this project gives EXEC@3 $=79/94=84.0\%$ and VFS $=59.4$, versus $83.2\%$ and $58.8$ with zero-imputation. GPT-5.4 remains third on EXEC@3, between Gemini and Qwen. The 79 executable apps have conditional VFS $70.7$.

\section{Conditional metrics: \vfsstar{} and \iisstar}
\label{app:cond_vfs}

\paragraph{Two denominator conventions.}
\bench{} reports two complementary scores for \vfs{} and \iis{}. The headline scores in the Overall block of the main table average over the full curated set $N{=}95$ with zero-imputation: an application that fails to build after three self-debug rounds ($\exect[3]{=}\mathrm{F}$) contributes $0$ to the model's average, penalizing buildability and the underlying capability jointly. This is the most direct measure of end-to-end usability. The conditional variants \vfsstar{} and \iisstar{} restrict the average to the $N^{\star}$ projects that pass executability ($\exect[3]{=}\mathrm{P}$, i.e., the build and home-route render both succeed), isolating visual-language understanding and interaction-inference capability from buildability. We report both because the two factor different concerns into the same score:

\begin{itemize}
\item Headline (zero-imputation, $N{=}95$): is the rendered output of a complete model run usable on average?
\item Conditional (\vfsstar / \iisstar, $N^{\star}{\le}N$): given the model produced something at all, how visually faithful or interaction-faithful is it?
\end{itemize}

Table~\ref{tab:cond_vfs} reports \vfs{}/\vfsstar{} and \iis{}/\iisstar{} per model alongside their headline$\to$conditional gaps. The headline$\to$conditional gap is exactly the contribution of build failures to the headline score. The model with no build failures (Sonnet 4.6 at \exect[3]{=}100\%) has $\Delta{=}0$ by construction; high-build-failure models open a large gap. GLM-4.6V's $+29.9$-point gap on \vfs{} ($33.2 \to 63.0$) is the most extreme: it has substantial visual-language ability ($N^{\star}{=}50$ projects average to $63.0$), but only $52.6\%$ of its outputs build at all, so its end-to-end \vfs{} sits last. On \iis{}, the conditional score \iisstar similarly lifts GLM-4.6V from headline-last to mid-pack ($4.6 \to 8.7$). The top three preserve their order (Claude $36.8 \to 36.8$, Kimi $19.7 \to 25.6$, Qwen3.5 $15.3 \to 18.7$); the bottom three reshuffle because GLM's $50$-app build-pass subset yields a higher per-app IIS than its full-set headline, moving it past Gemini ($8.1 \to 8.3$) and GPT-5.4 ($6.9 \to 8.3$) on the conditional view. The top-three preservation suggests that the \iis{} gaps among the strongest models in Table~\ref{tab:main_results} reflect interaction-inference capability rather than executability alone.

\begin{table}[!ht]
\centering
\caption{\textbf{Conditional metrics \vfsstar{} and \iisstar.} Both restrict the average to projects with \exect[3]{=}pass; $N^{\star}$ is the pass-conditioned project count. The headline$\to$conditional gap separates executability from intrinsic capability. Best per column in \textbf{bold}, worst \underline{underlined}; the largest absolute $\Delta$ (most build-constrained model) is bolded.}
\label{tab:cond_vfs}
\small
\setlength{\tabcolsep}{4pt}
\begin{tabular}{lrrrrrrrrr}
\toprule
Model & $N$ & \exect[3] (\%) & $N^{\star}$ & \vfs{} & \vfsstar{} & $\Delta_{\vfs}$ & \iis{} & \iisstar & $\Delta_{\iis}$ \\
\midrule
Claude Sonnet 4.6 & 95 & 100.0 & 95 & 71.3 & 71.3 & +0.0 & \textbf{36.8} & \textbf{36.8} & +0.0 \\
Gemini 3.1 Pro Preview & 95 & 97.9 & 93 & \textbf{73.2} & \textbf{74.8} & +1.6 & 8.1 & \underline{8.3} & +0.2 \\
Kimi K2.5 & 95 & 76.8 & 73 & 55.5 & 72.2 & +16.7 & 19.7 & 25.6 & \textbf{+5.9} \\
GPT-5.4 & 95 & 83.2 & 79 & 58.8 & 70.7 & +11.9 & 6.9 & \underline{8.3} & +1.4 \\
Qwen3.5-397B-A17B & 95 & 82.1 & 78 & 58.5 & 71.3 & +12.8 & 15.3 & 18.7 & +3.3 \\
GLM-4.6V & 95 & 52.6 & 50 & \underline{33.2} & \underline{63.0} & \textbf{+29.9} & \underline{4.6} & 8.7 & +4.1 \\
\bottomrule
\end{tabular}
\end{table}

\paragraph{Choice for the main table.}
We headline \vfs{} and \iis{} (zero-imputation) in Table~\ref{tab:main_results} because end-to-end usability is the property the benchmark targets: a generation run that does not build is unusable to a real user, and its visual or interaction quality on hypothetical-builds-only is not actionable. \vfsstar{} and \iisstar{} are reported as diagnostics that isolate visual-language and interaction-inference capability from build success: they measure output faithfulness over only the projects a model builds, independent of how often it builds.

\section{Conditional visual sub-metric breakdown}
\label{app:cond_submetric}

The main table reports the four zero-imputed VFS components, whose average reconstructs VFS up to stored-score rounding. In the Overall block, Gemini leads all four components, while GLM is lowest. Table~\ref{tab:cond_submetric} instead reports per-page means over matched pages in each model's executable subset; it diagnoses visual components without using the main table's app-level weighting.

\begin{table}[!ht]
\centering
\caption{Conditional sub-metric breakdown. Per-page average over matched pages of \exect[3]-pass projects only; $n_{\text{pages}}$ varies per model and is reported alongside. Best per column in \textbf{bold} (ties shown in bold for both rows), worst \underline{underlined}. GLM's scores draw from 235 pages (50 build-pass apps), versus Claude's 566 pages (95 apps); this diagnostic uses page weighting rather than the app weighting of \vfsstar{}.}
\label{tab:cond_submetric}
\small
\begin{tabular}{lrrrrrr}
\toprule
Model & $N^{\star}$ & $n_{\text{pages}}$ & Size & Text & Position & Color \\
\midrule
Gemini 3.1 Pro Preview & 93 & 564 & \textbf{73.1} & \textbf{84.8} & \textbf{82.5} & 69.1 \\
Claude Sonnet 4.6 & 95 & 566 & 72.5 & 82.4 & 80.0 & \underline{66.3} \\
Kimi K2.5 & 73 & 419 & 72.5 & 83.6 & 80.2 & 68.8 \\
GPT-5.4 & 79 & 477 & 72.5 & 84.7 & \textbf{82.5} & \textbf{72.6} \\
Qwen3.5-397B-A17B & 78 & 452 & 72.2 & 82.1 & 79.3 & 70.3 \\
GLM-4.6V & 50 & 235 & \underline{67.3} & \underline{77.9} & \underline{74.5} & 70.6 \\
\bottomrule
\end{tabular}
\end{table}

The conditional view reveals two patterns invisible in the zero-imputation main table:
(i) \emph{Sub-skill gaps narrow once buildability is controlled for}: Position spread shrinks from $44.1$ points (zero-impute) to $8.0$ points (conditional), Color from $34.0$ to $6.3$. Visual sub-skill differences across the six models are several-fold smaller than overall \vfs{} differences would suggest.
(ii) \emph{GLM-4.6V's signature is strong palette plus weak spatial composition}: it is last on Size and Position but second on Color ($70.6$, behind GPT-5.4's $72.6$). Its relatively strong Color is conditional on a selected subset ($n_{\text{pages}}{=}235$ from $50$ build-pass apps). These model-specific page populations do not identify performance on hypothetical repaired outputs.

\section{VFS algorithm: DOM-alignment matching and cascade fallback}
\label{app:vfs}

\paragraph{Overview.}
\vfs{} is a fully judge-free metric: no VLM, LLM, or pixel-classifier is invoked at any stage. Given a reference application $\mathcal{R}$ and a model-generated application $\mathcal{G}$ on the same input set $\mathcal{I}{=}\{(s_i,\,r_i)\}_{i=1}^{n^{\text{tot}}}$ (screenshot $s_i$ at reference route $r_i$), \vfs{} is produced by the pipeline of Algorithm~\ref{alg:vfs}: (A) extract the route set of $\mathcal{G}$ from \texttt{src/App.tsx}; (B) boot the generated dev server and render each candidate route in headless Chromium under an auth/gate-dismissal protocol; (C) extract reference-side DOM text-blocks from $\mathcal{R}$, falling back to OCR on the input PNG when $\mathcal{R}$ cannot be reached at $r_i$ (auth-locked routes, missing dependencies); (D) match input screenshots to generated routes by \emph{global Hungarian assignment} on the $M{\times}N$ matrix of reference-vs-generated DOM-block alignment scores, with a path-token Jaccard tie-breaker and an absolute floor; (E) score the aligned-block set of each matched pair on four sub-metrics (Size, Text, Position, Color); (F) aggregate to page-, application-, and model-level. A legacy cascade $T_1$--$T_5^b$ matcher is retained as a fallback for the rare case where $\mathcal{S}_{\text{ref}}$ fails to start (\S\ref{app:vfs_cascade}). All hyperparameters are held constant across all models and applications. Specific values are stated inline at each point of use.

\paragraph{Auth and gate handling.}
React applications in \bench{} routinely guard non-public routes behind authentication or splash gates that, if not dismissed, cause every captured route to render the same login/intro view. We apply three source-derived mitigations on the generated dev server (no per-app human configuration), invoked at Algorithm~\ref{alg:vfs} lines 2--4: (i) \textbf{\texttt{localStorage} auth seeding}: scan \texttt{src/**/*.[jt]s*} for storage call sites whose key matches the regex \texttt{/(auth|session|login|signin|loggedin|token|user)/i} and inject a permissive auth payload via Playwright \texttt{add\_init\_script}; (ii) \textbf{in-memory auth form-fill}: when (i) finds no localStorage hooks but a \texttt{useState(false)} auth flag exists, navigate to \texttt{/login}, fill synthetic credentials, submit, and switch subsequent navigation to \texttt{pushState} so React state survives; (iii) \textbf{splash/intro gate dismissal}: AST-detect the \texttt{useState(false) + early return} pattern in \texttt{App.tsx}, regex-extract the dismiss-button label, and dismiss on every non-\texttt{/} route (the home is captured with the gate visible to match the reference). The same localStorage seeding is also applied to $\mathcal{S}_{\text{ref}}$; routes that seeding cannot bypass fall through to OCR.

\paragraph{Render-validity probe.}
For each rendered route we evaluate a four-signal probe on \texttt{\#root}: \texttt{all} (descendant element count), \texttt{text} (\texttt{innerText} length), \texttt{visuals} (\texttt{svg/img/canvas/video} count), and \texttt{interactive} (\texttt{button/input/textarea/select/a[href]/[role="button"]} count). The route is accepted as rendered if \texttt{all}${\ge}3$ \emph{and} at least one of: \texttt{text}${\ge}20$ chars, \texttt{visuals}${\ge}2$, or \texttt{interactive}${\ge}2$. The any-of disjunction (rather than a stricter conjunction such as ``$\texttt{text}{>}10\,\wedge\,\texttt{all}{\ge}5$'') admits legitimate splash/CTA pages with shallow DOM but meaningful content (e.g., a \texttt{<div><button>Skip</button><p>...</p></div>} Intro component). Routes that fail the probe are recorded as \texttt{None} and become MISS candidates for the matcher.

\paragraph{Route-alias deduplication.}
After each successful screenshot we hash the page's \texttt{innerText}+child-count signature and consult a per-app \texttt{seen\_dom\_hashes} table. If the hash recurs we record the new route as an \emph{alias} of the canonical route and reuse the canonical screenshot rather than writing a duplicate. This collapses redirects (\texttt{<Route path="/" element={<Navigate to="/foo">}>}) and HashRouter mirror routes that would otherwise inflate $n^{\text{tot}}$ and double-count gen content.

\paragraph{DOM-alignment matching (default path).}
We extract DOM text-blocks for every rendered gen route (cached as $\mathcal{B}^{\text{gen}}_{\hat r}$) and, on the reference dev server, for every input route (cached as $\mathcal{B}^{\text{ref}}_{s_i}$). When the reference DOM at $r_i$ comes back empty (route auth-locked, dependency missing, build/dev-server failure on $\mathcal{R}$ at this route), we substitute OCR-derived blocks: Tesseract on the input PNG produces line-level boxes which are normalized to the same $\{\,\texttt{text},\,\texttt{bbox},\,\texttt{fg\_color},\,\texttt{tag}\,\}$ schema as the DOM extractor (Color is degraded for OCR-derived blocks since \texttt{fg\_color} defaults to black). Let $M$ be the number of inputs and $N$ the number of rendered gen routes. We build the similarity matrix
\[
\Sigma_{ij}\;=\;\mathrm{compute\_l3\_page}(\mathcal{B}^{\text{ref}}_{s_i},\,\mathcal{B}^{\text{gen}}_{\hat r_j})\,/\,100,
\qquad
\Sigma'_{ij}\;=\;\min\!\bigl(1,\;\Sigma_{ij}+\beta\cdot\mathrm{Jacc}(\tau(r_i),\tau(\hat r_j))\bigr),
\]
where $\mathrm{compute\_l3\_page}$ is the four-sub-metric scorer (\S\textit{Sub-metrics} below), $\tau(\cdot)$ tokenizes a path on \texttt{[-\_/]+}, and $\mathrm{Jacc}$ is set-Jaccard. The token bonus $\beta{=}0.08$ acts as a tie-breaker when DOM-alignment scores are within ${\sim}0.05$ of each other (e.g., gen \texttt{/leads} beats gen \texttt{/settings/team} for the input labeled \texttt{leads}, both candidates at $\Sigma{\approx}0.55$), without overriding strong DOM evidence. We solve the assignment globally with the Hungarian algorithm (\texttt{scipy.optimize.linear\_sum\_assignment}) on cost $C{=}1{-}\Sigma'$ padded to the square size $\max(M,N)$ with cost~$1$, then drop pairs whose underlying $\Sigma_{ij}$ falls below the absolute floor $\theta_{\text{DOM}}{=}0.20$ (the input becomes MISS and contributes $0$ to $n^{\text{matched}}$). Each gen route is used at most once across the assignment. This hard cap is equivalent to setting $K_{\max}{=}1$ (the same parameter that the cascade fallback uses, see \S\ref{app:vfs_cascade}) and prevents a single dynamic route (e.g., a \texttt{/post/:id} render) from absorbing three to five unrelated inputs and over-counting matched routes.

\paragraph{DOM block extraction.}
We navigate Playwright to the route under evaluation, wait until \texttt{domcontentloaded}, dismiss any detected gate (only on non-\texttt{/} routes; see auth-and-gate handling above), and idle for $\delta_{\text{ref}}{=}4{,}000$\,ms (reference) or $\delta_{\text{gen}}{=}3{,}000$\,ms (generated). Both renderings use viewport $1440{\times}900$ to match the reference capture. We then run an in-page \texttt{TreeWalker} over text nodes and emit one block per visible parent element with: (i) lower-cased text trimmed to $200$ characters; (ii) bounding box normalized to viewport so that $(x,y,w,h)\in[0,1]^2{\times}[0,1]^2$; (iii) the foreground color from \texttt{getComputedStyle().color}; (iv) the parent's tag name. Non-rendering elements (\texttt{display:none}, \texttt{visibility:hidden}, opacity $0$), boxes smaller than $2{\times}2$ pixels, content more than three viewports below the fold (pixel-space $y_{\text{px}}{>}3v_h$ where $v_h{=}900$\,px is the viewport height), and text strings longer than $500$ characters are filtered out before the bbox is normalized to $[0,1]^2$. Adjacent same-row same-tag fragments are merged when within $0.02$ normalized units, suppressing the spurious split of inline-styled text into multiple blocks.

\paragraph{Block alignment within a pair.}
Let $B^{\text{ref}}{=}\{b_i^{\text{ref}}\}_{i=1}^{n}$ and $B^{\text{gen}}{=}\{b_j^{\text{gen}}\}_{j=1}^{m}$. We build the cost matrix $C\in\mathbb{R}^{n\times m}$ with $C_{ij}{=}1{-}\rho(t_i^{\text{ref}},t_j^{\text{gen}})$ where $\rho{=}\,$\texttt{difflib.SequenceMatcher.ratio} is the Ratcliff--Obershelp similarity (longest common contiguous matching) $\rho(a,b){=}2L/(|a|{+}|b|)$ with $L$ the matched-character count. We pad $C$ to the square size $\max(n,m)$ with cost $1$, solve $(\sigma^*,\pi^*){=}\argmin\sum C_{\sigma\pi}$ via the Hungarian algorithm (\texttt{scipy.optimize.linear\_sum\_assignment}), and retain only assignments whose $\rho{\ge}\theta_{\text{text}}{=}0.3$. The retained assignment set $A$ is the basis for all four sub-metrics.

\paragraph{Sub-metrics.}
For a pair $p{=}(b^{\text{ref}}_p,b^{\text{gen}}_p)$ let $a(b)$ denote area, $c(b){=}(x{+}w/2,\,y{+}h/2)$ the center, $\mathrm{fg}(b)$ the foreground color, and $\Delta E_{00}$ the CIEDE2000 color difference computed in CIE Lab via \texttt{skimage.color}. The four sub-metrics are
\begin{align*}
SIZE\;&=\;\min\!\Bigg(1,\;\frac{\sum_{p\in A}\tfrac{1}{2}\!\left(a(b_p^{\text{ref}}){+}a(b_p^{\text{gen}})\right)}{\tfrac{1}{2}\!\left(\sum_{i}a(b_i^{\text{ref}}){+}\sum_{j}a(b_j^{\text{gen}})\right)}\Bigg),\\[2pt]
TEXT\;&=\;\frac{1}{|A|}\sum_{p\in A}\rho\bigl(t_p^{\text{ref}},t_p^{\text{gen}}\bigr),\\[2pt]
POSITION\;&=\;\frac{1}{|A|}\sum_{p\in A}\max\!\Bigl(0,\;1{-}\max\!\bigl(|\Delta c_x|_p,\,|\Delta c_y|_p\bigr)\Bigr),\\[2pt]
COLOR\;&=\;\frac{1}{|A|}\sum_{p\in A}\max\!\Bigl(0,\;1{-}\Delta E_{00}\!\bigl(\mathrm{fg}(b_p^{\text{ref}}),\mathrm{fg}(b_p^{\text{gen}})\bigr)/100\Bigr).
\end{align*}
SIZE measures matched-block area coverage. We use Chebyshev (max-norm) displacement for Position rather than Euclidean because the four-metric average should treat horizontal and vertical drift symmetrically while remaining bounded in $[0,1]$ without an additional normalization constant. CIEDE2000 is divided by $100$ since saturated complementary colors produce $\Delta E_{00}{\approx}100$ in the Lab gamut. Each sub-metric defaults to $0$ if $|A|{=}0$ (no aligned blocks).

\paragraph{Aggregation.}
The page-, application-, and model-level scores are
\begin{equation*}
\ell^{\text{page}}_p \;=\; 25\,(SIZE{+}TEXT{+}POSITION{+}COLOR),
\quad
L^{\text{matched}}_{3,a} \;=\; \frac{1}{|P_a|}\sum_{p\in P_a} \ell^{\text{page}}_p,
\end{equation*}
\begin{equation}
\vfs_{a,m}\;=\;L^{\text{matched}}_{3,a,m}\cdot\frac{n^{\text{matched}}_{a,m}}{n^{\text{tot}}_{a}},
\qquad
\vfs_m\;=\;\frac{1}{N_{\text{app}}}\sum_{a=1}^{N_{\text{app}}}\vfs_{a,m},
\label{eq:vfs}
\end{equation}
with $N_{\text{app}}{=}95$ (the curated app count; the per-app local $N$ in Algorithm~\ref{alg:vfs} denotes the number of rendered gen routes and is unrelated). The coverage factor $n^{\text{matched}}/n^{\text{tot}}$ penalizes models that achieve a high per-page score on a small subset of routes. For example, $1/12$ matched at $L^{\text{matched}}_{3}{=}31.7$ degrades to $\vfs{}{=}2.6$ rather than retaining the per-page score. Consistent with the scoring policy of \S\ref{sec:pipeline}, applications with $\exect[3]{=}\mathrm{F}$ are zero-imputed at the application level. The conditional variant \vfsstar{} (Appendix~\ref{app:cond_vfs}) drops the zero-imputation, averaging only over $\exect[3]{=}\mathrm{P}$ applications.

\begin{algorithm}[t]
\caption{Computation of \vfs{} for a single (application, model) pair (DOM-alignment default; cascade fallback in \S\ref{app:vfs_cascade}).}
\label{alg:vfs}
\small
\begin{algorithmic}[1]
\Require Input set $\mathcal{I}{=}\{(s_i,r_i)\}_{i=1}^{n^{\text{tot}}}$; reference source $\mathcal{R}$; generated source $\mathcal{G}$.
\State $\mathcal{T} \gets$ \Call{ExtractRoutes}{$\mathcal{G}/\texttt{src/App.tsx}$} \Comment{indent-aware nested-\texttt{<Route>} parser; concretize \texttt{:id}$\to$\texttt{1}}
\State $g \gets$ \Call{DetectGates}{$\mathcal{G}$};\quad $\sigma_g \gets$ \Call{DetectLocalStorageSeeds}{$\mathcal{G}$};\quad $\iota \gets$ \Call{DetectInMemoryAuth}{$\mathcal{G}$} when $\sigma_g{=}\emptyset$
\State $\sigma_r \gets$ \Call{DetectLocalStorageSeeds}{$\mathcal{R}$}
\State $\mathcal{S}_{\text{gen}} \gets$ \Call{StartDevServer}{$\mathcal{G}$};\quad apply seeds $\sigma_g$ via \texttt{add\_init\_script}
\State $H_{\text{gen}},\,\mathcal{B}^{\text{gen}},\,\mathrm{aliases},\,\mathrm{hashes} \gets \emptyset,\,\emptyset,\,\emptyset,\,\emptyset$
\If{$\iota$} capture pre-auth Login at \texttt{\_\_login\_\_}; \Call{FormFillLogin}{$\mathcal{S}_{\text{gen}}$}; switch to SPA \texttt{pushState}
\EndIf
\For{$\hat r \in \mathrm{sort}(\mathcal{T})$ \Comment{\texttt{/} first, then by $|\hat r.\texttt{url}|$}}
  \State navigate $\mathcal{S}_{\text{gen}}{\to}\hat r.\texttt{url}$;\, if $\hat r.\texttt{url}{\ne}/$ \textbf{and} $g.\mathrm{has\_gate}$ \textbf{then} \Call{TryDismiss}{$\mathcal{S}_{\text{gen}}, g.\mathrm{selectors}$}
  \State probe $\gets \{\texttt{all},\texttt{text},\texttt{visuals},\texttt{interactive}\}$ on \texttt{\#root}
  \If{\textbf{not} \Call{IsRendered}{probe}} $H_{\text{gen}}[\hat r]\gets$\texttt{None}; \textbf{continue} \Comment{$\texttt{all}{\ge}3 \,\wedge\, (\texttt{text}{\ge}20 \vee \texttt{visuals}{\ge}2 \vee \texttt{interactive}{\ge}2)$}
  \EndIf
  \State $h\gets\mathrm{md5}(\texttt{innerText}\,|\,\texttt{all})$;\, \textbf{if} $h\in\mathrm{hashes}$ \textbf{then} $\mathrm{aliases}[\hat r]\gets\mathrm{canon}(h)$; $H_{\text{gen}}[\hat r]\gets H_{\text{gen}}[\mathrm{canon}(h)]$; \textbf{continue}
  \State $H_{\text{gen}}[\hat r]\gets$\Call{Screenshot}{$\mathcal{S}_{\text{gen}}$};\, $\mathcal{B}^{\text{gen}}_{\hat r}\gets$\Call{ExtractBlocks}{$\mathcal{S}_{\text{gen}}$};\, $\mathrm{hashes}[h]\gets\hat r$
\EndFor
\If{$\sigma_g{\ne}\emptyset$ \textbf{and} \texttt{\_\_login\_\_}${\notin}H_{\text{gen}}$} render \texttt{/} in a \emph{fresh non-seeded} context, save as \texttt{\_\_login\_\_}
\EndIf
\State \Call{StopDevServer}{$\mathcal{S}_{\text{gen}}$}
\State $\mathcal{S}_{\text{ref}}\gets$\Call{StartDevServer}{$\mathcal{R}$};\, apply seeds $\sigma_r$;\, \textbf{if} fail \textbf{then} \textbf{goto} \emph{cascade fallback} (\S\ref{app:vfs_cascade})
\State $\mathcal{B}^{\text{ref}}\gets\emptyset$
\For{$(s_i,r_i)\in\mathcal{I}$}
  \State $b\gets$\Call{ExtractBlocks}{$\mathcal{S}_{\text{ref}}, r_i$ with $g$-aware dismissal}
  \State \textbf{if} $b{=}\emptyset$ \textbf{then} $b\gets$\Call{OCRBlocks}{$s_i$} \Comment{Tesseract on input PNG; COLOR degraded to placeholder}
  \State $\mathcal{B}^{\text{ref}}_{s_i}\gets b$
\EndFor
\State \Call{StopDevServer}{$\mathcal{S}_{\text{ref}}$}
\State $\Sigma\in\mathbb{R}^{M\times N}$ with $\Sigma_{ij}\gets\mathrm{compute\_l3\_page}(\mathcal{B}^{\text{ref}}_{s_i},\mathcal{B}^{\text{gen}}_{\hat r_j})/100$ (rendered routes only)
\State $\Sigma'_{ij}\gets\min(1,\,\Sigma_{ij}+\beta\cdot\mathrm{Jacc}(\tau(r_i),\tau(\hat r_j)))$ with $\beta{=}0.08$
\State $(\sigma^{\!*}\!,\pi^{\!*})\gets$\Call{Hungarian}{$1{-}\Sigma'$} on the $\max(M,N)$-square padded matrix (pad cost $1$)
\State $P\gets\{(s_i,\hat r_{j^*}):\,\Sigma_{ij^*}\ge\theta_{\text{DOM}}{=}0.20\}$;\, $n^{\text{matched}}\gets|P|$ \Comment{each $\hat r$ used at most once}
\State $\mathcal{L}\gets[\,]$
\For{$(s_i,\hat r^*)\in P$}
  \State $A\gets$\Call{Hungarian}{$\mathcal{B}^{\text{ref}}_{s_i},\mathcal{B}^{\text{gen}}_{\hat r^*};\,\theta_{\text{text}}{=}0.3$} \Comment{within-pair text-dissimilarity alignment}
  \State $\ell^{\text{page}}\gets 25\,(SIZE{+}TEXT{+}POSITION{+}COLOR)$;\quad \textbf{push} $\ell^{\text{page}}$ to $\mathcal{L}$ \Comment{Eq.~(\ref{eq:vfs})}
\EndFor
\State $L^{\text{matched}}_{3}\gets\mathrm{mean}(\mathcal{L})$;\quad $\vfs_{a,m}\gets L^{\text{matched}}_{3}\cdot n^{\text{matched}}/n^{\text{tot}}$
\State \Return $\vfs_{a,m}$
\end{algorithmic}
\end{algorithm}

\subsection{Cascade fallback ($T_1$--$T_5^b$, used only when $\mathcal{S}_{\text{ref}}$ fails to start)}
\label{app:vfs_cascade}

When the reference dev server $\mathcal{S}_{\text{ref}}$ cannot be started for an application (missing \texttt{node\_modules}, install timeout, port collision under heavy parallelism), the DOM-alignment matrix in Algorithm~\ref{alg:vfs} is undefined and we fall back to a path-name + visual cascade. Each input is matched to at most one rendered gen route ($K_{\max}{=}1$, including dynamic routes) and a hit at tier $T_t$ skips $T_{t+1},\dots,T_5^b$.
\begin{itemize}\itemsep1pt
\item[$T_1$] \textbf{Exact route match.} The reference route $r_i$ (lower-cased, slashes/query stripped) equals the generated route's literal definition or its concretized URL.
\item[$T_2$] \textbf{Dynamic-route regex match.} Generated routes containing path parameters (\texttt{:slug}, \texttt{:id}) are converted to \texttt{[\^{}/]+} regex. Catches \texttt{/product/1} $\leftrightarrow$ \texttt{/product/:id}.
\item[$T_3$] \textbf{Path-suffix match.} Largest common trailing-segment overlap $k{\ge}1$. Admits \texttt{/dashboard} $\leftrightarrow$ \texttt{/app/dashboard}.
\item[$T_4$] \textbf{Synonym structural match.} Tokenize on \texttt{[-\_/]} and expand by a fixed table of $32$ synonym groups (e.g.\ \{\texttt{home, landing, index, overview}\}; \{\texttt{contact, support, etc, misc, links, elsewhere}\}; \{\texttt{signin, signup, forgot, otp, verify}\}); a hit requires non-empty intersection with at least one cross-direction unexpanded-vs-expanded overlap.
\item[$T_5^a$] \textbf{Visual-first phash Hungarian.} Build the $M{\times}N$ \texttt{phash}-similarity matrix on $16{\times}16$ perceptual hashes of input PNG vs gen screenshot, solve Hungarian on $1-\mathrm{sim}$, and accept pairs with $\mathrm{sim}{\ge}\theta_{\text{phash}}{=}0.45$.
\item[$T_5^b$] \textbf{Visual fallback (legacy SSIM+hist).} For inputs unmatched after $T_1$--$T_5^a$, compute $0.7\,\mathrm{SSIM}+0.3\,\mathrm{HistCorr}$ on $256{\times}160$ resized RGB; accept the best candidate \emph{only if} it satisfies both the relative margin $\mathrm{best}{\ge}1.2\bar\mu$ and the absolute floor $\theta_{\text{vis}}^{\text{abs}}{=}0.55$. The absolute floor blocks ``least bad'' assignments from a uniformly poor candidate set. For instance, it rejects a \texttt{zen-keyboard} screenshot pairing to \texttt{/player} at SSIM+hist${\approx}0.45$.
\end{itemize}

\paragraph{Implementation notes and known limitations.}
The DOM-alignment default is robust to \emph{route renaming} (gen \texttt{/} that actually serves Blog content correctly pairs with the reference's \texttt{/blog} input via DOM evidence rather than path string) and rejects \emph{path-name false positives} (gen \texttt{/} that serves Blog cannot pair with a reference \texttt{/} that serves a music landing). The token-Jaccard bonus $\beta{=}0.08$ is small enough that DOM evidence dominates whenever it is decisive, and large enough to break ties in the noisy $0.5$--$0.65$ similarity band typical of dark-theme/icon-heavy apps. The OCR fallback for $\mathcal{B}^{\text{ref}}$ degrades Color (no real foreground color information) but preserves Size, Text, and Position. This is the maximal signal recoverable for routes the reference dev server cannot reach. Because input screenshots in \bench{} are URL-reachable static routes (\S\ref{sec:benchmark_dataset}), the pipeline does not replay click sequences before \texttt{ExtractBlocks}. Interaction-state fidelity is captured by \iis{} (\S\ref{sec:iis}) rather than \vfs{}. We report \vfs{} as observed under this protocol. An interaction-aware variant is left to future work.

\begin{table}[h]
\centering
\caption{\textbf{Application-category composition by subcategory.} The four application categories resolve into the subcategories listed below, spanning the realized dataset diversity. Subcategories are descriptive and not used as analytical strata.}
\label{tab:category_summary}
\small
\begin{tabular}{lrp{0.62\textwidth}}
\toprule
Category & Share & Subcategories \\
\midrule
Content & $24.2\%$ & Personal Sites, Publishing \& Docs \\
Commerce & $25.3\%$ & Marketing Sites, Online Stores, Markets \& Booking \\
Admin & $31.6\%$ & Business Management, Analytics \& Monitoring, General Administration \\
Specialty & $18.9\%$ & Integrated Workspaces, Task-focused Apps \\
\bottomrule
\end{tabular}
\end{table}

\begin{table}[!ht]
\centering\scriptsize
\caption{IIS by application category. Mean and spread summarize the six model means.}
\label{tab:category_iis}
\setlength{\tabcolsep}{4pt}
\begin{tabular}{lrrrrrrrrr}
\toprule
Category & Apps & Mean & Spread & Claude & Kimi & Qwen & Gemini & GPT & GLM \\
\midrule
Content & 23 & 26.1 & 54.7 & 57.3 & 37.2 & 29.4 & 13.7 & 16.3 & 2.6 \\
Commerce & 24 & 15.4 & 28.9 & 36.7 & 12.5 & 15.0 & 10.2 & 7.8 & 10.3 \\
Admin & 30 & 8.6 & 20.3 & 21.3 & 13.9 & 6.9 & 4.9 & 1.1 & 3.5 \\
Specialty & 18 & 12.3 & 35.3 & 36.7 & 16.5 & 11.9 & 3.7 & 3.5 & 1.4 \\
\bottomrule
\end{tabular}
\end{table}

\section{Failure taxonomy details}
\label{app:failures}

This appendix expands the failure analysis of §\ref{sec:exp_failures} into a breakdown across the benchmark's three scoring axes. We first decompose the $101$ \exect[3] build/runtime failures into a nine-mode taxonomy (Table~\ref{tab:failure_modes}) with per-model incidence, prompt-clause attribution, and representative code anchors, then trace how the dominant mode shifts along the Qwen2.5-VL size ladder. We next specify the fixed self-debug feedback protocol under which \exect[3] is measured, and the recurring patterns that reduce \vfs{} and \iis{}. A closing set of case studies illustrates the latent-affordance inference that \iis{} targets. Counts are over the $570$ nominal (app, model) generation pairs spanning $95$ apps and $6$ baselines; one additional (GPT-5.4, app) pair rejected by the vendor's content-safety filter is treated separately (Appendix~\ref{app:provider_rejection}) and is excluded from these $101$ failures.

\subsection{Generation-level failures}
\label{app:failures-gen}

The $101$ \exect[3] failures decompose into nine fine-grained modes (Table~\ref{tab:failure_modes}), which collapse into five high-level causes: hallucinated imports (C1, C6); scaffold-file overwrites (C2, C8); undeclared / wrong-name source dependencies (C3); broken import paths (C4, C5); and parse-level or runtime errors (C7, C9).

\begin{table}[!ht]
\centering
\caption{\textbf{Build-failure modes across $101$ \exect[3] failures.} The \emph{Share} column gives the mode's percentage of all $101$ failures; per-model columns give the mode's incidence as a percentage of that model's $95$ apps. ``IFV'' marks instruction-following violations against the four explicit plan-step prompt constraints (Appendix~\ref{app:failures-gen}). Models: \textit{C}=Claude Sonnet~4.6, \textit{Ge}=Gemini~3.1 Pro Preview, \textit{Gp}=GPT-5.4, \textit{K}=Kimi K2.5, \textit{Q}=Qwen3.5-397B-A17B, \textit{GL}=GLM-4.6V. One provider rejection is excluded from attribution and scored zero in the main results. Percentages are rounded to the nearest integer, so the \emph{Share} column need not sum to exactly $100\%$ and a model's per-mode entries need not sum exactly to its \emph{Failure rate}.}
\label{tab:failure_modes}
\small
\setlength{\tabcolsep}{3.9pt}
\begin{tabular}{@{}llrcrrrrrr@{}}
\toprule
ID & Failure mode & Share & IFV & C & Ge & Gp & K & Q & GL \\
\midrule
C1 & Hallucinated icon export (\texttt{lucide-react}) & $35\%$ &  & 0 & $1\%$ & $11\%$ & $11\%$ & $11\%$ & $4\%$ \\
C2 & Scaffold \texttt{package.json} overwritten & $25\%$ & \checkmark & 0 & $1\%$ & 0 & $1\%$ & $4\%$ & $20\%$ \\
C3 & Pkg used but not in \texttt{extra\_dependencies} & $11\%$ & \checkmark & 0 & 0 & 0 & $3\%$ & $3\%$ & $5\%$ \\
C4 & \texttt{@/*} path alias used & $2\%$ & \checkmark & 0 & 0 & 0 & 0 & 0 & $2\%$ \\
C5 & Wrong relative path & $5\%$ &  & 0 & 0 & 0 & 0 & 0 & $5\%$ \\
C6 & Cross-file import/export mismatch & $6\%$ &  & 0 & 0 & 0 & $1\%$ & 0 & $5\%$ \\
C7 & JSX / TypeScript parse error & $10\%$ &  & 0 & 0 & $1\%$ & $6\%$ & 0 & $3\%$ \\
C8 & Scaffold \texttt{src/index.css} overwritten & $1\%$ & \checkmark & 0 & 0 & 0 & 0 & 0 & $1\%$ \\
C9 & Runtime API misuse & $6\%$ &  & 0 & 0 & $4\%$ & $1\%$ & 0 & $1\%$ \\
\midrule
\multicolumn{3}{@{}l}{\emph{Failure rate} (failures / $95$ apps)} & --- & \textbf{0} & 2\% & 16\% & 23\% & 18\% & \underline{47\%} \\
\multicolumn{3}{@{}l}{\emph{ICS} $= 1 - n_{\mathrm{IFV}} / n_{\mathrm{apps}}$} & --- & \textbf{100\%} & 99\% & \textbf{100\%} & 96\% & 93\% & \underline{72\%} \\
\bottomrule
\end{tabular}
\end{table}

\paragraph{C1: hallucinated \texttt{lucide-react} icon export ($34.7\%$ of failures).}
The model imports a symbol not exported by the scaffold-pinned \texttt{lucide-react@\^{}0.460.0}: typically a brand icon (\texttt{Tiktok}, \texttt{Spotify}, \texttt{Weibo}, \texttt{Alipay}, etc.), a future-renamed name (\texttt{Funnel}, the post-\texttt{0.460} canonical name introduced when the icon previously exported as \texttt{Filter} was renamed), or a foreign-library identifier (\texttt{DashboardOutlined} from Ant Design). Anchor: GPT-5.4 on \texttt{55\_shadcn-nextjs-dashboard} writes \texttt{import \{ Funnel \} from "lucide-react"}, but in \texttt{0.460.0} the icon is still exported as \texttt{Filter}.

\paragraph{C2: scaffold \texttt{package.json} overwritten ($24.8\%$ of failures).}
The model emits a \texttt{package.json} in violation of the ``Do NOT include scaffold files'' clause, pinning unpublished versions or inventing package names (e.g., \texttt{mantine} as a top-level package; registry name is \texttt{@mantine/core}). \texttt{pnpm install} aborts with \texttt{ERR\_PNPM\_NO\_MATCHING\_VERSION} / \texttt{ERR\_PNPM\_NO\_VERSIONS}. Anchor: GLM-4.6V on \texttt{60\_mantine-dashboard}.

\paragraph{C3: source uses pkg not in \texttt{extra\_dependencies} ($10.9\%$).}
The plan-step prompt requires a structured \texttt{extra\_dependencies} field for non-scaffold packages, but the source imports a package neither pinned in the scaffold nor declared. Declaring non-scaffold dependencies is the model's responsibility under the plan-step contract, and the Self-Debug stage does not backfill the missing declaration by scanning generated source. Anchor: GLM-4.6V on \texttt{54\_slash-admin} declares \texttt{react-calendar} but imports \texttt{react-big-calendar}.

\paragraph{C4: \texttt{@/*} path alias used ($2.0\%$).}
The model writes shadcn-style imports (\texttt{import X from "@/components/X"}) although the prompt requires relative imports and the scaffold's \texttt{tsconfig.json} declares no such alias.

\paragraph{C6: cross-file export name mismatch ($5.9\%$).}
The importer references an identifier whose source declaration uses a different name (default vs.\ named, casing mismatch, or omitted altogether). Anchor: GLM-4.6V on \texttt{38\_dexkit-nft-marketplace} uses \texttt{<CollectionCard>} at \texttt{Home.tsx:31} without a matching \texttt{import}. \texttt{vite build} passes (esbuild does not type-check undefined identifiers) but the home route throws \texttt{ReferenceError} on first render, failing \exect{}'s home-route render gate (\S\ref{sec:metrics}).

\paragraph{Long-tail modes C5, C7, C8, C9 ($21.8\%$ combined).}
C5 (wrong relative path): file-root-relative specifier instead of \texttt{../}. C7 (JSX/TS parse error): unbalanced closing tag, stray \texttt{>}. C8 (scaffold \texttt{src/index.css} overwritten): \texttt{@layer base} without \texttt{@tailwind base}. C9 (runtime API misuse): GPT-5.4 on \texttt{18\_myPortfolio} destructures a non-existent \texttt{location} field from the \texttt{<NavLink className=\{...\}>} render-prop, which in \texttt{react-router-dom@6} provides \texttt{\{ isActive, isPending, isTransitioning \}}.

\paragraph{Per-model signature and prompt-clause attribution.}
Per-model totals (Table~\ref{tab:failure_modes}) reveal distinct signatures: Claude produces zero failures, while Gemini has two; GPT-5.4 ($15.8\%$ of its $95$ apps) is dominated by C1 ($66.7\%$ of its failures, all factuality); GLM-4.6V ($47.4\%$ of its apps) is dominated by scaffold-respect violations (C2+C3+C4+C8 $60.0\%$). The plan-step prompt (Appendix~\ref{app:prompts}) states four mechanically-checkable constraints: (i) source files under \texttt{src/}; (ii) relative imports; (iii) every non-scaffold import declared in \texttt{extra\_dependencies}; (iv) no regeneration of scaffold files. Of the 101 failures, 39 ($38.6\%$) are instruction-following violations (IFV column of Table~\ref{tab:failure_modes}): clause (iv) accounts for $66.7\%$ of IFVs via C2 + C8, clause (iii) for $28.2\%$ (C3), clause (ii) for $5.1\%$ (C4), and clause (i) has zero violations. The remaining 62 failures (C1, C5, C6, C7, C9) are coding-quality or factuality errors not pre-empted by the prompt. The Instruction-Compliance Score (ICS) row shows the resulting model-level split.

\subsection{Per-size failure-mode shifts on the Qwen2.5-VL ladder}
\label{app:failures-scaling}
On the 45-app scaling subset, Figure~\ref{fig:qwen_scaling_failures} composes the build-failure category at each size, supporting the dominant-shift narrative in §\ref{sec:exp_scaling}: the dominant category shifts from \emph{syntax} (3B) to \emph{inconsistency} (7B) to a mixed \emph{syntax}~$+$~\emph{hallucination} regime (32B). The post-transition $72$B residual is dominated by \emph{syntax} at deeper build stages plus the figure's \emph{other} bucket, here \emph{runtime} errors (e.g., importing TypeScript-only types as values). These four buckets correspond to the C1--C9 taxonomy of \S\ref{app:failures-gen} as: \emph{syntax}$\supseteq$C7, \emph{inconsistency}$\supseteq$\{C2,C3,C4,C5,C8\}, \emph{hallucination}$\supseteq$\{C1,C6\}, \emph{other}$\supseteq$C9. We caution that the breakdown at $\leq 32$B is over a near-fully-failing population ($\geq 44/45$ failures per size), so the precise within-size category proportions are noisy; the identity of the dominant (mode) category, taken over $44$--$45$ failures, is well determined and is what the cross-size dominant-shift claim relies on.

\subsection{Self-debug feedback protocol}
\label{app:failures-pipeline}

\exect[3] is measured under a Self-Debug stage configured identically for every model: each failing app is retried for up to three rounds with the same model, and each round forwards the toolchain's own diagnostics back to the model for repair. \exect[3] therefore reflects iterate-and-fix ability under this fixed feedback protocol rather than the model in isolation; how reliably a model localizes and repairs the root cause from standard build output is part of what it measures, with weaker self-debuggers returning \texttt{NO\_FILE}.

\subsection{Visual-fidelity failures}
\label{app:failures-nav-vis}

Among $\exect[3]$-pass apps, the sub-skill view is best read off the conditional breakdown Table~\ref{tab:cond_submetric} (the zero-imputation columns of Table~\ref{tab:main_results} are dominated by build-success differences, not sub-skill). Conditional spreads are tight: Color $66.3$--$72.6$ ($6.3$), Position $74.5$--$82.5$ ($8.0$), Size $67.3$--$73.1$ ($5.8$), Text $77.9$--$84.8$ ($6.9$). The GLM case illustrates reference-subset rendering rather than horizontal grid collapse: the generated DOM contains a strict subset of the reference's blocks, so missing reference blocks reduce visual coverage (see the GLM-4.6V case study in Fig.~\ref{fig:case_grid}). Color drops by mis-derived theme palettes when screenshots use a non-default brand color. GLM-4.6V is the salient outlier: last on Size and Position but second on Color among build-pass apps. Palette extraction on the matched subset is strong, but the generated DOM covers only part of the reference.

\subsection{Interaction-inference failures}
\label{app:failures-int}

Figure~\ref{fig:int_failure_cart} anchors a representative \iis{} failure on a shared input. Two models receive the same five e-commerce screenshots of \texttt{48\_shopco-ecommerce}; an annotator then performs the same Add-to-Cart action on the product page and navigates to \texttt{/cart}. Claude~Sonnet~4.6 (left) threads the action through a \texttt{CartContext} and the cart route renders the added product (\texttt{result}=\texttt{W}). Gemini~3.1~Pro~Preview (right) emits a visually-faithful product page with an \texttt{Add to Cart} button whose handler does not propagate to any cross-route store, so the cart route still renders its empty-state placeholder (\texttt{result}=\texttt{F}). The cart route's static screenshot at protocol time is therefore indistinguishable from the reference's empty-cart capture.

\par\medskip
{\centering
\begin{minipage}[t]{0.49\linewidth}\centering
  \includegraphics[width=\linewidth]{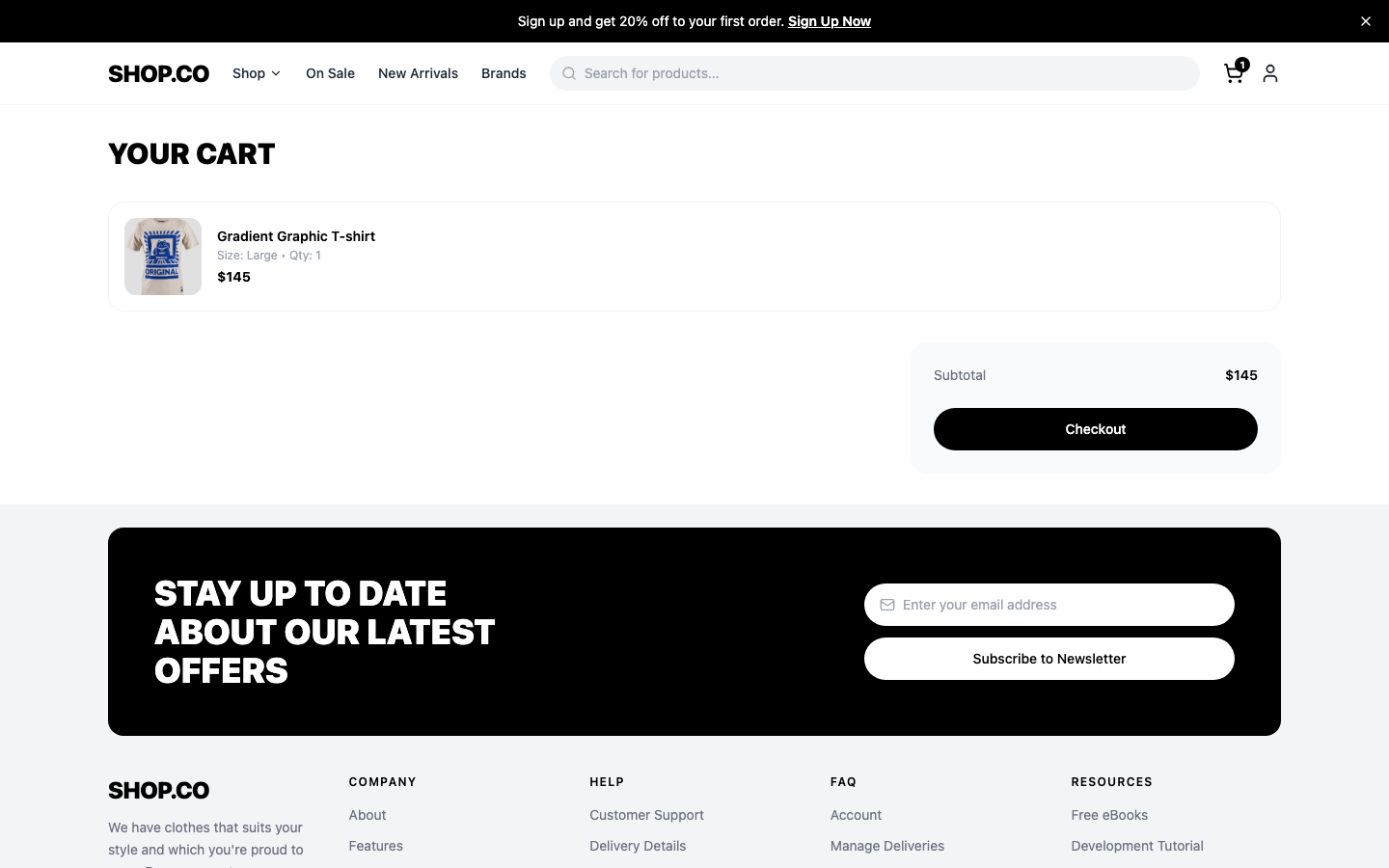}\\[2pt]
  {\footnotesize (a) Claude Sonnet~4.6, \texttt{/cart}: \textcolor{blue}{product persists} across the route boundary.}
\end{minipage}\hfill
\begin{minipage}[t]{0.49\linewidth}\centering
  \includegraphics[width=\linewidth]{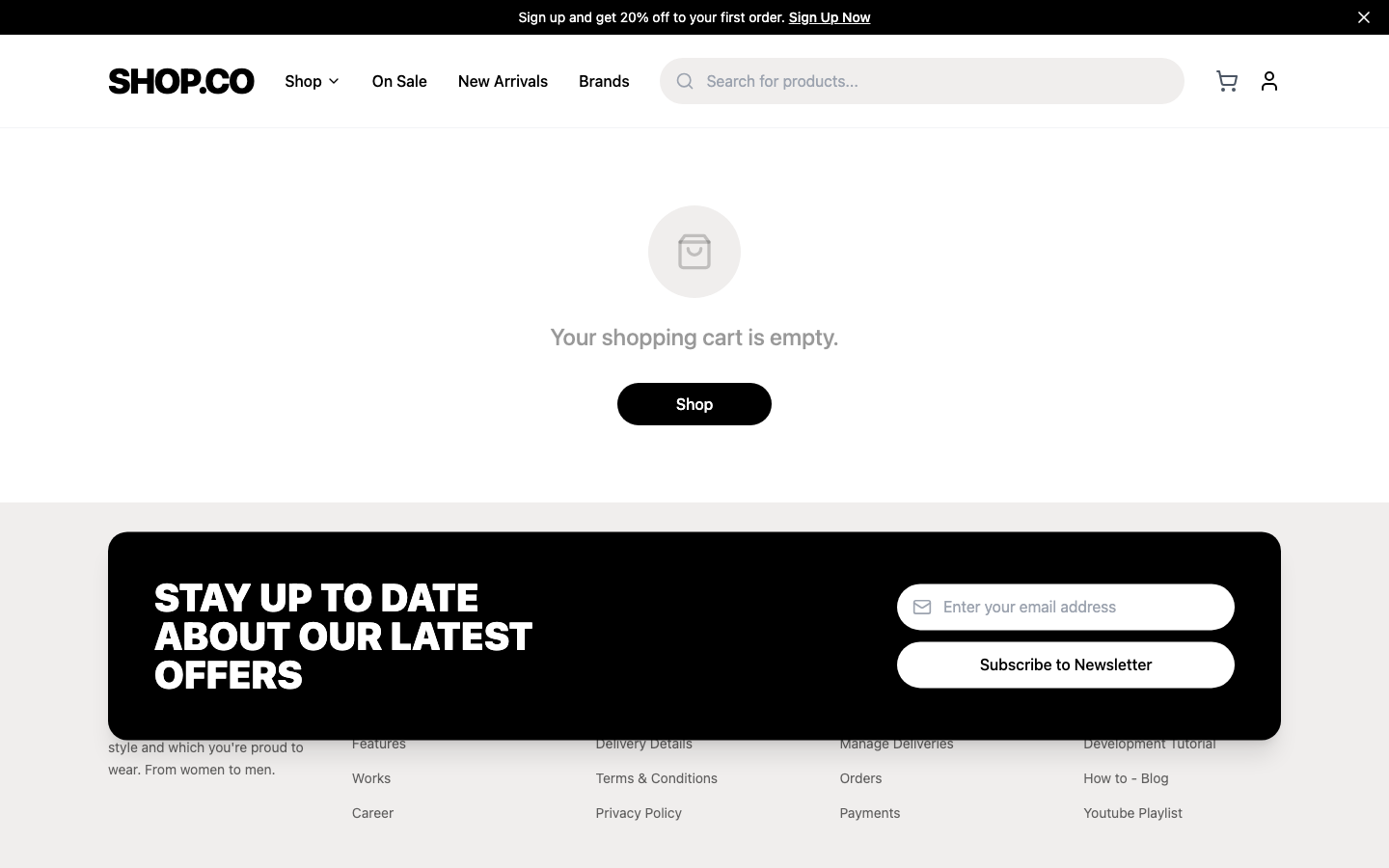}\\[2pt]
  {\footnotesize (b) Gemini~3.1~Pro~Preview, \texttt{/cart}: \textcolor{gray}{empty-state placeholder}; handler did not reach a cross-route store.}
\end{minipage}\par\medskip
\captionof{figure}{\textbf{Interaction-inference failure on \texttt{48\_shopco-ecommerce}.} Identical input screenshots and identical click sequence; the divergence is the \emph{Broken shared state} pattern. \vfs{} on the static cart-route capture cannot distinguish the two outputs (both render an empty cart at protocol time), whereas \iis{} flags the missing \sthree{} affordance.}
\label{fig:int_failure_cart}
\par}
\medskip

\subsection{Case studies: latent-affordance inference}
\label{app:case-genshin}

The eight case studies below \emph{illustrate} four claims from the main text: differences in interaction implementation, domain-level behavioral inference, build-time failure modes, and \vfs{} sub-metric divergence; they are chosen to span these four claim types, and the aggregate evidence for each lives in the corresponding section and table (\S\ref{sec:exp_iis}, Tables~\ref{tab:failure_modes} and~\ref{tab:cond_submetric}).

\paragraph{Audio synthesis (\texttt{31\_genshin-music}).}
The \texttt{31\_genshin-music} task supplies $9$ screenshots of a stylized in-browser music synthesizer (Fig.~\ref{fig:case_genshin_renders}): a hexagonal note pad, a key-binding overlay, a composer timeline, and a full-screen ``Zen mode'' player. None of the screenshots state explicitly that the application produces sound. The audio capability must be inferred from layout cues alone: musical symbols on each pad, MIDI-style key labels (\texttt{Q}--\texttt{U} / \texttt{A}--\texttt{J} / \texttt{Z}--\texttt{M}), an instrument-picker dropdown, and a decay-envelope visualization behind each note.

Of the six baselines, four reproduce the interface as a runnable static visual replica: pads animate on hover, but \texttt{onMouseDown} handlers are absent or no-ops and the emitted \texttt{package.json} declares no audio dependencies; GLM-4.6V fails to build. Only Kimi~K2.5 instantiates a complete Web-Audio signal chain (\texttt{AudioContext} $\to$ triangle-wave \texttt{OscillatorNode} $\to$ \texttt{GainNode} with a $500$\,ms exponential-decay envelope) and wires keyboard events together with pointer events on every pad through a MIDI-to-frequency converter $f = 440 \cdot 2^{(n-69)/12}$. The artifact is a functionally playable synthesizer rather than a screenshot.

The contrast illustrates the regime \iis{} (\S\ref{sec:iis}) is designed to expose. The rendered interfaces are visually comparable and \vfs{} cannot separate them, yet only one artifact is functionally playable. Crucially, no input screenshot can be flagged as ``wrong'' for the four UI-only models, since their renderings reproduce the visual targets' structure: grid, labels, and layout. The gap is one of behavior inferred from cues that the visual modality under-specifies, precisely the regime \bench{} is built to expose for multi-modal code-generation agents beyond surface mimicry. Figure~\ref{fig:case_genshin} contrasts the source for one pad cell across the two regimes (the UI-only listing is from Gemini~3.1~Pro Preview, representative of the four visually-faithful-but-silent baselines).

\par\medskip
{\centering
  \begin{minipage}[t]{0.40\linewidth}\centering
    \includegraphics[width=\linewidth,keepaspectratio]{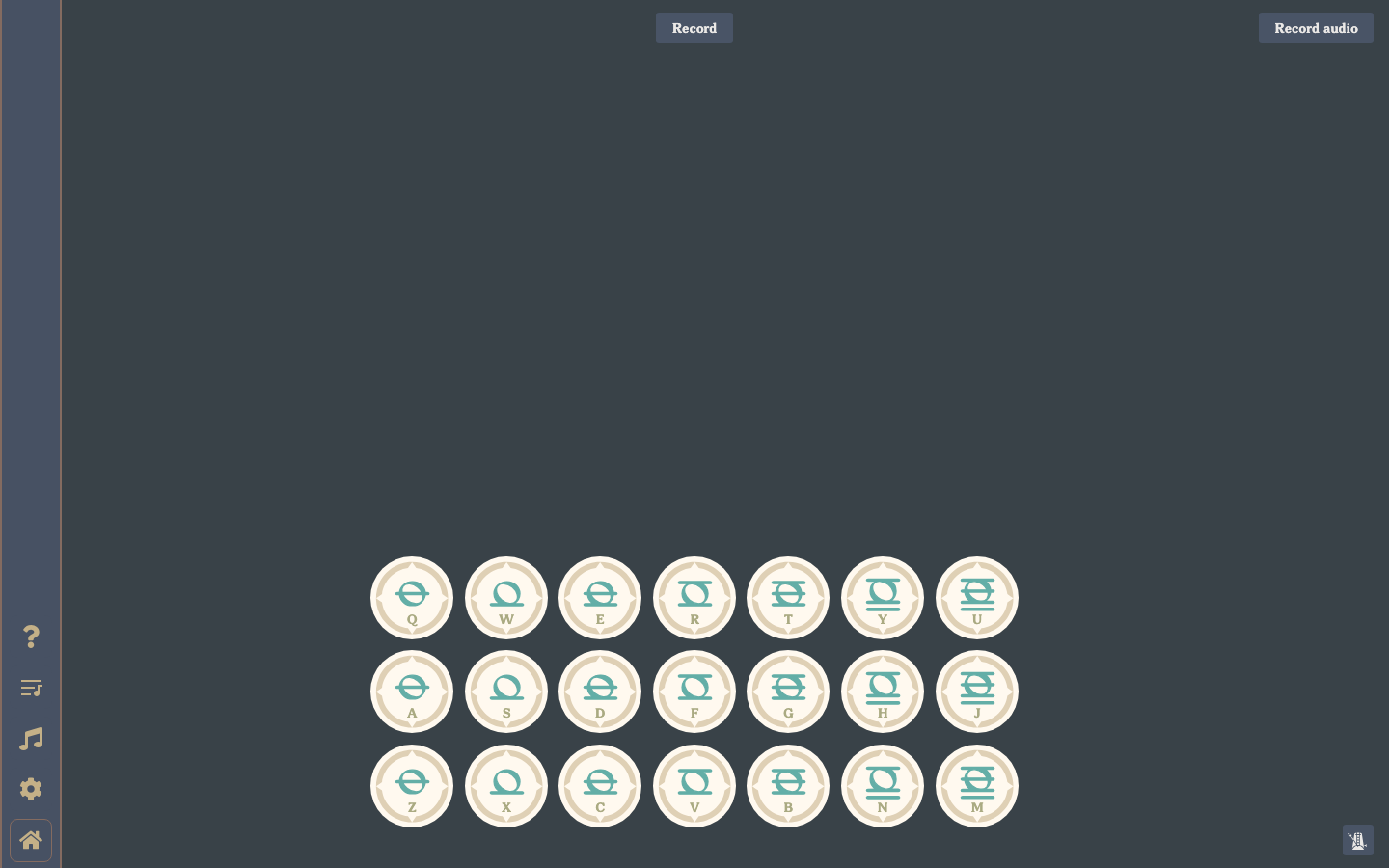}\\[2pt]
    {\footnotesize \texttt{/}}
  \end{minipage}\hfill
  \begin{minipage}[t]{0.40\linewidth}\centering
    \includegraphics[width=\linewidth,keepaspectratio]{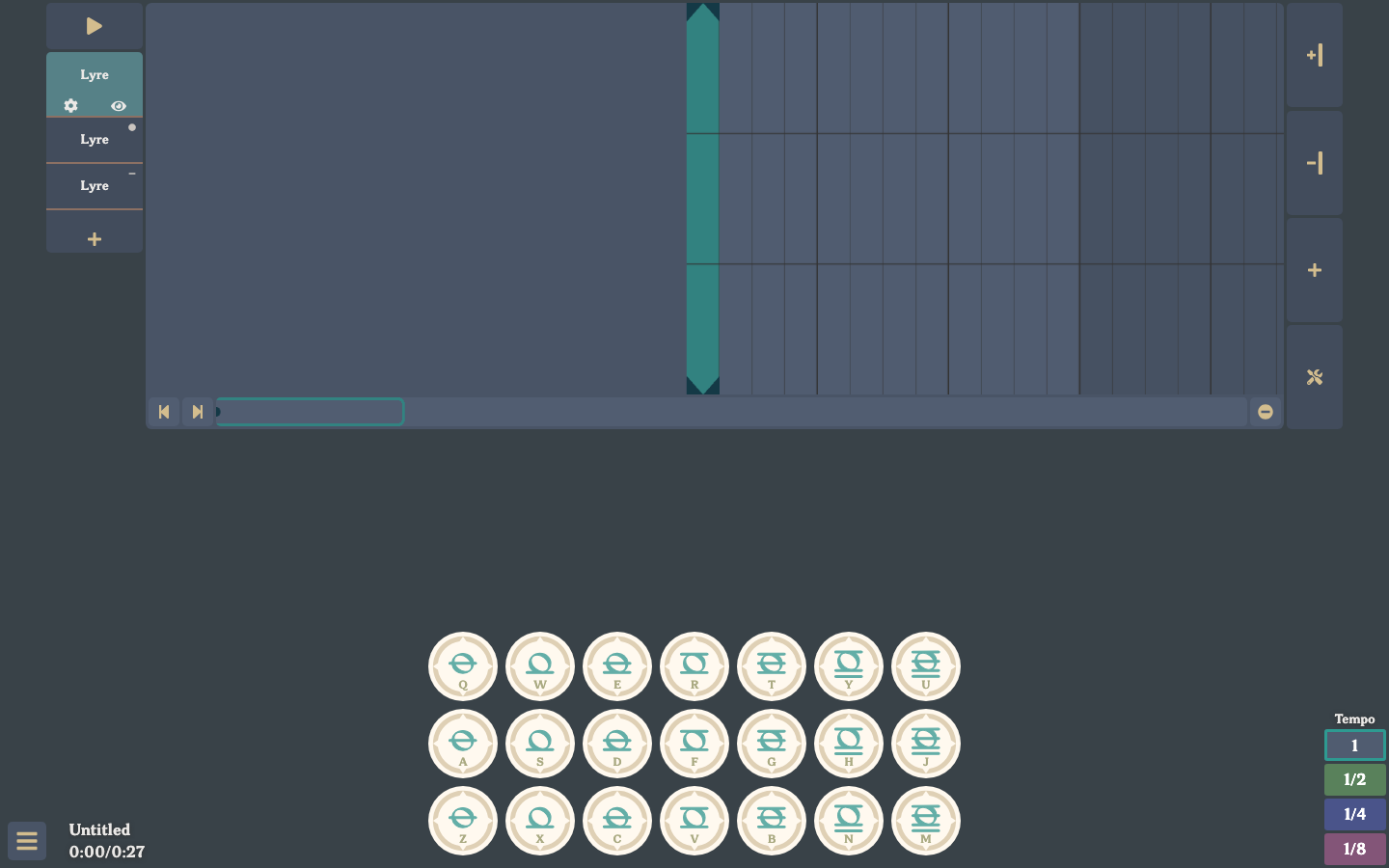}\\[2pt]
    {\footnotesize \texttt{/composer}}
  \end{minipage}

  \vspace{6pt}
  \begin{minipage}[t]{0.40\linewidth}\centering
    \includegraphics[width=\linewidth,keepaspectratio]{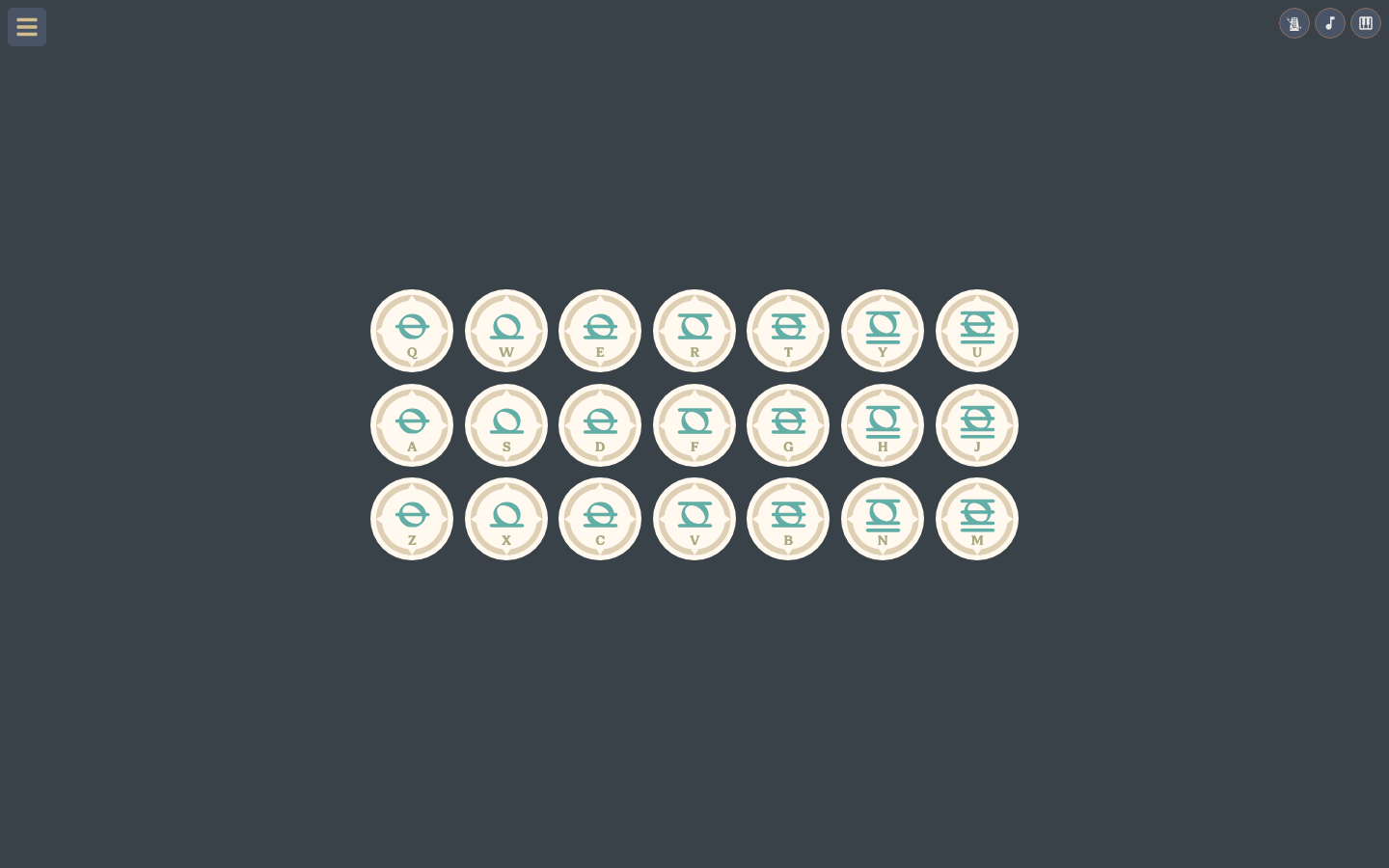}\\[2pt]
    {\footnotesize \texttt{/zen-keyboard}}
  \end{minipage}\hfill
  \begin{minipage}[t]{0.40\linewidth}\centering
    \includegraphics[width=\linewidth,keepaspectratio]{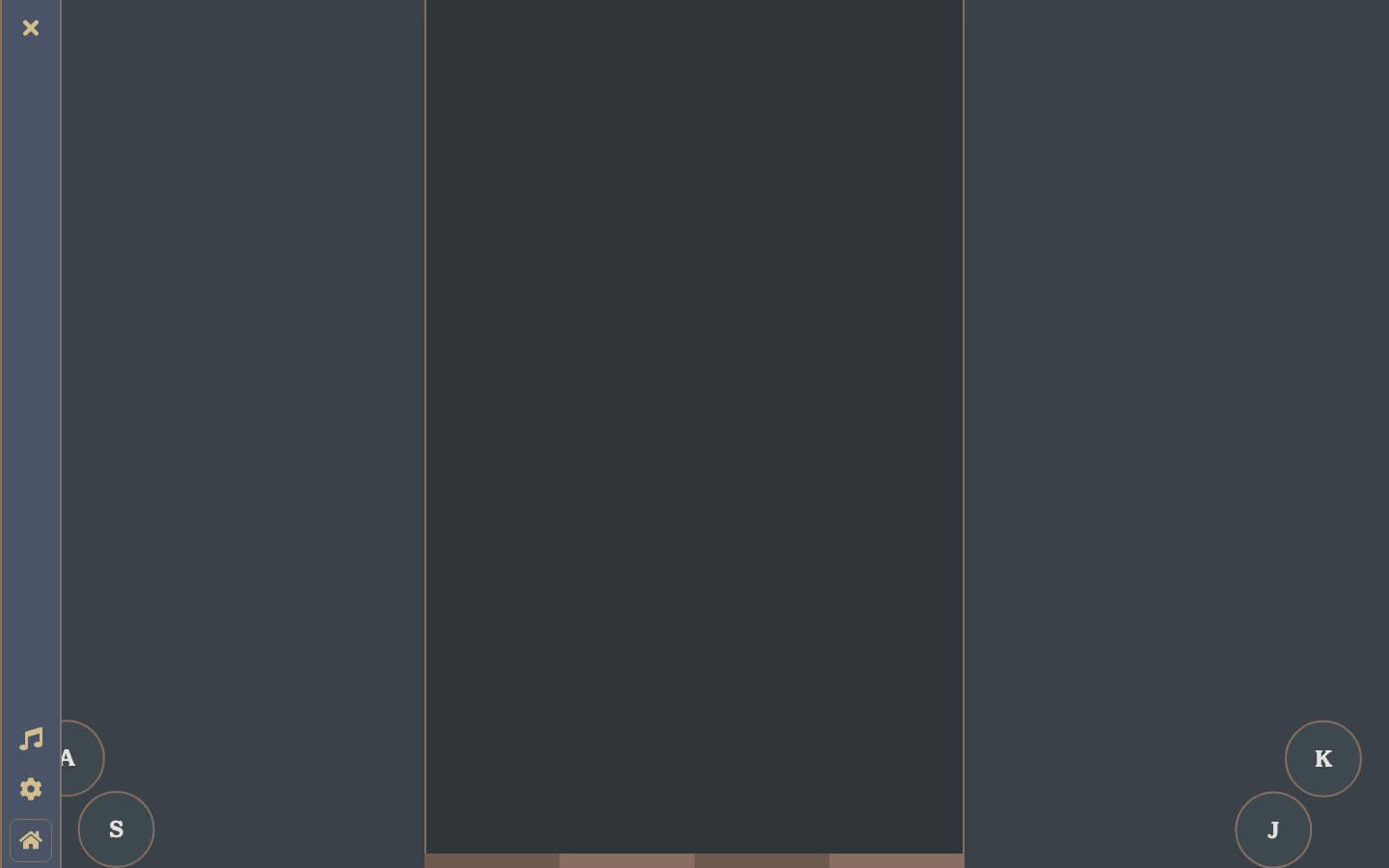}\\[2pt]
    {\footnotesize \texttt{/vsrg-player}}
  \end{minipage}

  \vspace{6pt}
  \begin{minipage}[t]{0.40\linewidth}\centering
    \includegraphics[width=\linewidth,keepaspectratio]{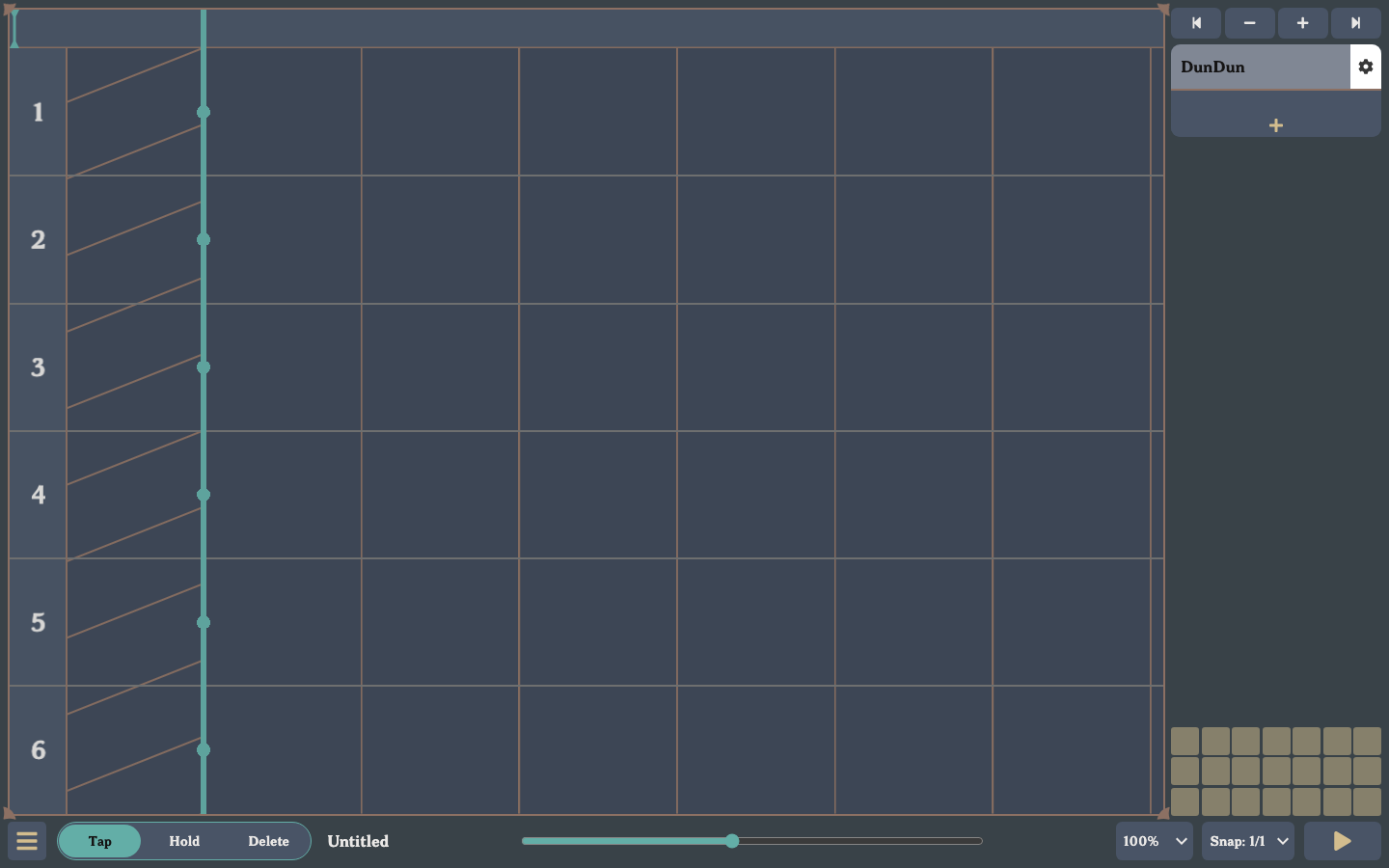}\\[2pt]
    {\footnotesize \texttt{/vsrg-composer}}
  \end{minipage}\hfill
  \begin{minipage}[t]{0.40\linewidth}\centering
    \includegraphics[width=\linewidth,keepaspectratio]{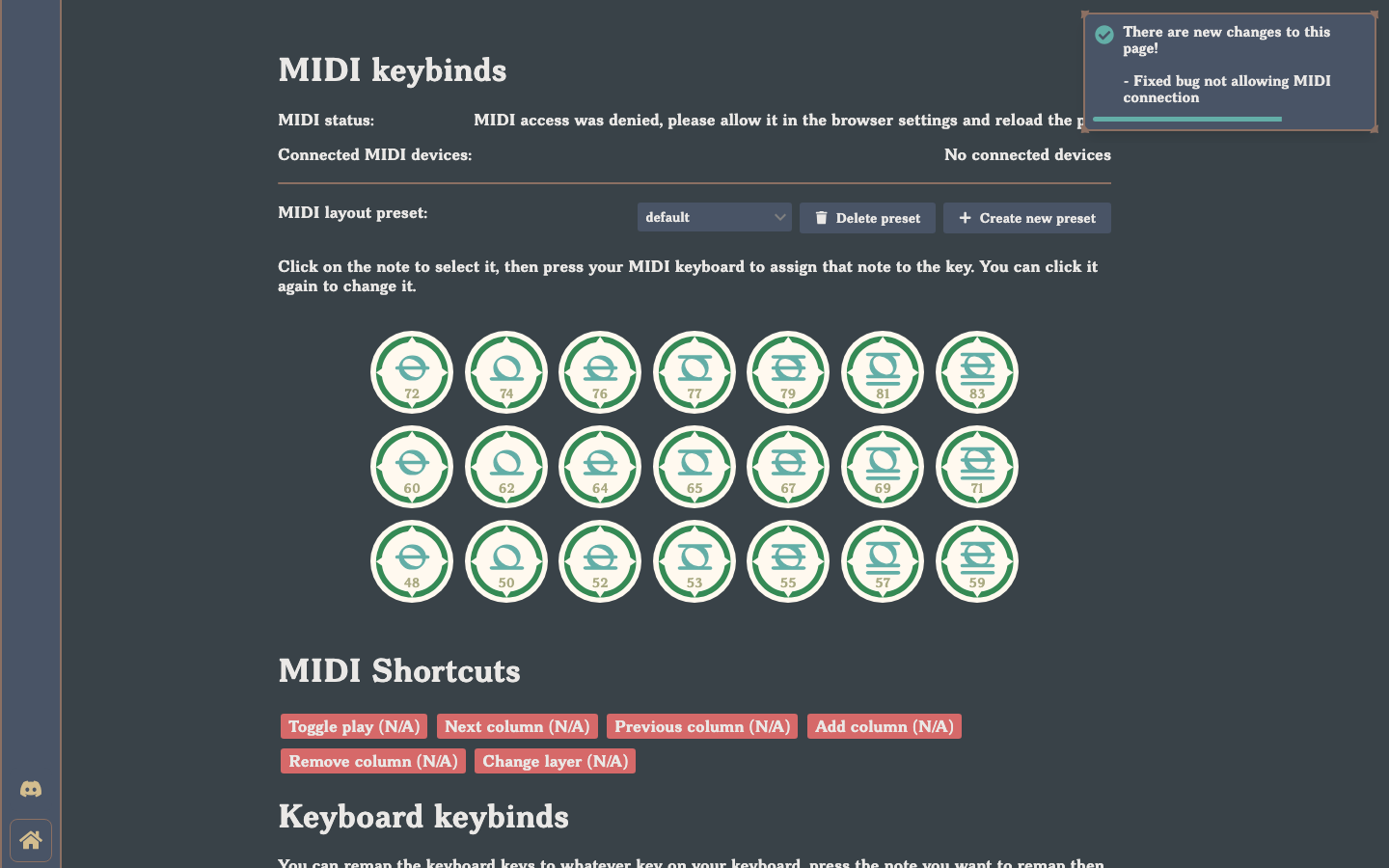}\\[2pt]
    {\footnotesize \texttt{/keybinds}}
  \end{minipage}

  \vspace{6pt}
  \begin{minipage}[t]{0.40\linewidth}\centering
    \includegraphics[width=\linewidth,keepaspectratio]{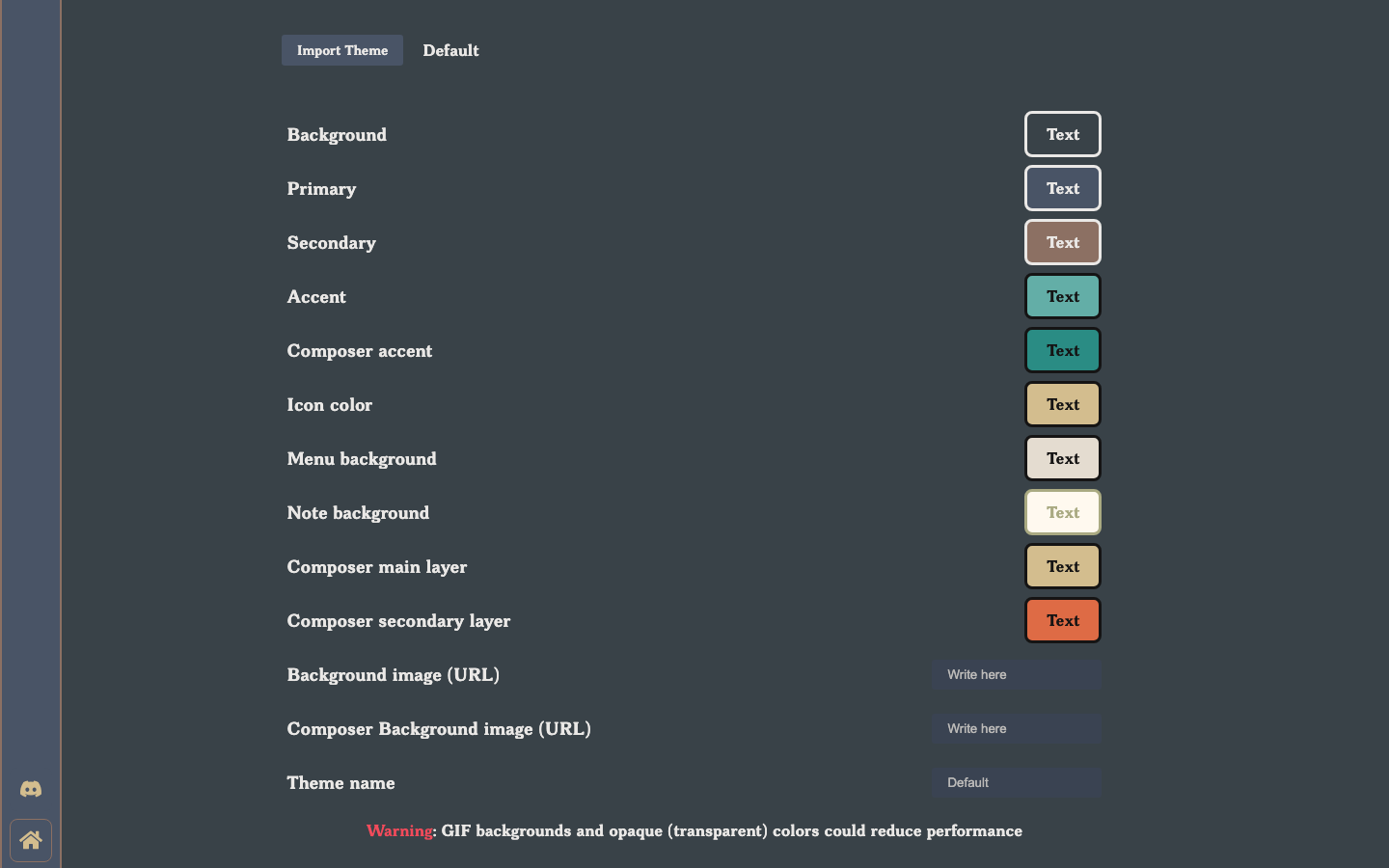}\\[2pt]
    {\footnotesize \texttt{/theme}}
  \end{minipage}\hfill
  \begin{minipage}[t]{0.40\linewidth}\centering
    \includegraphics[width=\linewidth,keepaspectratio]{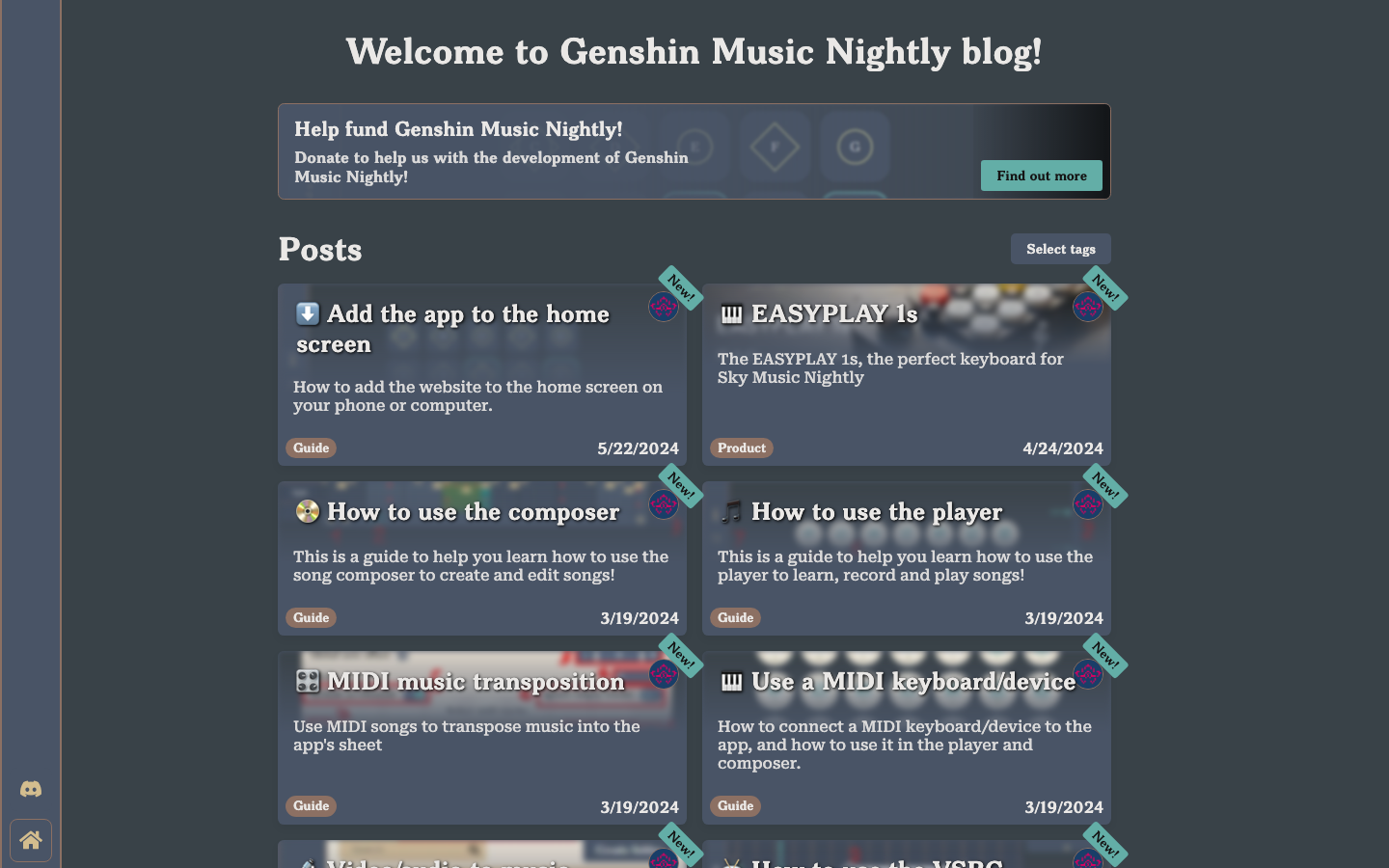}\\[2pt]
    {\footnotesize \texttt{/blog}}
  \end{minipage}

  \vspace{6pt}
  \begin{minipage}[t]{0.40\linewidth}\centering
    \includegraphics[width=\linewidth,keepaspectratio]{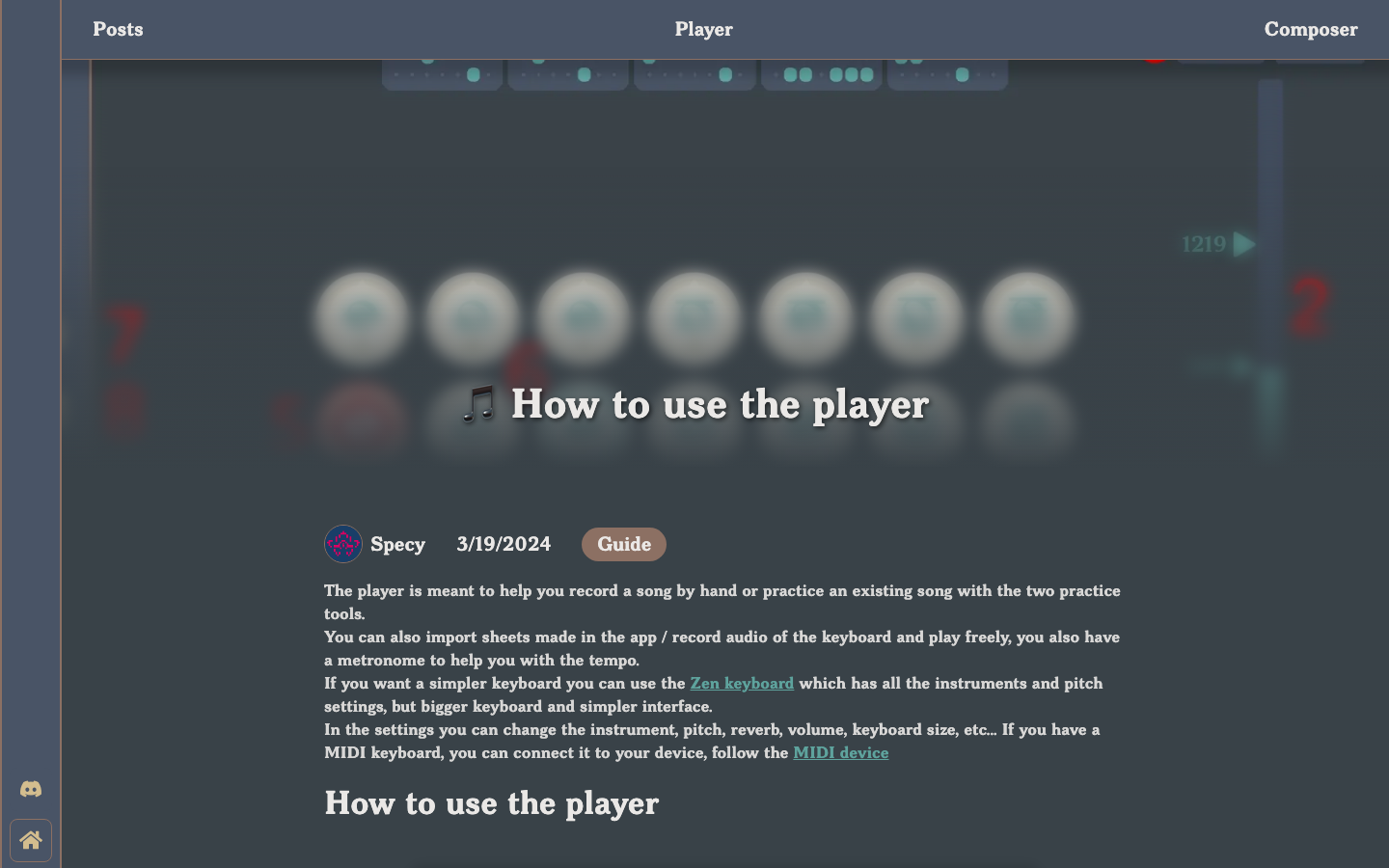}\\[2pt]
    {\footnotesize \texttt{/blog/posts/\dots}}
  \end{minipage}\par\medskip
\captionof{figure}{\textbf{Input set for \texttt{31\_genshin-music} (9 screenshots, one per route).} The audio capability must be inferred from these static cues alone: note pads with musical glyphs and MIDI key labels (\texttt{Q}--\texttt{U}/\texttt{A}--\texttt{J}/\texttt{Z}--\texttt{M}) on \texttt{/zen-keyboard} and \texttt{/keybinds}, a composer timeline (\texttt{/composer}, \texttt{/vsrg-composer}), and a full-screen player (\texttt{/vsrg-player}).}
\label{fig:case_genshin_renders}
\par}

\par\medskip
{\centering
  \begin{minipage}[t]{0.49\linewidth}
\begin{codecard}{src/components/Keyboard.tsx}
function KeyButton({ label, ... }) {
  return (
    <button className="w-[72px] h-[72px]
                       rounded-full bg-[#ebdcb8]
                       active:scale-95 ...">
      <div className="rounded-full
                      bg-[#f8f5ea] border-2
                      flex flex-col ...">
        <svg width="24" height="24" ...>
          <circle cx="12" cy="12" r="9" />
          <line x1="4" y1="12"
                x2="20" y2="12" />
        </svg>
        <span className="...">
          {label}
        </span>
      </div>
    </button>
  );
  // no onClick / onMouseDown
}
\end{codecard}
\captionof{lstlisting}{(b) Gemini 3.1 Pro Preview: \texttt{src/components/Keyboard.tsx} (extract). Pad is a styled \texttt{<button>} with no \texttt{onClick}/\texttt{onMouseDown}; the SVG and key label are decorative.}
  \end{minipage}\hfill
  \begin{minipage}[t]{0.49\linewidth}
\begin{codecard}{src/hooks/useAudio.ts}
const playNote = useCallback(
  (midiNote: number) => {
    initAudio();
    const ctx = audioContextRef.current;
    if (!ctx) return;
    stopNote(midiNote);
    const osc = ctx.createOscillator();
    const gain = ctx.createGain();
    const freq = 440 * Math.pow(
      2, (midiNote - 69) / 12);
    osc.type = 'triangle';
    osc.frequency.setValueAtTime(
      freq, ctx.currentTime);
    gain.gain.setValueAtTime(
      0.3, ctx.currentTime);
    gain.gain.exponentialRampToValueAtTime(
      0.01, ctx.currentTime + 0.5);
    osc.connect(gain);
    gain.connect(ctx.destination);
    osc.start();
    oscillatorsRef.current.set(midiNote, osc);
    setTimeout(() => stopNote(midiNote), 500);
  }, [initAudio]);
\end{codecard}
\captionof{lstlisting}{(c) Kimi K2.5: \texttt{src/hooks/useAudio.ts} (extract). \texttt{playNote} builds the Web-Audio chain (\texttt{OscillatorNode}\,$\to$\,\texttt{GainNode}, 500\,ms decay) with a MIDI-to-frequency converter.}
  \end{minipage}\par\medskip
\captionof{figure}{\textbf{Latent-affordance inference on \texttt{31\_genshin-music}.} Four of six baselines emit (b)-style code: visually-faithful pads with no audio. Only Kimi~K2.5 instantiates the (c)-style Web-Audio signal chain, \texttt{AudioContext}~$\to$~\texttt{OscillatorNode}~$\to$~\texttt{GainNode} with a $500$\,ms exponential decay, and a MIDI-to-frequency converter $f = 440\cdot 2^{(n-69)/12}$. \vfs{} cannot read this gap from the rendered DOM; the missing audio runtime is a latent affordance that surfaces only under behavioral inspection, the inference regime \iis{} targets (\S\ref{sec:iis}).}
\label{fig:case_genshin}
\par}
\medskip

\paragraph{State-machine inference (\texttt{01\_fashion-ecommerce}, \texttt{48\_shopco-ecommerce}).}
\label{app:case-ecommerce}
Two e-commerce tasks, \texttt{01\_fashion-ecommerce} ($6$ screenshots: home, shop, product detail, cart, checkout, search) and \texttt{48\_shopco-ecommerce} ($5$ screenshots: home, shop, cart, and two product details), require an analogous inference along an orthogonal direction. Rather than an audio runtime, the model must infer the underlying \emph{state machine} that backs the interface: a global cart that survives route changes (\sthree{}), variant selection on the product page that propagates into add-to-cart payloads (\stwo{}), search and sort on the shop (\stwo{}), and controlled inputs on a checkout form. As with audio, the screenshots never declare these affordances explicitly: a ``\$0.00'' subtotal hints at a reactive total, a size grid at selectable variants, a search bar at client-side filtering.

Across interaction items spanning the two projects, Claude~Sonnet~4.6 produces lots of interaction components that work end-to-end: a \texttt{CartContext} de-duplicating items by the composite key $(\textit{id},\,\textit{size},\,\textit{color})$, auto-clearing entries when quantity drops to zero, a controlled \texttt{CheckoutPage} form, and a search page with submit-driven filtering on product name and category. The next-best on these two tasks, GPT-5.4, supplies a partially wired \texttt{ShopContext} on \texttt{48\_shopco-ecommerce}, but its \texttt{addToCart} mutator is never read by the cart page, which renders a static empty-state placeholder, dead code in the literal sense. The remaining four baselines (Gemini~3.1~Pro Preview, GLM-4.6V, Qwen3.5-397B-A17B, Kimi~K2.5) systematically emit visually faithful but inert components: add-to-cart buttons without \texttt{onClick}, cart pages reduced to ``\textit{Intentionally left blank}'' placeholders, and no cross-component state container at all. Figure~\ref{fig:case_ecommerce} contrasts the \texttt{/cart} render across the three regimes.

\par\medskip
{\centering
\begin{minipage}[t]{0.49\linewidth}\centering
  \includegraphics[width=\linewidth,keepaspectratio]{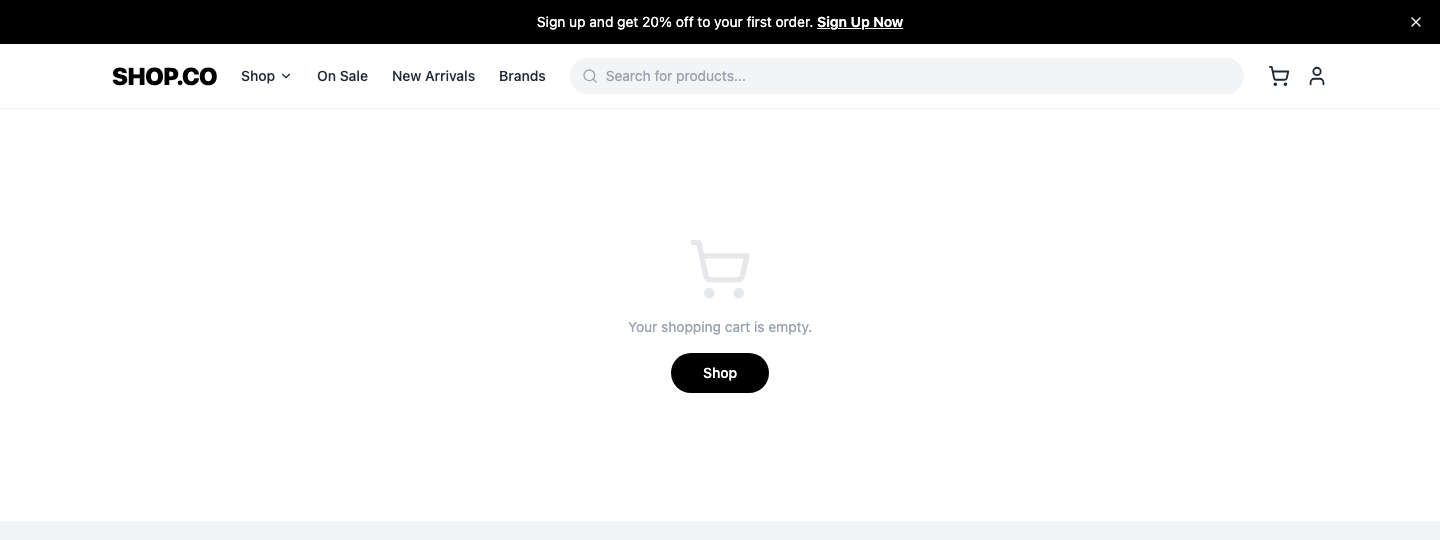}\\[2pt]
  {\footnotesize (a$_1$) Sonnet, \textbf{before} click}
\end{minipage}\hfill
\begin{minipage}[t]{0.49\linewidth}\centering
  \includegraphics[width=\linewidth,keepaspectratio]{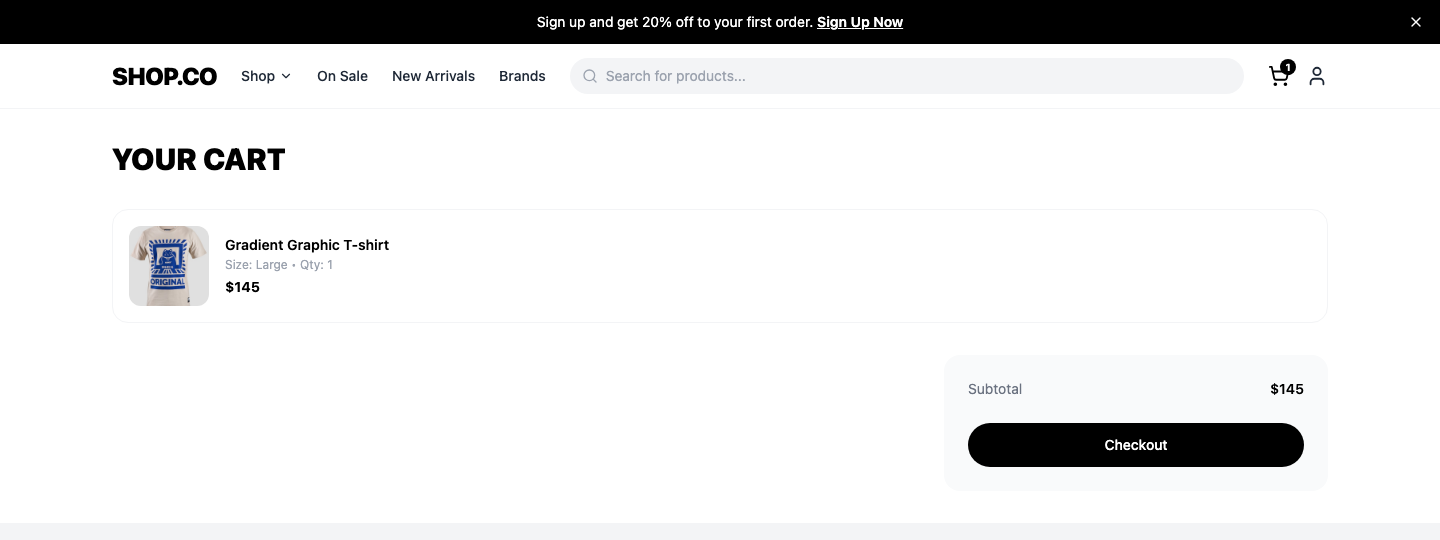}\\[2pt]
  {\footnotesize (a$_2$) Sonnet, \textbf{after} \textcolor{blue}{wired}: item, \$145, badge \textbf{1}}
\end{minipage}\par\medskip
\begin{minipage}[t]{0.49\linewidth}\centering
  \includegraphics[width=\linewidth,keepaspectratio]{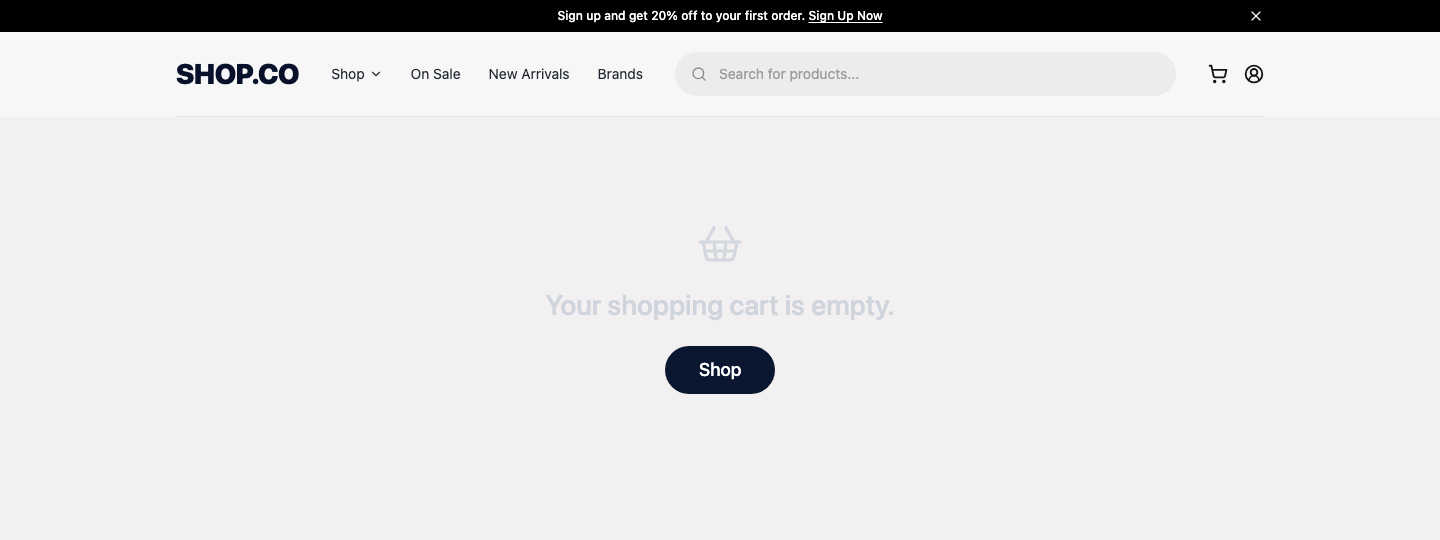}\\[2pt]
  {\footnotesize (b$_1$) GPT-5.4, \textbf{before} click}
\end{minipage}\hfill
\begin{minipage}[t]{0.49\linewidth}\centering
  \includegraphics[width=\linewidth,keepaspectratio]{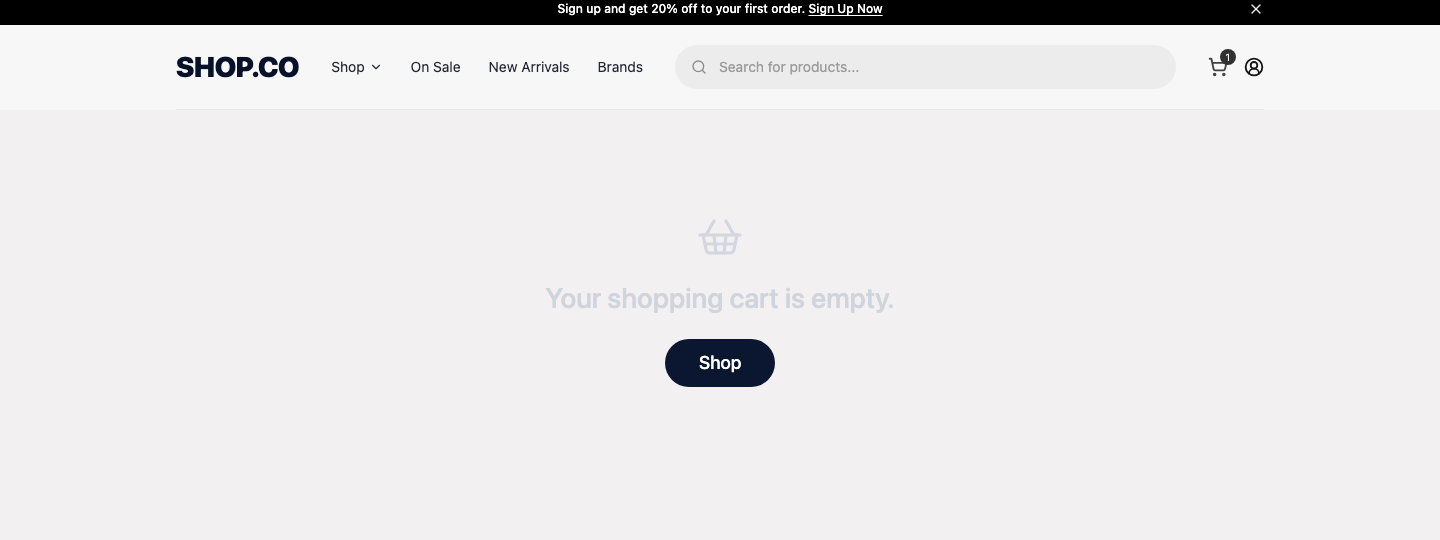}\\[2pt]
  {\footnotesize (b$_2$) GPT-5.4, \textbf{after} \textcolor{red}{dead code}: cart empty, badge \textbf{1}}
\end{minipage}\par\medskip
\begin{minipage}[t]{0.49\linewidth}\centering
  \includegraphics[width=\linewidth,keepaspectratio]{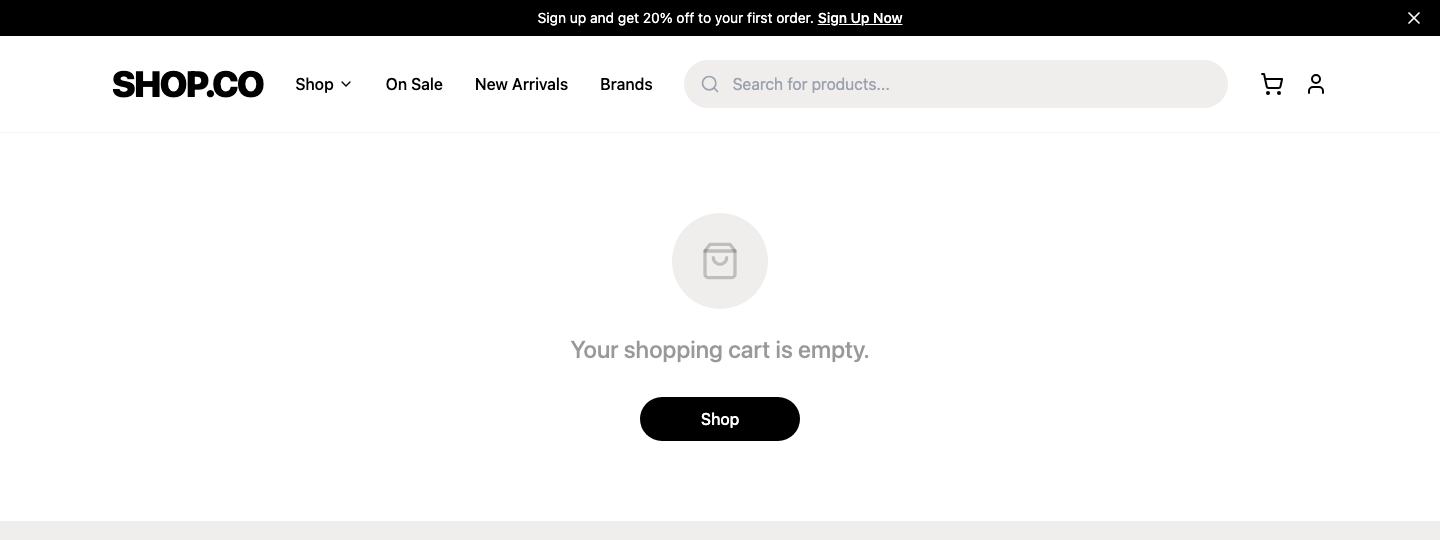}\\[2pt]
  {\footnotesize (c$_1$) Gemini, \textbf{before} click}
\end{minipage}\hfill
\begin{minipage}[t]{0.49\linewidth}\centering
  \includegraphics[width=\linewidth,keepaspectratio]{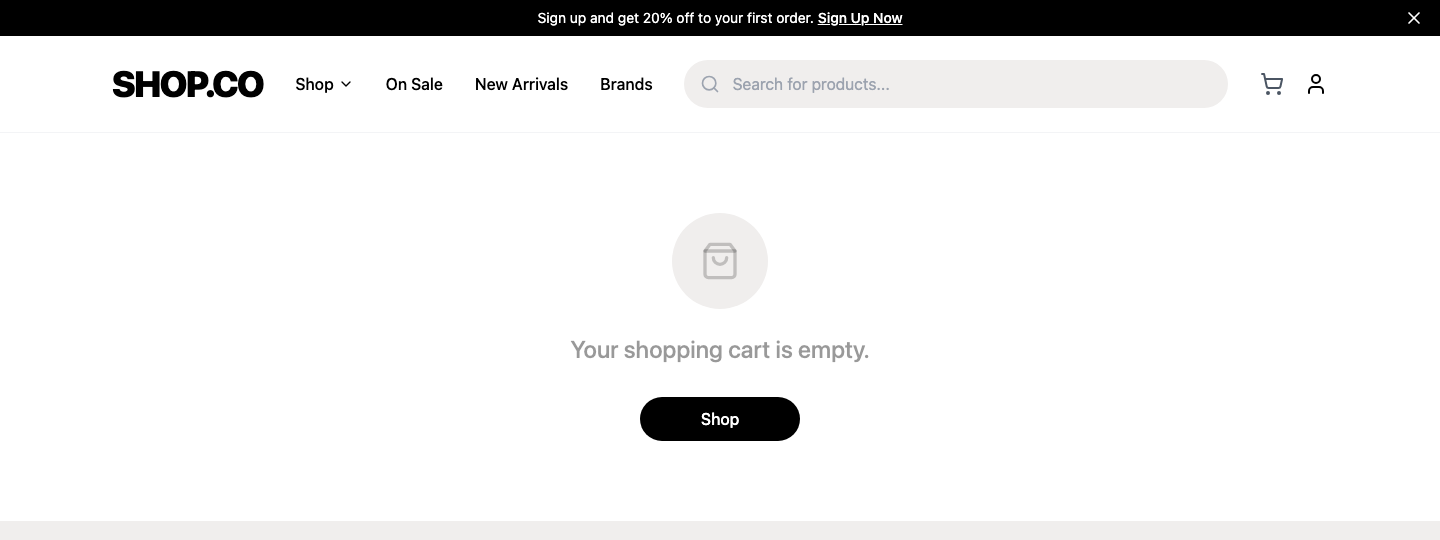}\\[2pt]
  {\footnotesize (c$_2$) Gemini, \textbf{after} \textcolor{gray}{inert}: no change, no badge}
\end{minipage}\par\medskip
\captionof{figure}{\textbf{State-machine inference on \texttt{48\_shopco-ecommerce}: before/after live capture across an identical click sequence.} Each row is one model, before/after an identical add-to-cart\,$\to$\,cart-navigation sequence (SPA \texttt{<Link>}, no hard reload, so in-memory state must survive). Left column (before): all three carts are empty baselines. Right column (after): (a$_2$) Sonnet's \texttt{CartContext} propagates end-to-end: \textsc{Gradient Graphic T-shirt} (Size~Large, \$145) appears, subtotal recomputes, header badge reads \textbf{1}; (b$_2$) GPT-5.4's badge \emph{also} reads \textbf{1} (\texttt{addToCart} did mutate \texttt{cartCount}), yet the cart route still renders ``Your shopping cart is empty.'': \texttt{CartPage} never reads \texttt{cartItems} from \texttt{ShopContext}, a literal dead-code branch; (c$_2$) Gemini's render is visually identical to its pre-click state (no badge change, no list item), no state container at all. All six panels (three models $\times$ before/after) are comparable in \vfs{}, yet \iis{} on the \sthree{} cross-route scope splits them sharply.}
\label{fig:case_ecommerce}
\par}
\medskip

\paragraph{Persistent task state (\texttt{39\_maciekt07-todoapp}).}
\label{app:case-todoapp}
The todo task supplies $5$ screenshots of a single-page task manager with categories, multi-select, completion toggles, and JSON import/export. Of the six models that generated this app, only Kimi~K2.5 persists task state, lifting it into a single \texttt{AppContext} that mirrors itself into \texttt{localStorage} on every mutation (\texttt{STORAGE\_KEY = "taskapp\_data"}) and re-hydrates on mount. None of the other five survives a reload: Claude implements working in-memory CRUD in an app-level Context, while GPT-5.4, Gemini, and Qwen render the task list as static mock data with no mutable state; GLM-4.6V's generated source holds tasks in component-local \texttt{useState}, but the app fails to build ($\exect[3]{=}\mathrm{F}$). Figure~\ref{fig:case_todoapp} verifies the re-hydration half directly: three tasks seeded into \texttt{localStorage} survive a hard reload and re-appear on mount. This cross-session (\texttt{localStorage}-backed) persistence is strictly stronger than the \sthree{} cross-route criterion (\S\ref{app:iis_handbook}), which several models satisfy only in memory (e.g., Sonnet's cart in Fig.~\ref{fig:case_ecommerce}).

\par\medskip
{\centering
\begin{minipage}{\linewidth}\centering
  \includegraphics[width=\linewidth,keepaspectratio]{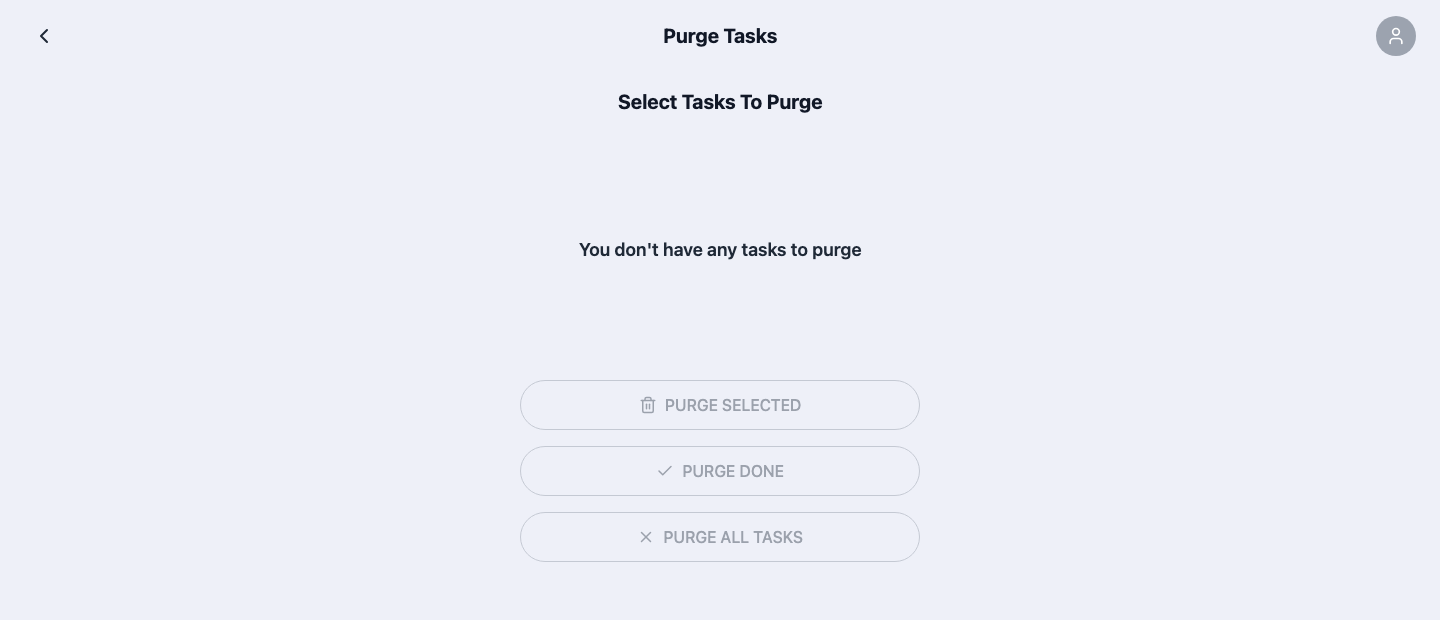}\\[2pt]
  {\footnotesize (a) Kimi \texttt{/purge}, fresh ctx \textcolor{gray}{``no tasks''}}
\end{minipage}\par\medskip
\begin{minipage}{\linewidth}\centering
  \includegraphics[width=\linewidth,keepaspectratio]{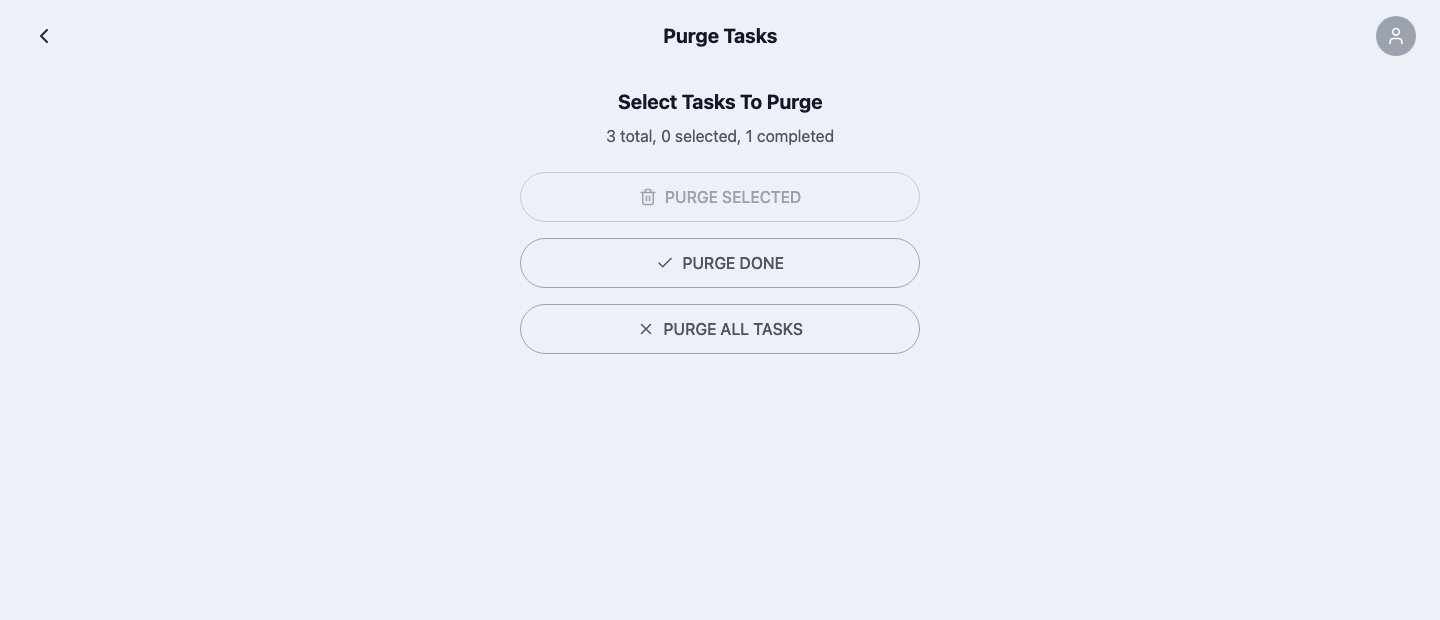}\\[2pt]
  {\footnotesize (b) Kimi \texttt{/purge}, after seeding \texttt{localStorage} + reload \textcolor{blue}{``3 total, 0 selected, 1 completed''}}
\end{minipage}\par\medskip
\captionof{figure}{\textbf{Cross-session state persistence on \texttt{39\_maciekt07-todoapp}.} (a) the \texttt{/purge} route on a fresh context shows ``no tasks to purge'' and disabled action buttons; (b) after seeding three tasks (one completed) into \texttt{localStorage[\textquotesingle taskapp\_data\textquotesingle]} and triggering a hard reload, the same route reports ``3~total, 0~selected, 1~completed'' and re-enables the \textsc{Purge~Done}/\textsc{Purge~All~Tasks} buttons, proving that Kimi's \texttt{AppContext} re-hydrates from \texttt{localStorage} on mount. None of the remaining five baselines survives a reload: Claude holds tasks in an app-level Context (in memory only) while GPT-5.4, Gemini, and Qwen render the list as static mock data; GLM-4.6V's source uses component-local \texttt{useState} but the app fails to build.}
\label{fig:case_todoapp}
\par}
\medskip

\paragraph{Drag-and-drop file upload (\texttt{39\_maciekt07-todoapp}).}
\label{app:case-dnd}
The same task's \texttt{/transfer} screen shows a dashed-border ``Drop JSON file here to import tasks'' import region. The cue is purely visual: a dashed rectangle with an upload icon and a brief instruction label, with no API specification. Kimi~K2.5 and Claude~Sonnet~4.6 both extract a \texttt{FileDropZone} component bound to native HTML5 \texttt{onDragOver}/\texttt{onDrop} events that pre-fills the JSON-import field on drop. Gemini~3.1~Pro Preview, GPT-5.4, and Qwen3.5-397B-A17B render the dashed rectangle as decoration plus a click-only fallback button with no drop handlers; GLM-4.6V's build of this app fails ($\exect[3]{=}\mathrm{F}$).

\par\medskip
{\centering
\begin{minipage}{0.66\linewidth}\centering
  \includegraphics[width=\linewidth,keepaspectratio]{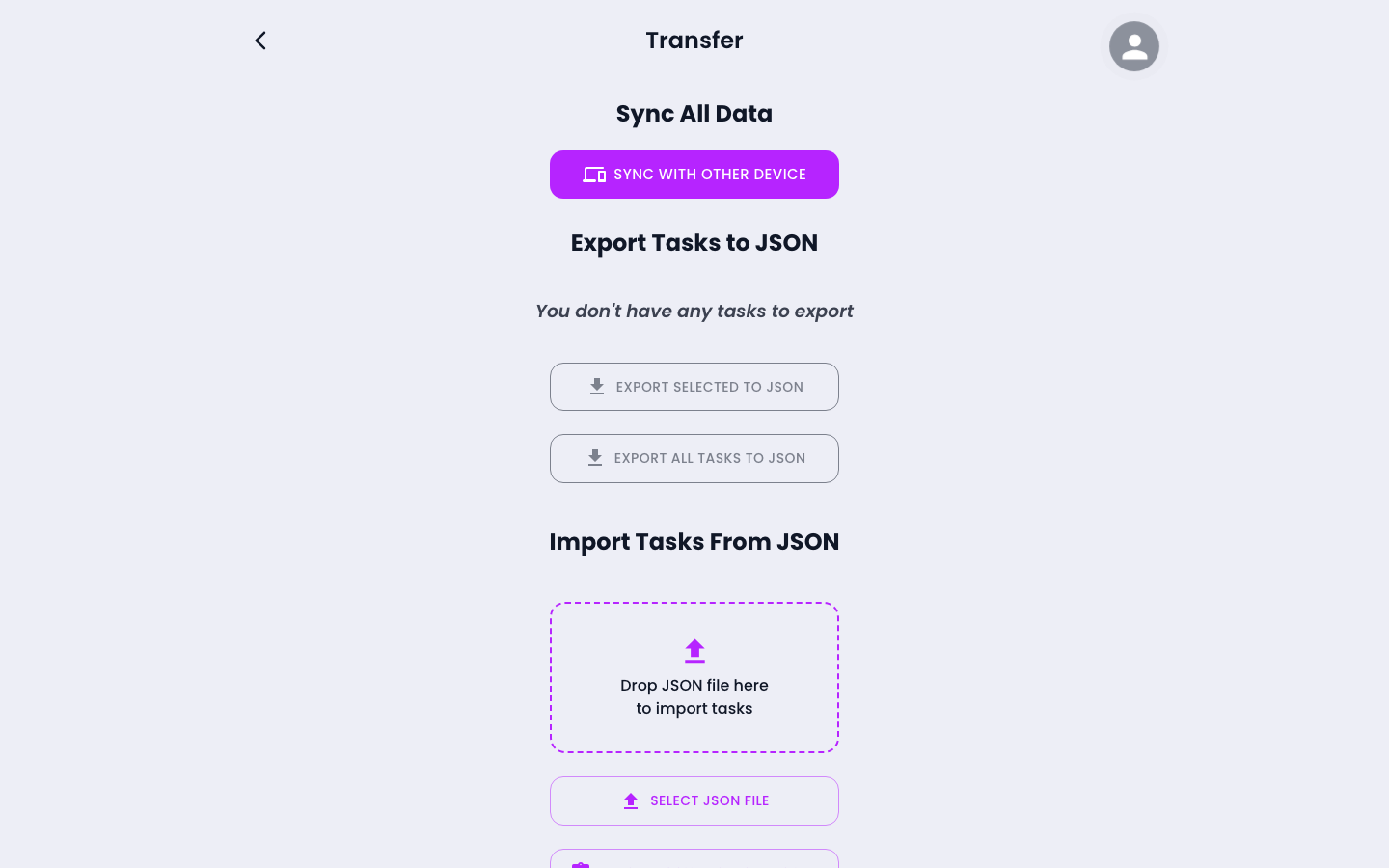}\\[2pt]
  {\footnotesize (a) input \texttt{/transfer}}
\end{minipage}\par\medskip
\begin{minipage}[t]{0.32\linewidth}\centering
  \includegraphics[width=\linewidth,keepaspectratio]{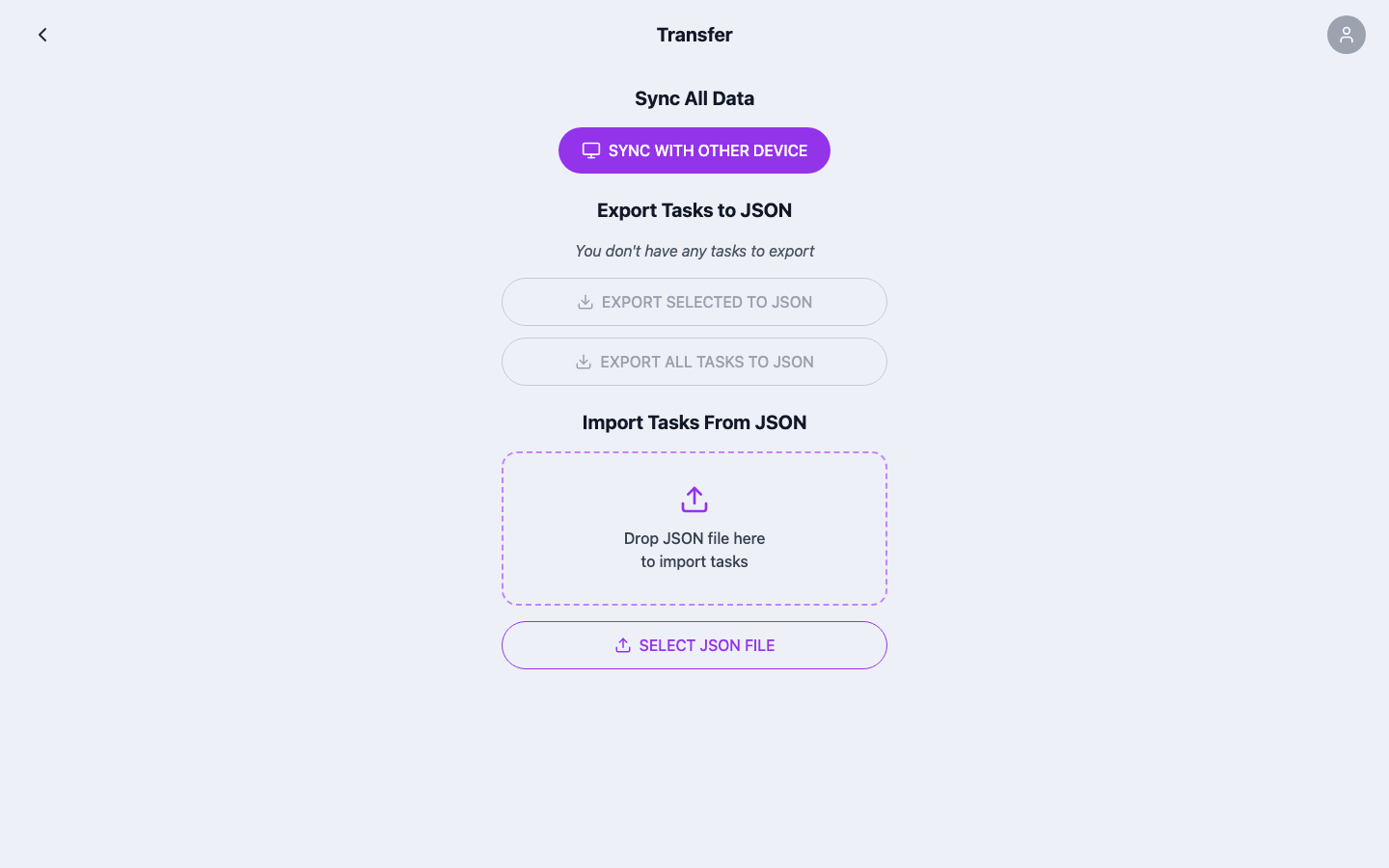}\\[2pt]
  {\footnotesize (b) Kimi K2.5 \textcolor{blue}{onDrop wired}}
\end{minipage}\hfill
\begin{minipage}[t]{0.32\linewidth}\centering
  \includegraphics[width=\linewidth,keepaspectratio]{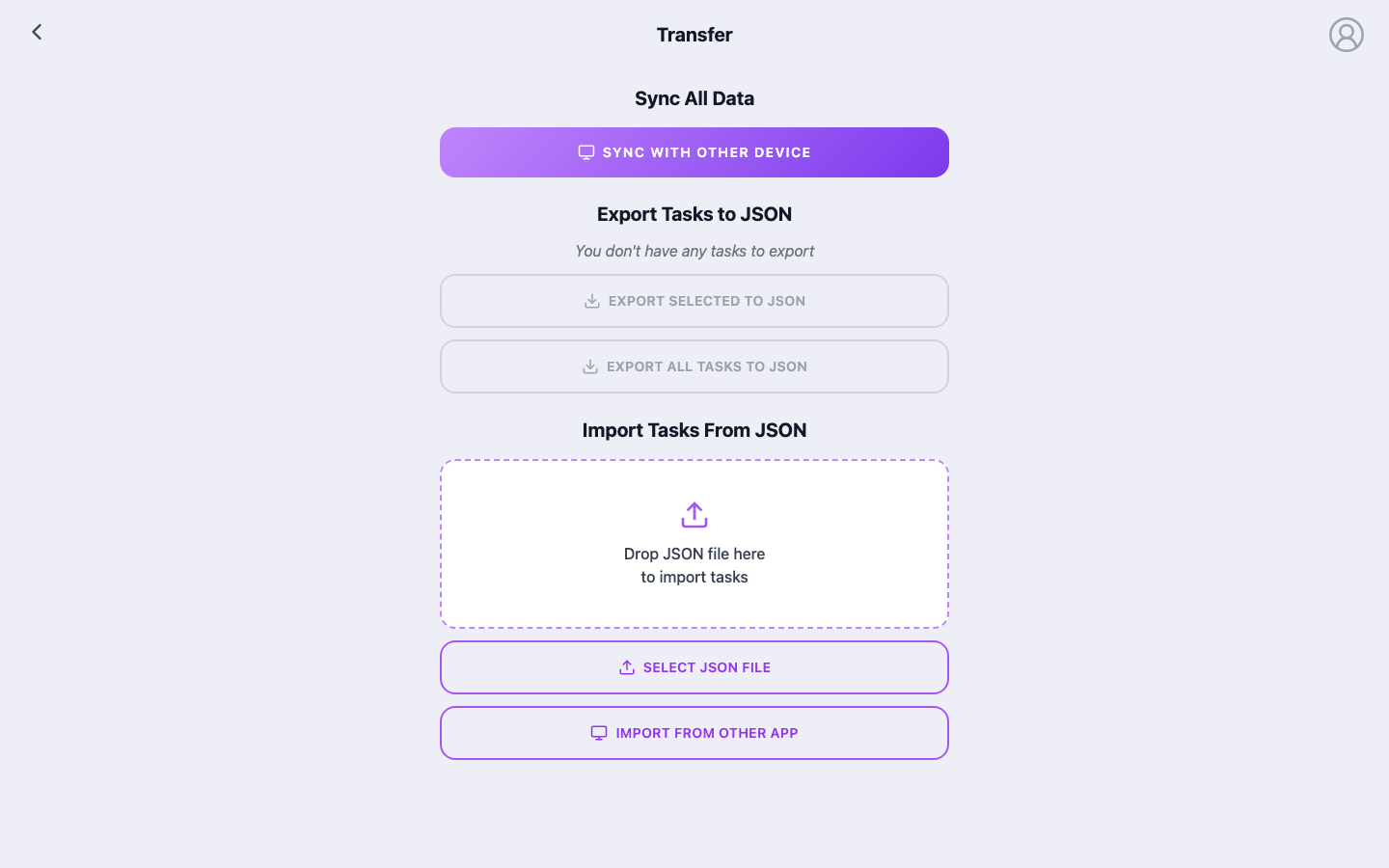}\\[2pt]
  {\footnotesize (c) Sonnet 4.6 \textcolor{blue}{onDrop wired}}
\end{minipage}\hfill
\begin{minipage}[t]{0.32\linewidth}\centering
  \includegraphics[width=\linewidth,keepaspectratio]{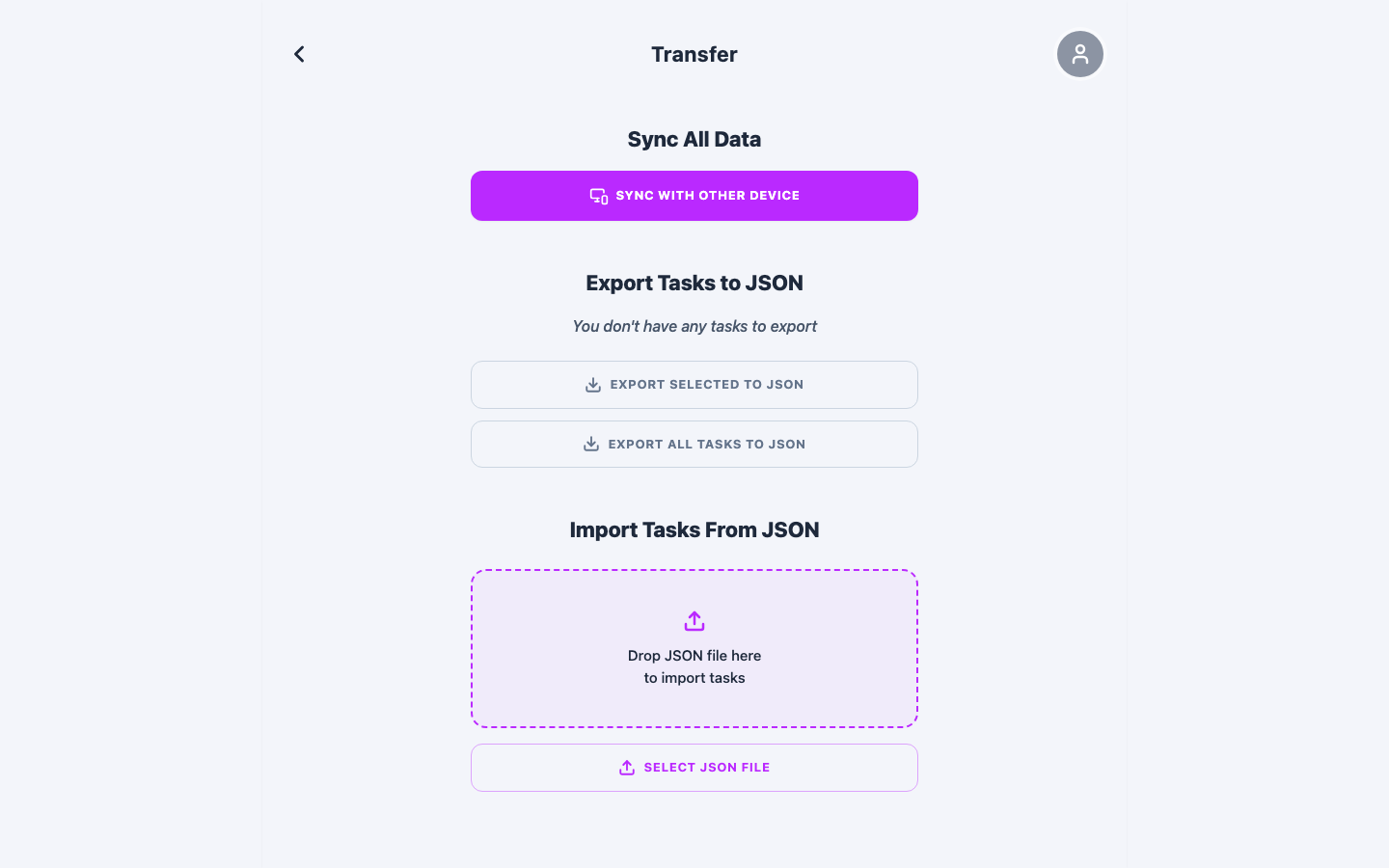}\\[2pt]
  {\footnotesize (d) Gemini 3.1 \textcolor{gray}{decoration only}}
\end{minipage}\par\medskip
\captionof{figure}{\textbf{HTML5 DnD primitive reach on \texttt{39\_maciekt07-todoapp}.} Kimi/Sonnet bind \texttt{onDragOver}+\texttt{onDrop} to a real file-import flow; Gemini produces the same rectangle as decoration. Identical visual targets, divergent event topology. Pairs with the audio case (Fig.~\ref{fig:case_genshin}) to illustrate that frontier models repeatedly reach for domain primitives invisible at the pixel level.}
\label{fig:case_dnd}
\par}
\medskip

\paragraph{Rich-text editing (\texttt{67\_shadboard}).}
\label{app:case-editor}
\texttt{67\_shadboard} ships an editor screen whose only behavioral cue is a formatting toolbar (bold, italic, list, link) above an empty document area. Claude~Sonnet~4.6 and Qwen3.5-397B-A17B mark the document area with \texttt{contentEditable}, allowing genuine cursor-driven typing on the rendered page. Kimi~K2.5 does the same and additionally binds the toolbar buttons to \texttt{document.execCommand}. GPT-5.4 is the sharper negative: it renders the full editor UI --- a formatting toolbar above a document pane --- that \emph{looks} like a working editor, yet the pane is a static \texttt{<div>} (the toolbar icons carry no handler and there is no \texttt{<textarea>} or \texttt{contentEditable} element), so typing produces no change. The failure is indistinguishable from a real editor at idle and surfaces only under the live-typing test (Fig.~\ref{fig:case_editor}). The remaining two models land between these poles: GLM-4.6V wires a plain \texttt{<textarea>} that accepts text but leaves its formatting toolbar inert, and Gemini~3.1~Pro Preview emits no \texttt{/editor} route at all.

\par\medskip
{\centering
\begin{minipage}{\linewidth}\centering
  \includegraphics[width=\linewidth,keepaspectratio]{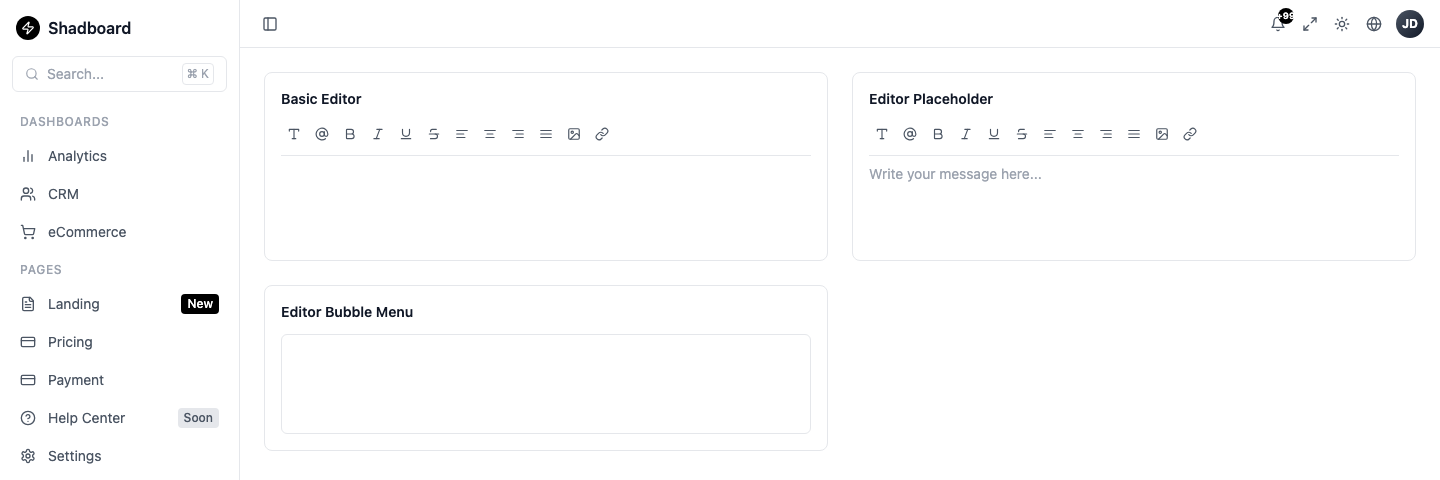}\\[2pt]
  {\footnotesize (a$_1$)~Sonnet \texttt{/editor}, \textbf{idle}}
\end{minipage}\par\smallskip
\begin{minipage}{\linewidth}\centering
  \includegraphics[width=\linewidth,keepaspectratio]{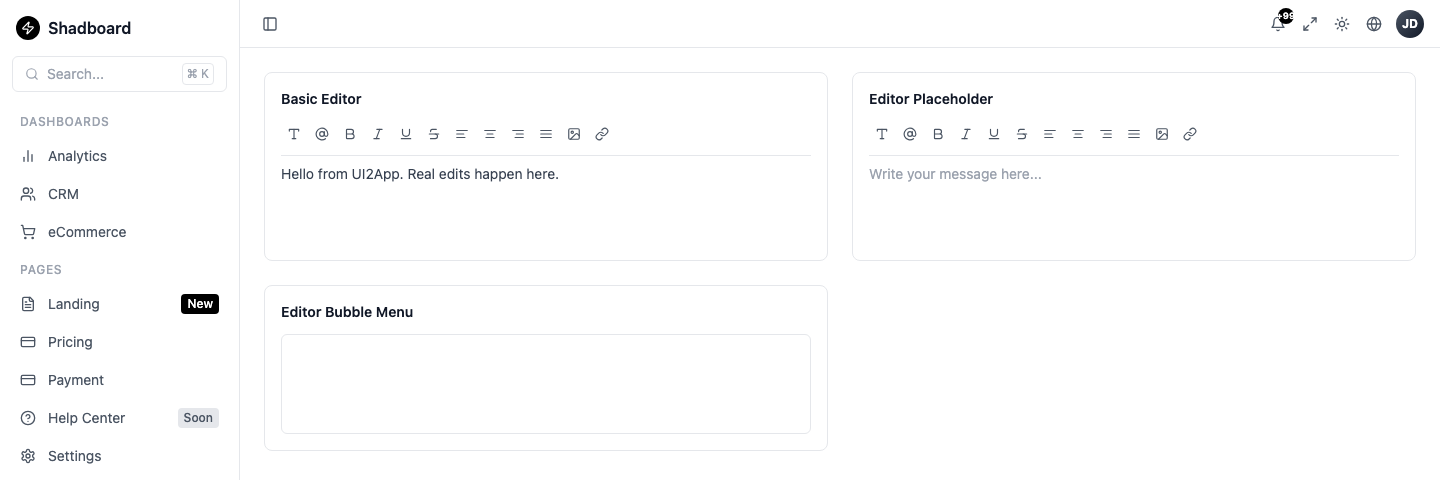}\\[2pt]
  {\footnotesize (a$_2$)~Sonnet, \textbf{after typing}: \textcolor{blue}{text appears}}
\end{minipage}\par\medskip
\begin{minipage}{\linewidth}\centering
  \includegraphics[width=\linewidth,keepaspectratio]{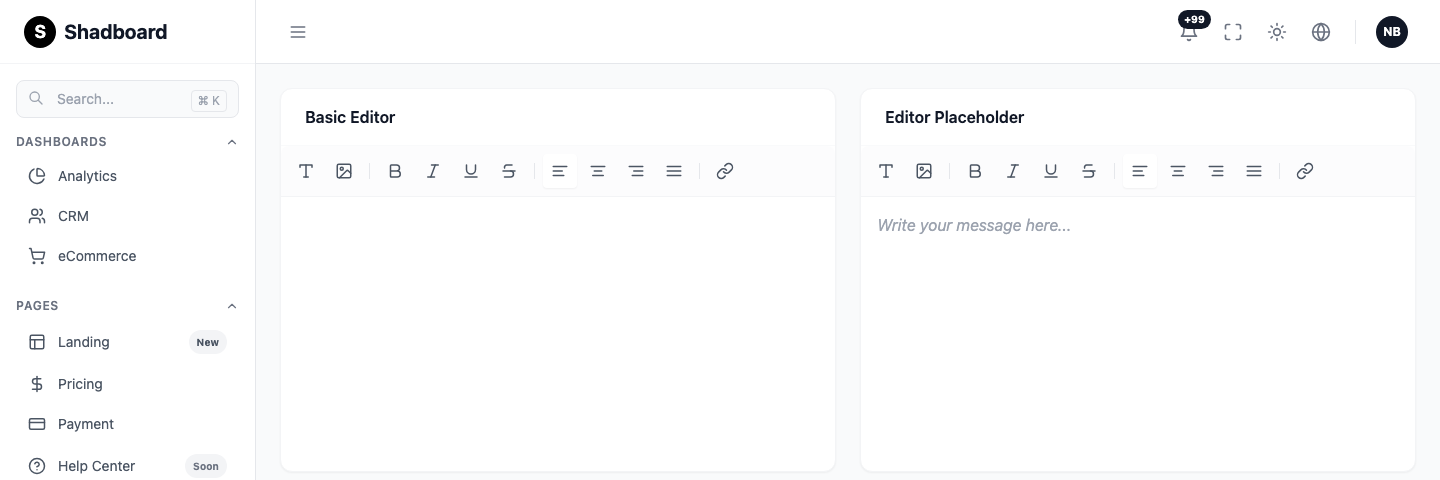}\\[2pt]
  {\footnotesize (b$_1$)~Qwen3.5 \texttt{/editor}, \textbf{idle}}
\end{minipage}\par\smallskip
\begin{minipage}{\linewidth}\centering
  \includegraphics[width=\linewidth,keepaspectratio]{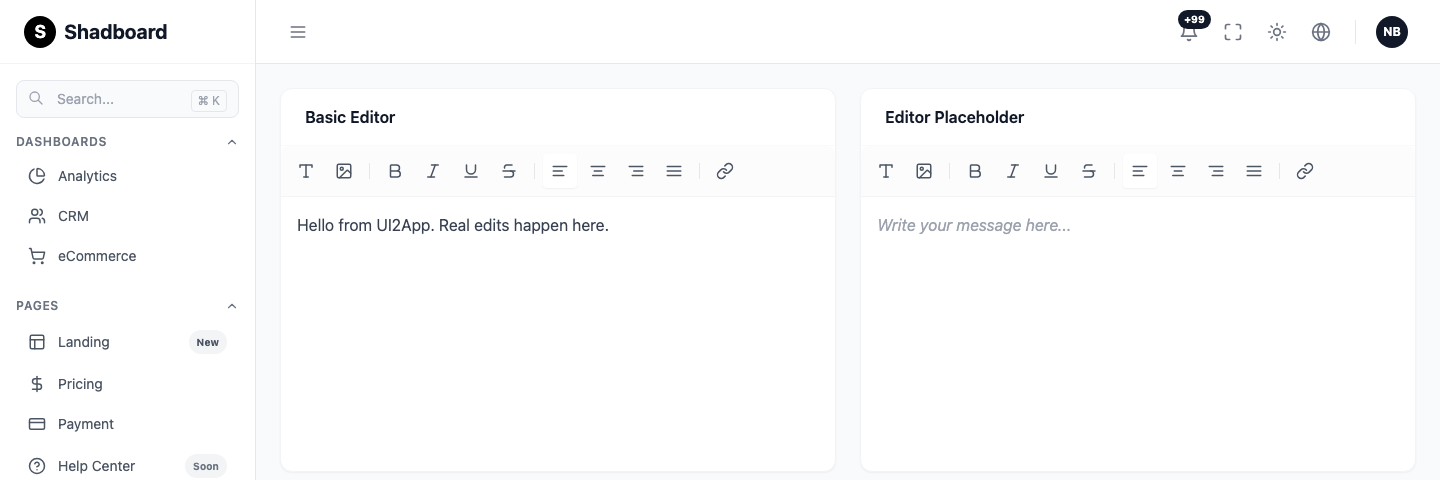}\\[2pt]
  {\footnotesize (b$_2$)~Qwen3.5, \textbf{after typing}: \textcolor{blue}{text appears}}
\end{minipage}\par\medskip
\begin{minipage}{\linewidth}\centering
  \includegraphics[width=\linewidth,keepaspectratio]{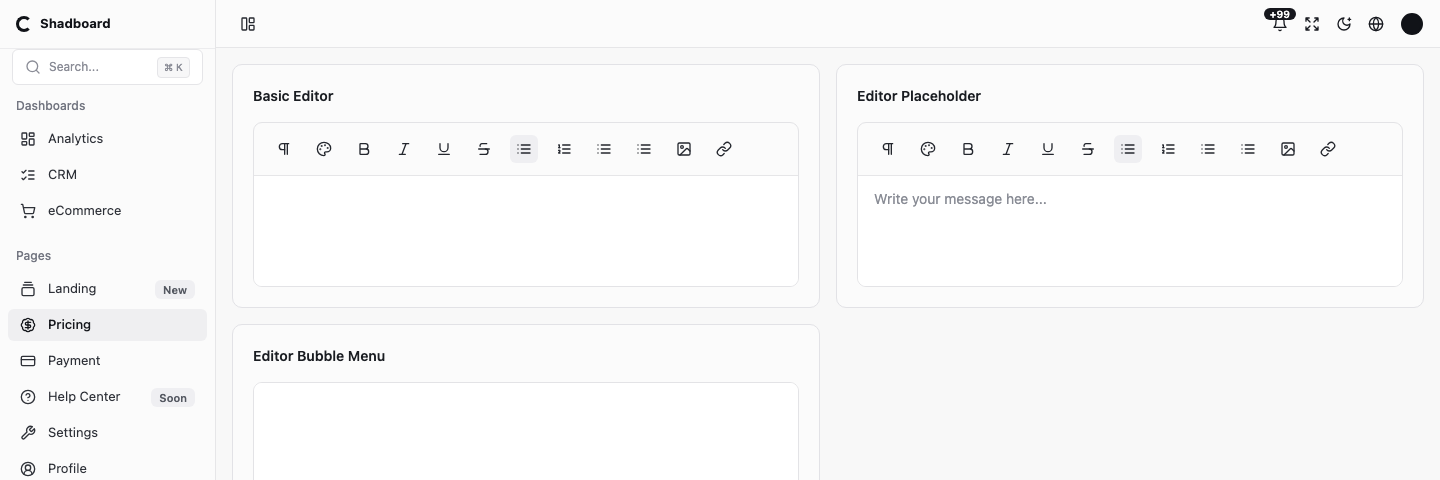}\\[2pt]
  {\footnotesize (c$_1$)~GPT-5.4 \texttt{/editor}, \textbf{idle}: \textcolor{gray}{looks like an editor}}
\end{minipage}\par\smallskip
\begin{minipage}{\linewidth}\centering
  \includegraphics[width=\linewidth,keepaspectratio]{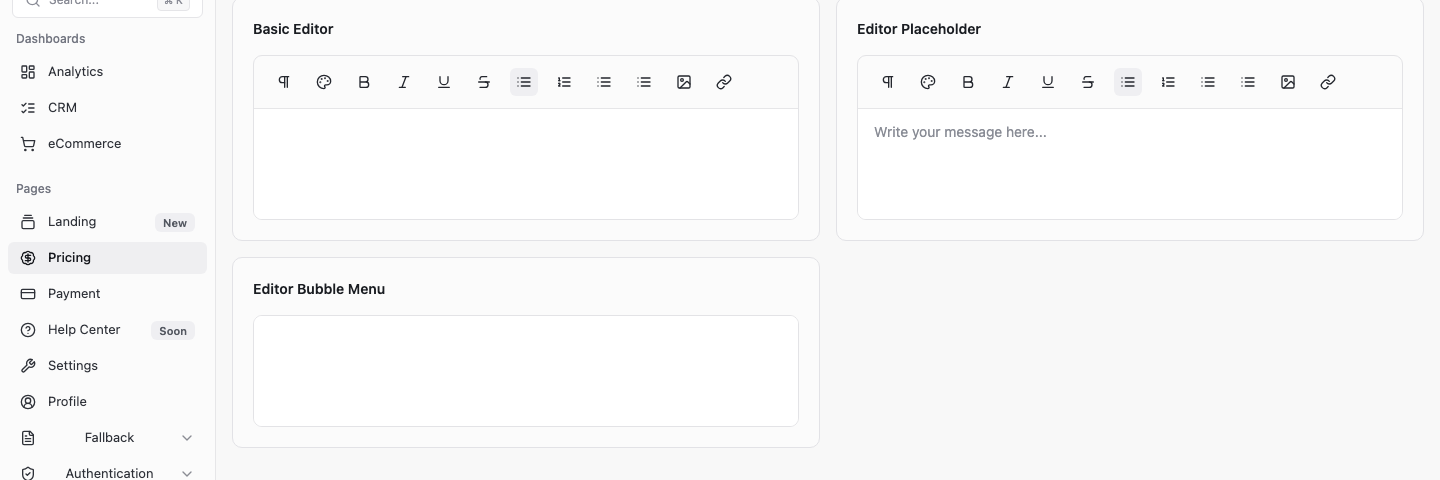}\\[2pt]
  {\footnotesize (c$_2$)~GPT-5.4, \textbf{after typing}: \textcolor{red}{no change} --- static \texttt{<div>}, no editable element}
\end{minipage}\par\medskip
\captionof{figure}{\textbf{Editor primitive on \texttt{67\_shadboard}: before/after typing into the document area.} Top of each pair (idle): each model immediately after navigating to its editor route (Sonnet: 3 \texttt{[contenteditable]} regions detected; Qwen: 1; GPT-5.4: 0 --- the document panes are static \texttt{<div>}s under a decorative toolbar). Bottom (after typing): the same test string is typed into the first document pane. (a$_2$, b$_2$) Sonnet and Qwen show the typed text inside the Basic Editor card --- the toolbar+document affordance wired into a real editing surface. (c$_1$, c$_2$) GPT-5.4 renders a visually faithful editor whose panes are inert: the after-typing capture is identical to idle (no text appears). The live-typing test exposes the divergence \iis{} captures and pixel fidelity cannot.}
\label{fig:case_editor}
\par}
\medskip

\paragraph{Reference-subset rendering (\texttt{62\_tokyo-mui-dashboard}).}
\label{app:case-grid}
The Tokyo crypto-dashboard organizes content into two vertical sections: a top section (Welcome header, \textsc{Account Balance} card, donut chart with per-asset breakdown) and a continuation (\textsc{Wallets} row of three asset cards plus the \textsc{Account Security} side card). Gemini~3.1~Pro Preview reproduces both sections: the \textsc{Wallets} header is visible at the bottom of the captured viewport, with the first row of asset-card frames already laid out. GLM-4.6V's render is a \emph{strict subset} of the reference: the top section is reproduced faithfully (the donut, the \textsc{Account Balance} card with the same numeric value and Send/Receive controls, the per-asset legend), while the \textsc{Wallets} row and \textsc{Account Security} card have \emph{no counterpart} in the generated DOM. We verify the subset relation rather than infer it from a viewport screenshot: a full-page Playwright capture across seven candidate routes (\texttt{/dashboard}, \texttt{/crypto}, \texttt{/cryptocurrency}, \texttt{/dashboards/crypto}, \texttt{/dashboard-crypto}, \texttt{/overview}, \texttt{/}) returns \texttt{document.documentElement.scrollHeight=900\,px} (exactly the viewport height, so no overflow exists where missing blocks could hide), and \texttt{document.body.innerText} on every probed route contains none of the \textsc{Wallets}/\textsc{Security} section keywords. The Wallets/Security blocks are absent from the rendered DOM, not pushed past the viewport.

\par\medskip
{\centering
\begin{minipage}{0.82\linewidth}\centering
  \includegraphics[width=\linewidth,keepaspectratio]{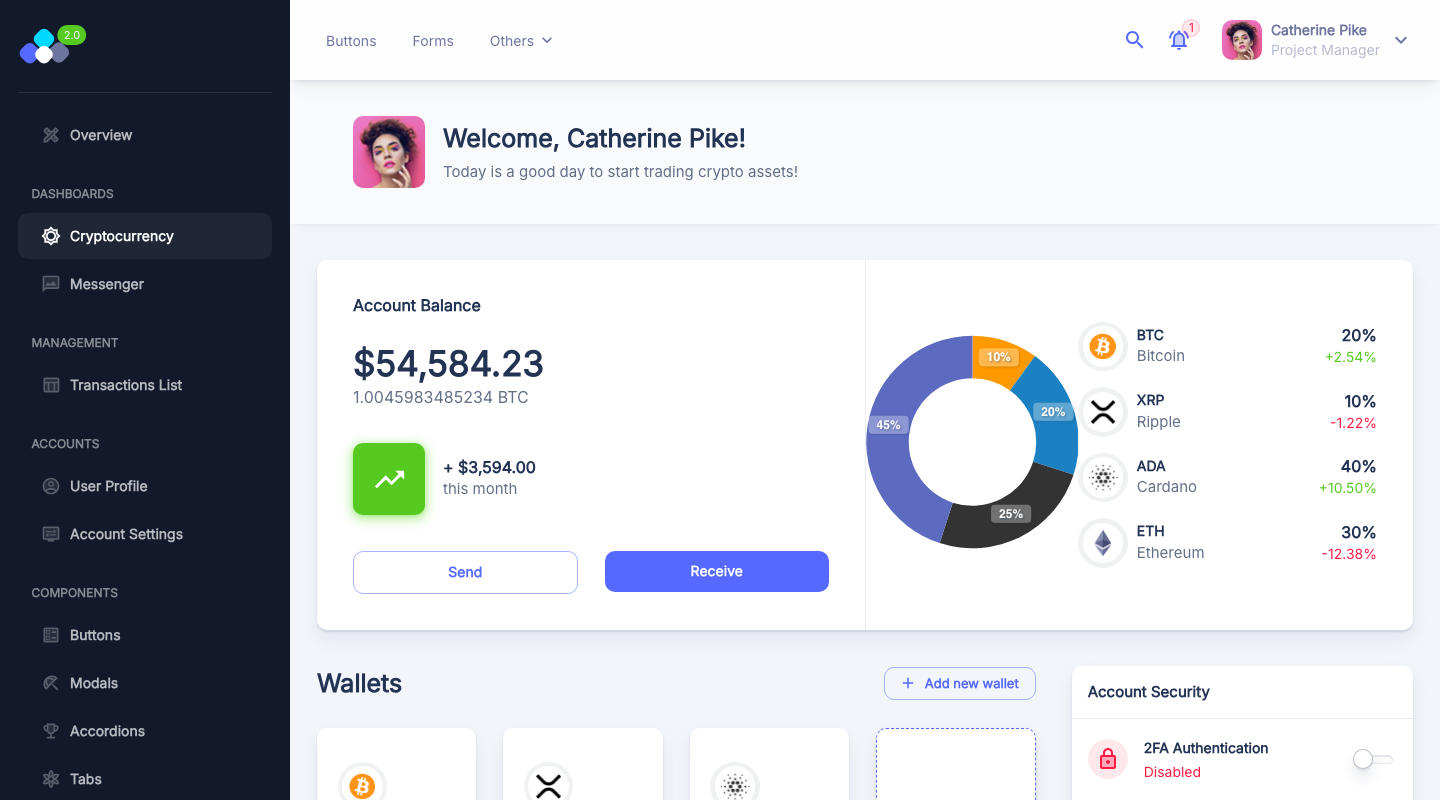}\\[2pt]
  {\footnotesize (a) input \texttt{/crypto}: top row + \textsc{Wallets} header + \textsc{Account Security} card visible}
\end{minipage}\par\medskip
\begin{minipage}[t]{0.49\linewidth}\centering
  \includegraphics[width=\linewidth,keepaspectratio]{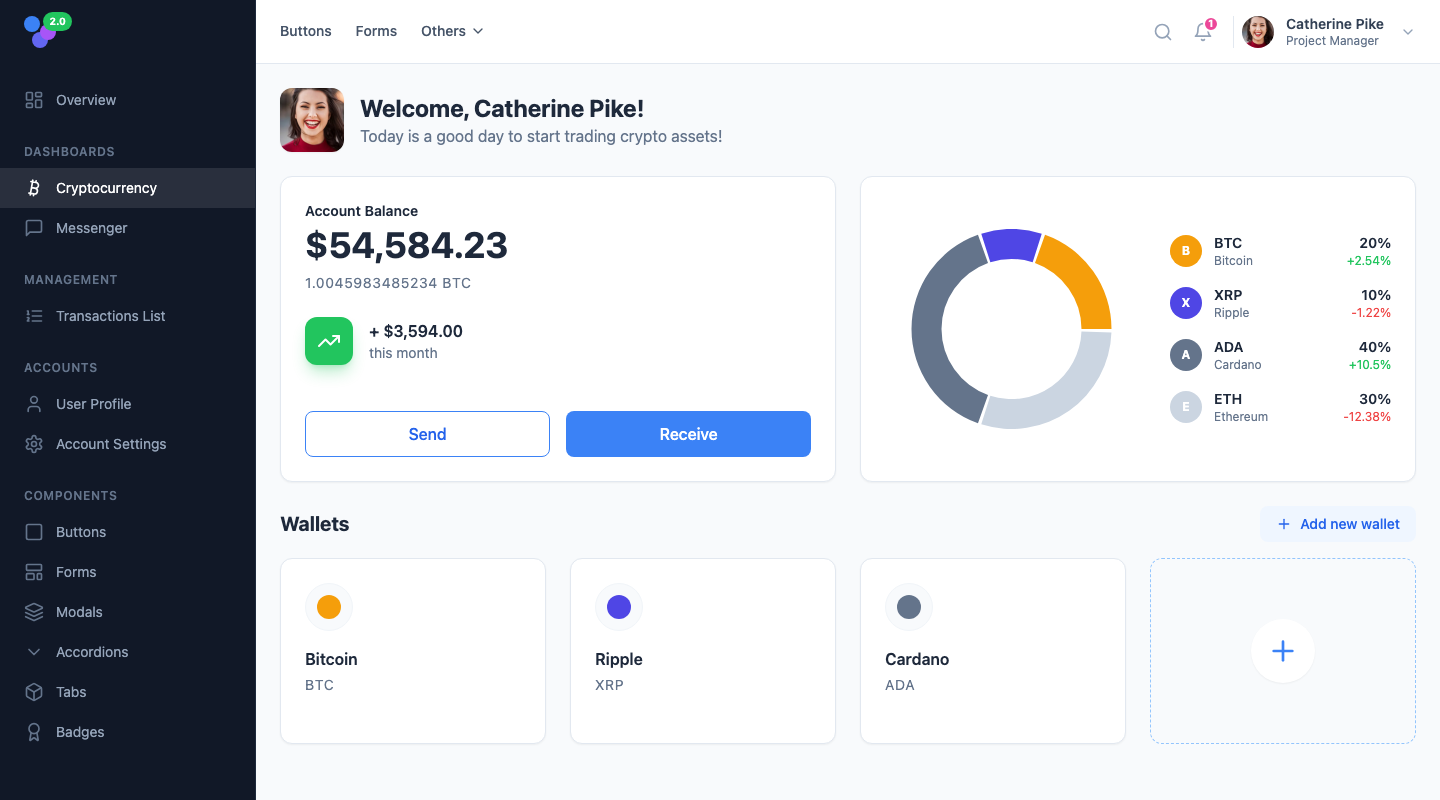}\\[2pt]
  {\footnotesize (b) Gemini 3.1 \textcolor{blue}{below-fold preserved}: top row + \textsc{Wallets} row}
\end{minipage}\hfill
\begin{minipage}[t]{0.49\linewidth}\centering
  \includegraphics[width=\linewidth,keepaspectratio]{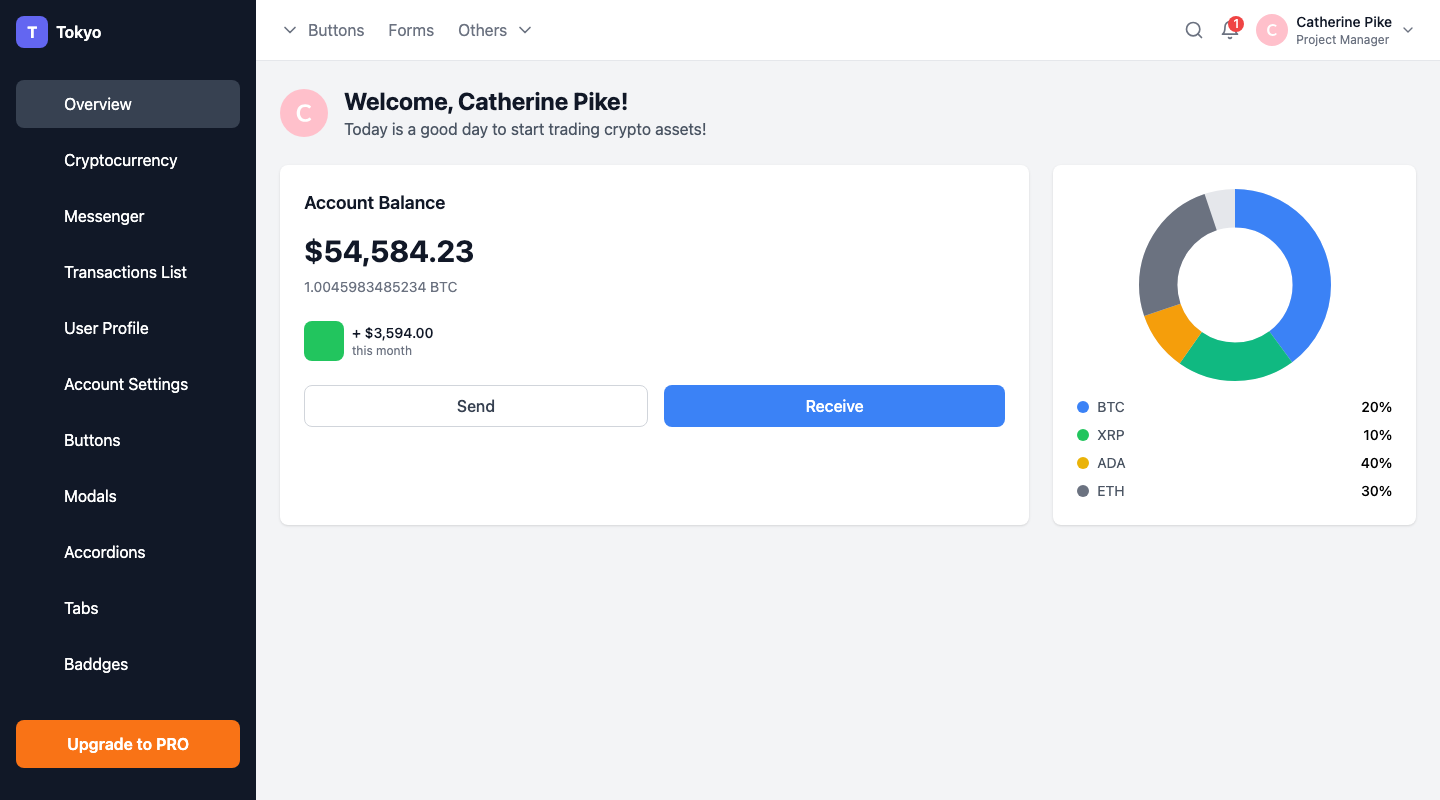}\\[2pt]
  {\footnotesize (c) GLM-4.6V \textcolor{red}{strict reference subset}: top section only; Wallets/Security blocks have no DOM counterpart (\texttt{scrollHeight=900}, probed keywords absent)}
\end{minipage}\par\medskip
\captionof{figure}{\textbf{Reference-subset rendering on \texttt{62\_tokyo-mui-dashboard/crypto}.} Pictorial counterpart to \S\ref{app:cond_submetric}. GLM-4.6V's rendered DOM is a strict subset of the reference dashboard's DOM: the top section is reproduced faithfully, while the reference's subsequent \textsc{Wallets}+\textsc{Security} blocks have no counterpart in the generated DOM. We verify the subset relation, not just a viewport screenshot: a full-page Playwright capture across seven candidate routes returns \texttt{scrollHeight}\,=\,$900$\,px (no overflow) and \texttt{innerText} contains none of the \textsc{Wallets}/\textsc{Security} section keywords. Gemini preserves both sections.}
\label{fig:case_grid}
\par}
\medskip

\paragraph{Build-time failures (\texttt{55\_shadcn-nextjs-dashboard}, \texttt{60\_mantine-dashboard}).}
\label{app:case-build}
The two dominant \exect[3] modes (Table~\ref{tab:failure_modes}) have orthogonal etiologies. C1 (hallucinated icon) is a factuality error: GPT-5.4 imports \texttt{Funnel} from \texttt{lucide-react}, but in the scaffold-pinned \texttt{0.460.0} the icon is still exported as \texttt{Filter} (\texttt{Funnel} is the post-\texttt{0.460} canonical name); \texttt{vite build} aborts with ``\texttt{"Funnel" is not exported by \dots/lucide-react.js}''. C2 (\texttt{package.json} overwrite) is an instruction-compliance error: GLM-4.6V regenerates the scaffold's \texttt{package.json} despite the explicit ``Do NOT include scaffold files'' clause (Appendix~\ref{app:prompts}, clause~iv), pinning the unpublished version \texttt{"mantine":"\^{}1.0.0"}; \texttt{pnpm install} aborts with \texttt{ERR\_PNPM\_NO\_VERSIONS}.

\par\medskip
{\centering
  \begin{minipage}[t]{0.48\linewidth}
\begin{codecard}{AnalyticsPage.tsx}
import {
  Download, CalendarDays,
  Funnel              // <-- not exported
} from "lucide-react";
\end{codecard}
\captionof{lstlisting}{(a) C1, GPT-5.4 on \texttt{55\_shadcn-nextjs-dashboard}: hallucinated \texttt{lucide-react} export. \textcolor{red}{Factuality error}.}
  \end{minipage}\hfill
  \begin{minipage}[t]{0.48\linewidth}
\begin{codecard}{package.json (regenerated)}
{
  "dependencies": {
    "mantine": "^1.0.0"   // pkg name is
                          // @mantine/core
  }
}
\end{codecard}
\captionof{lstlisting}{(b) C2, GLM-4.6V on \texttt{60\_mantine-dashboard}: regenerated \texttt{package.json} (clause-iv violation). \textcolor{red}{Instruction-compliance error}.}
  \end{minipage}\par\medskip
\captionof{figure}{\textbf{Factuality vs instruction-compliance failures.} C1 (left) is a symbol-existence hallucination that the prompt does not pre-empt. C2 (right) violates clause (iv) of the plan-step prompt. The orthogonality of the two error modes motivates reporting an instruction-compliance rate alongside \exect[3] (\S\ref{app:failures-gen}).}
\label{fig:case_build}
\par}
\medskip

\paragraph{Functional quiz engine (\texttt{61\_react-duolingo}).}
The gamified learning app's $10$-screenshot input set (Fig.~\ref{fig:case_duolingo_input}) spans a lesson map, a lesson exercise, a leaderboard, a shop, a profile, and settings. The lesson exercise --- a prompt, three answer tiles, and a \textsc{Check} button --- has behavior the screenshots never state: selecting a tile, judging it against the correct answer, and awarding XP on success must all be inferred. Across the six baselines the rendered exercise is near-identical, yet only Claude~Sonnet~4.6 wires a working quiz engine --- \texttt{handleCheck} compares the selection to \texttt{question.correctId}, sets correct/incorrect feedback, and on a correct answer calls \texttt{incrementLessons} (advancing lesson progress and adding $10$ XP). Kimi~K2.5 only partially wires it (its \textsc{Check} handler navigates without validating or scoring), and the remaining four (Gemini~3.1~Pro Preview, GPT-5.4, Qwen3.5-397B-A17B, GLM-4.6V) render a static mockup whose \textsc{Check} button carries no handler. The $1$-functional / $1$-partial / $4$-static split is invisible to \vfs{} --- every render shows the same tiles and button --- and is the answer-validation analogue of the audio case (Fig.~\ref{fig:case_genshin}).

\par\medskip
{\centering
  \begin{minipage}[t]{0.41\linewidth}\centering\includegraphics[width=\linewidth,keepaspectratio]{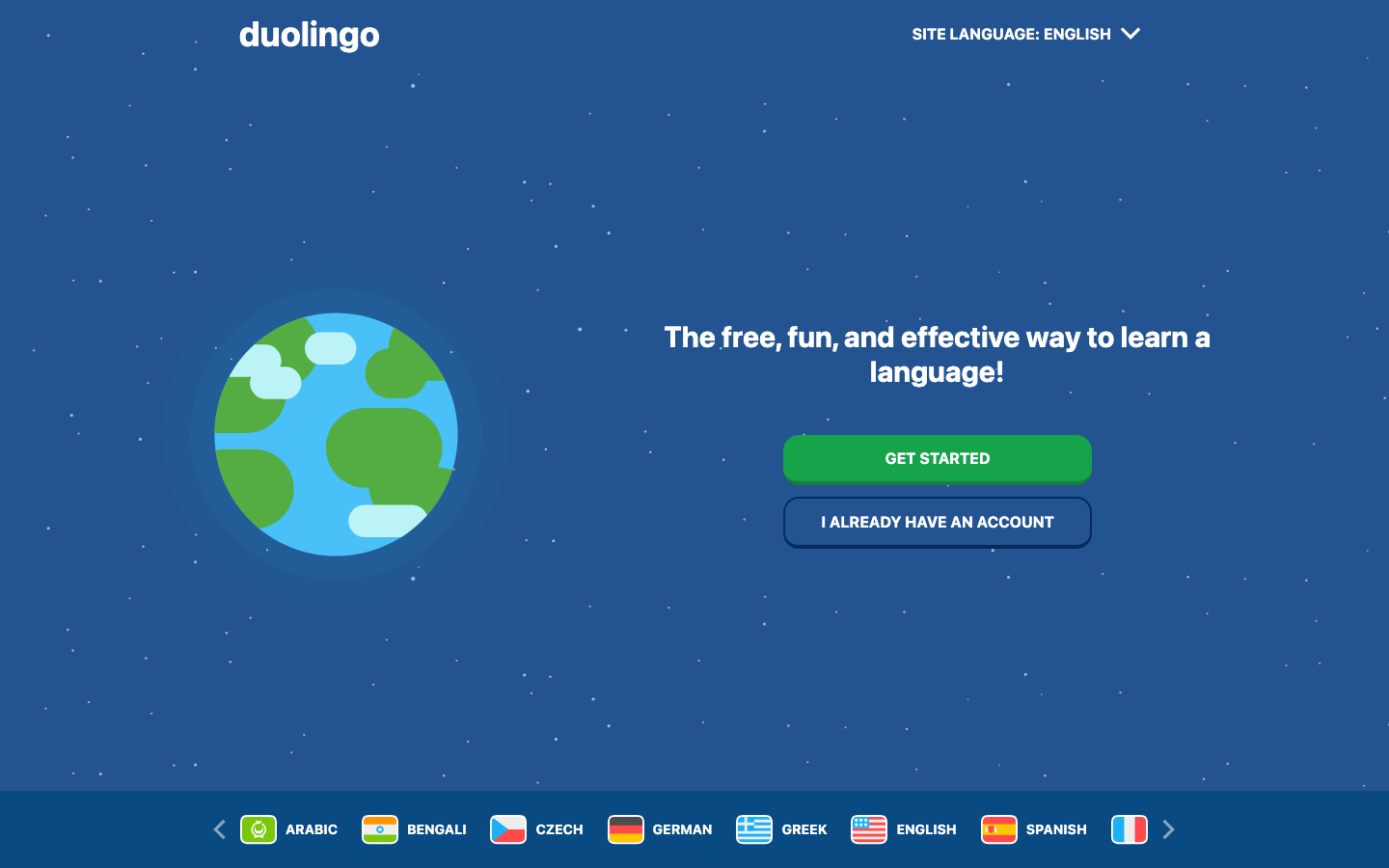}\\[2pt]{\footnotesize\texttt{/}}\end{minipage}\hfill
  \begin{minipage}[t]{0.41\linewidth}\centering\includegraphics[width=\linewidth,keepaspectratio]{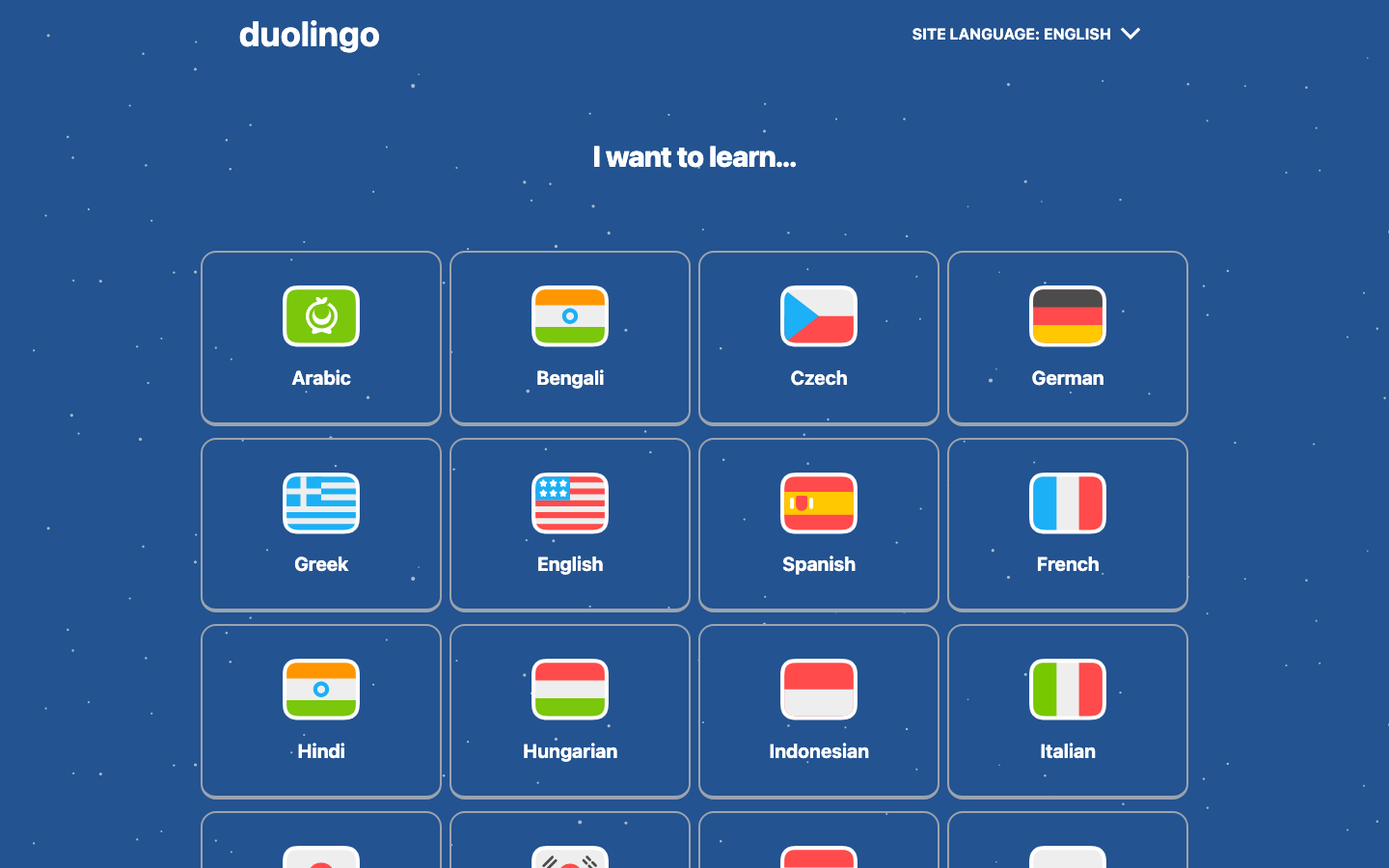}\\[2pt]{\footnotesize\texttt{/register}}\end{minipage}

  \vspace{4pt}
  \begin{minipage}[t]{0.41\linewidth}\centering\includegraphics[width=\linewidth,keepaspectratio]{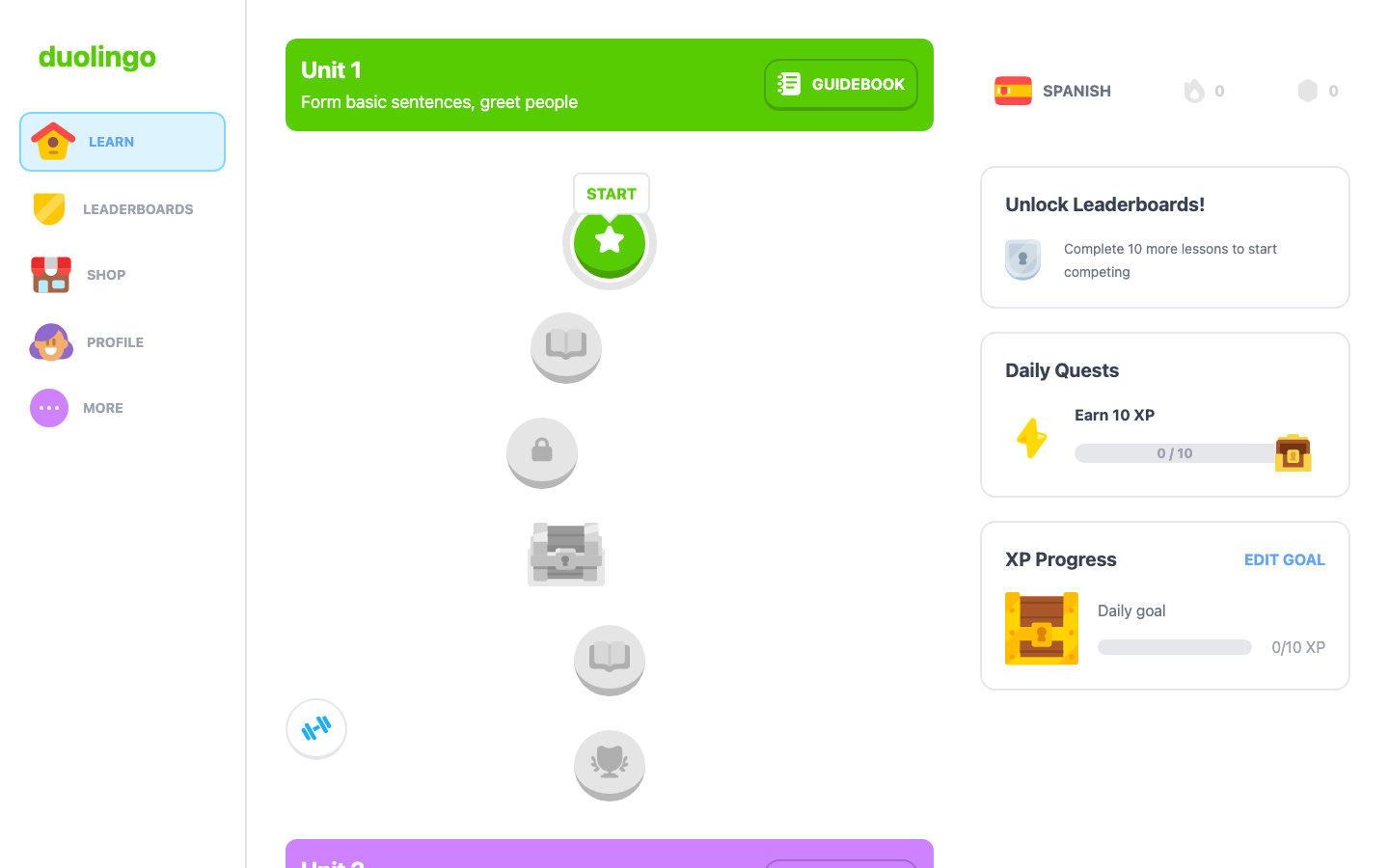}\\[2pt]{\footnotesize\texttt{/learn}}\end{minipage}\hfill
  \begin{minipage}[t]{0.41\linewidth}\centering\includegraphics[width=\linewidth,keepaspectratio]{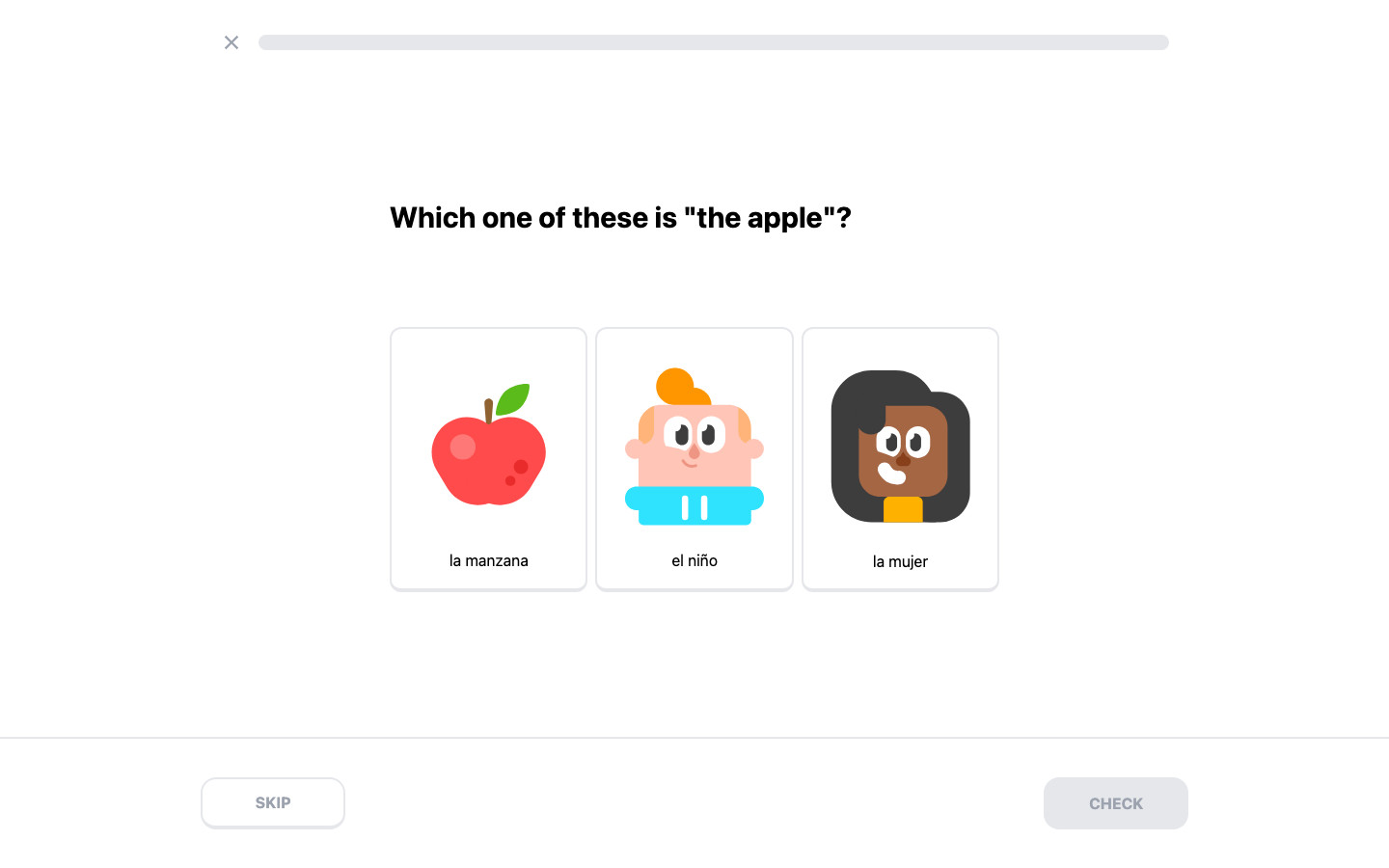}\\[2pt]{\footnotesize\texttt{/lesson}}\end{minipage}

  \vspace{4pt}
  \begin{minipage}[t]{0.41\linewidth}\centering\includegraphics[width=\linewidth,keepaspectratio]{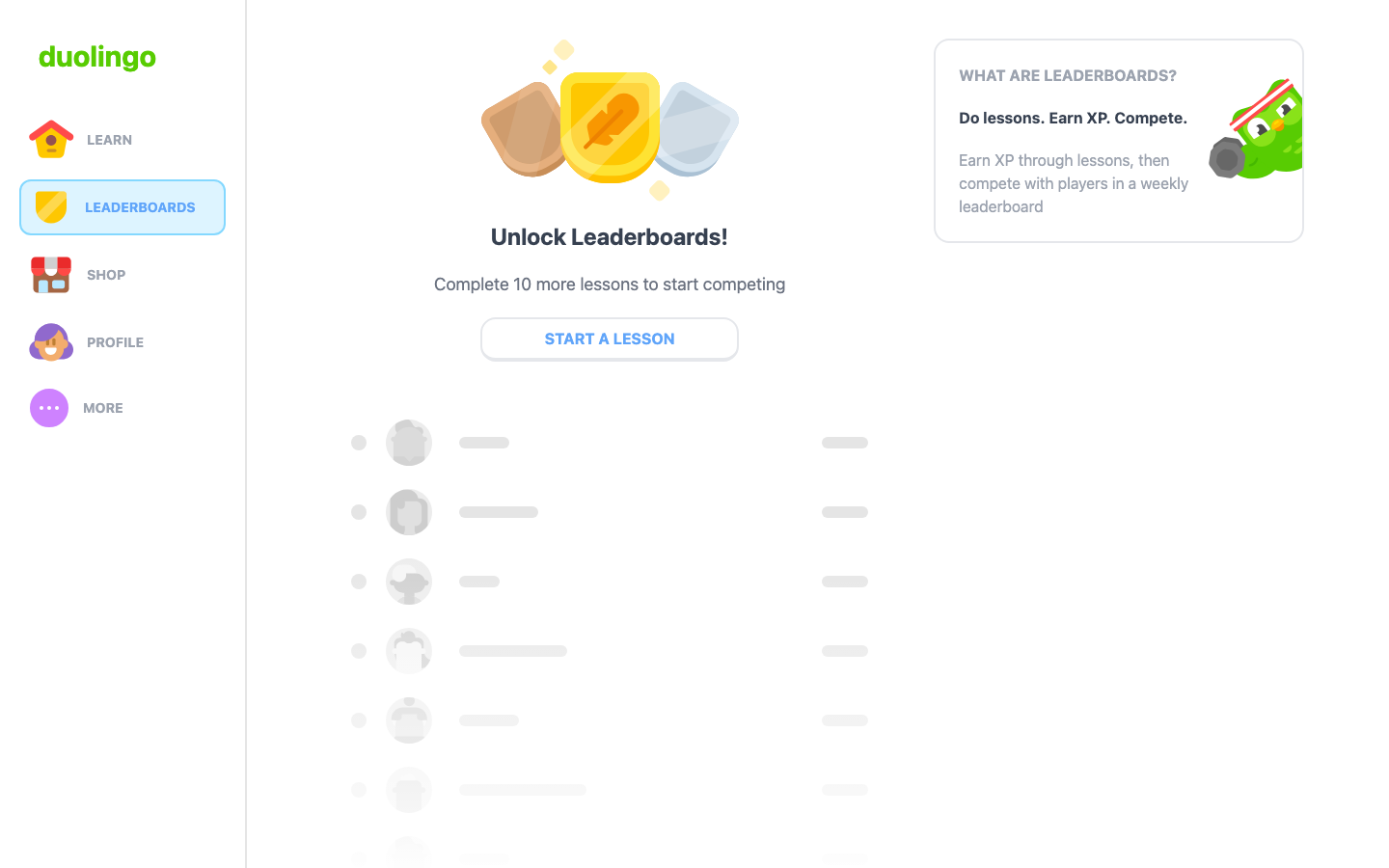}\\[2pt]{\footnotesize\texttt{/leaderboard}}\end{minipage}\hfill
  \begin{minipage}[t]{0.41\linewidth}\centering\includegraphics[width=\linewidth,keepaspectratio]{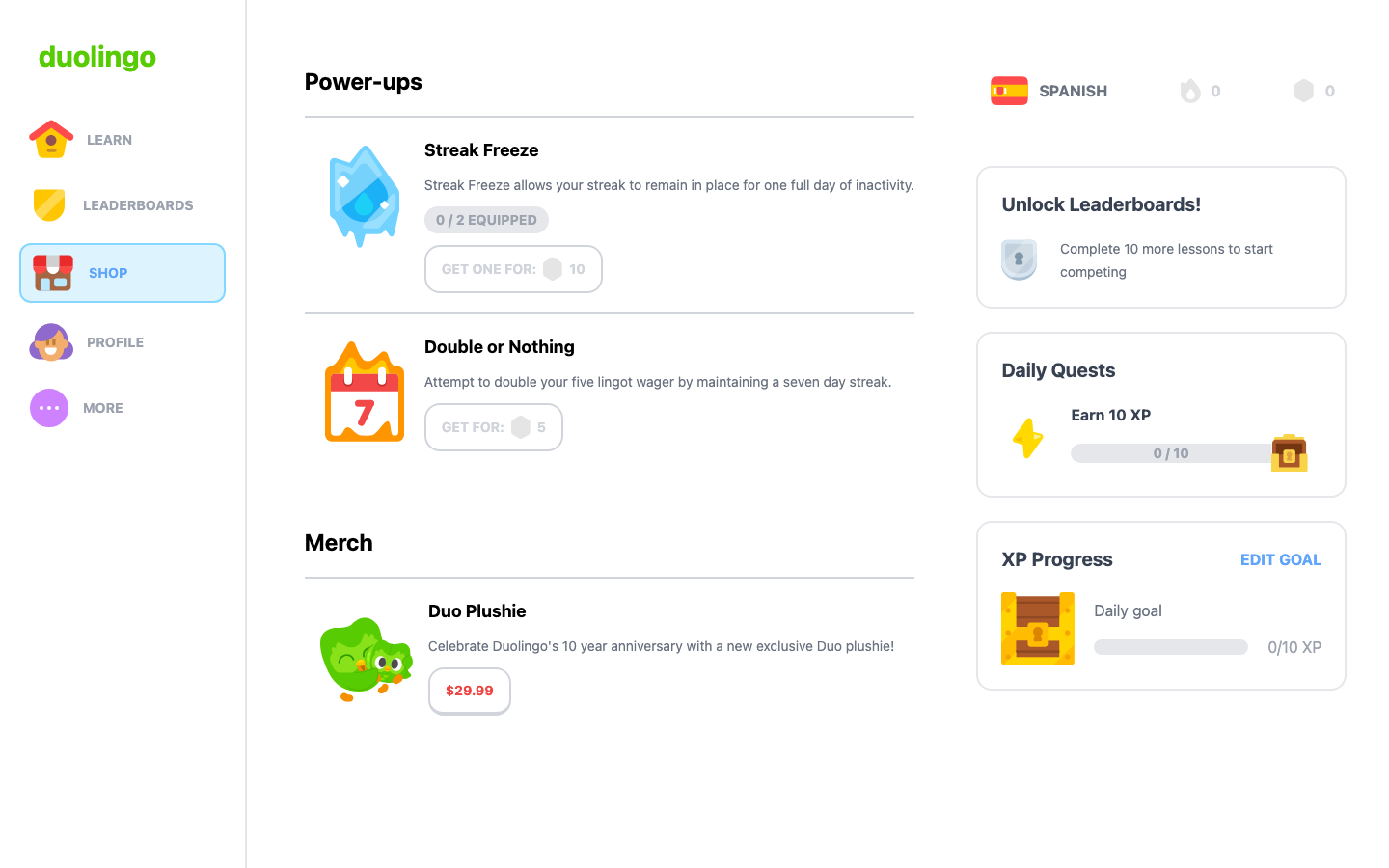}\\[2pt]{\footnotesize\texttt{/shop}}\end{minipage}

  \vspace{4pt}
  \begin{minipage}[t]{0.41\linewidth}\centering\includegraphics[width=\linewidth,keepaspectratio]{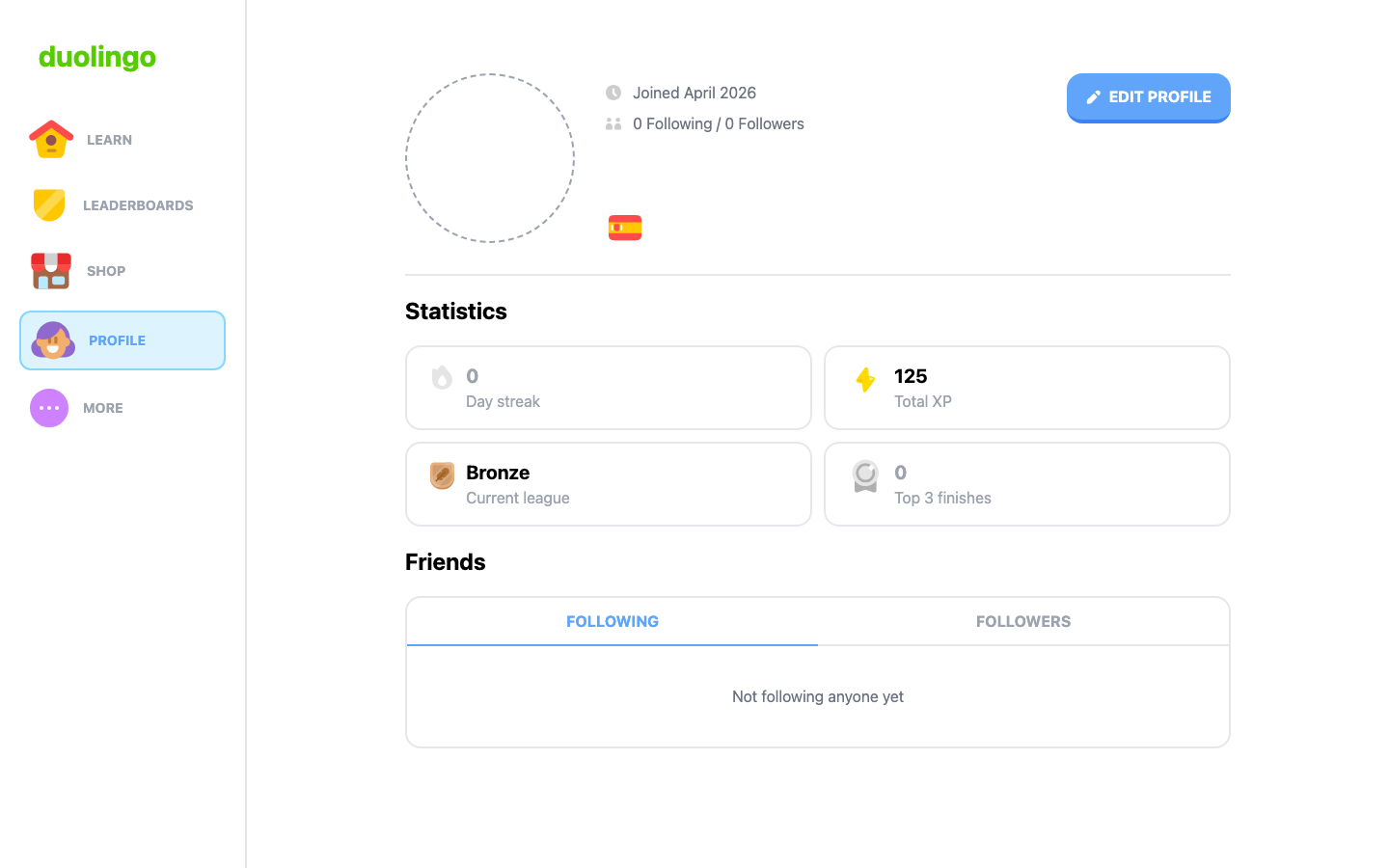}\\[2pt]{\footnotesize\texttt{/profile}}\end{minipage}\hfill
  \begin{minipage}[t]{0.41\linewidth}\centering\includegraphics[width=\linewidth,keepaspectratio]{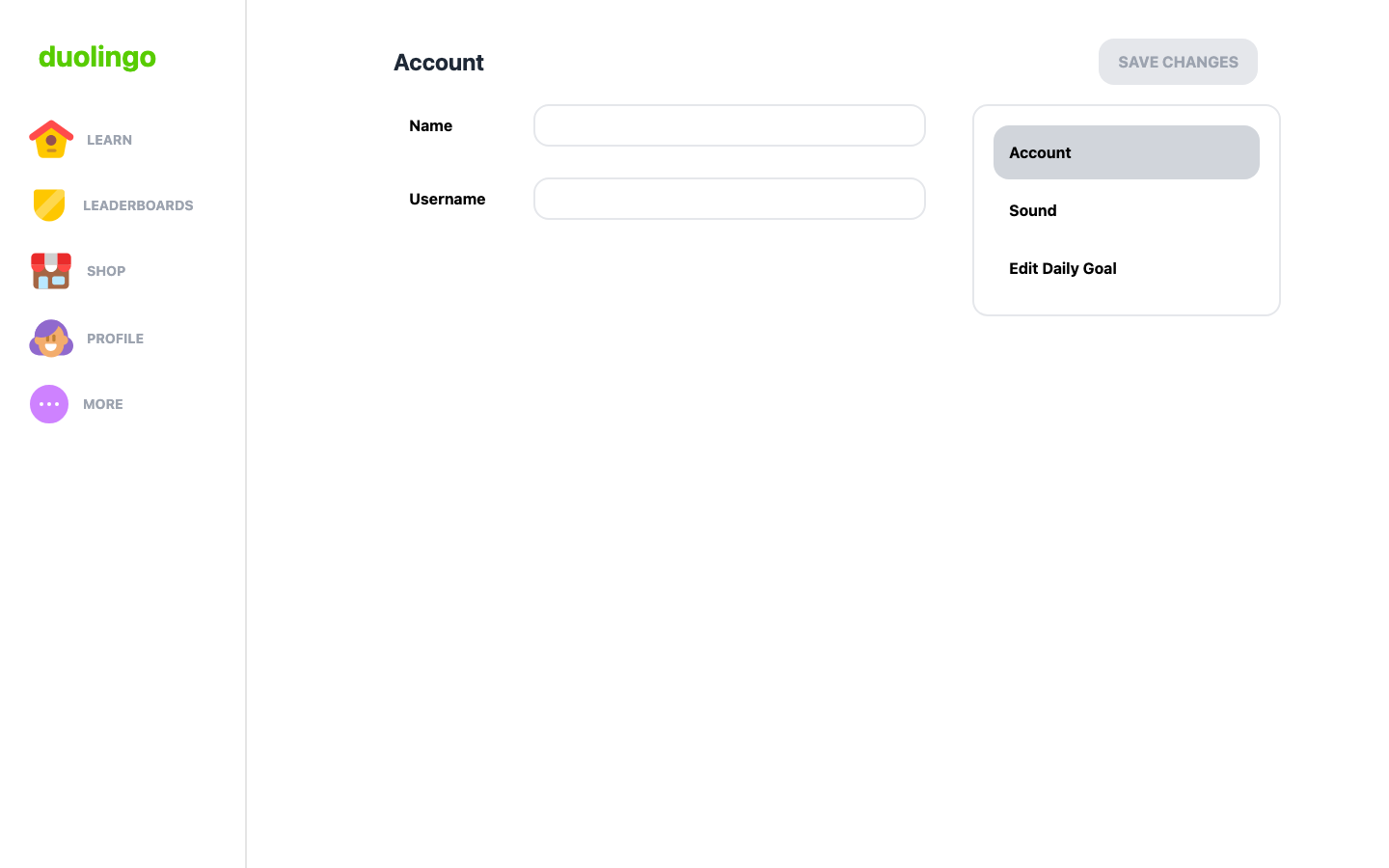}\\[2pt]{\footnotesize\texttt{/settings/account}}\end{minipage}

  \vspace{4pt}
  \begin{minipage}[t]{0.41\linewidth}\centering\includegraphics[width=\linewidth,keepaspectratio]{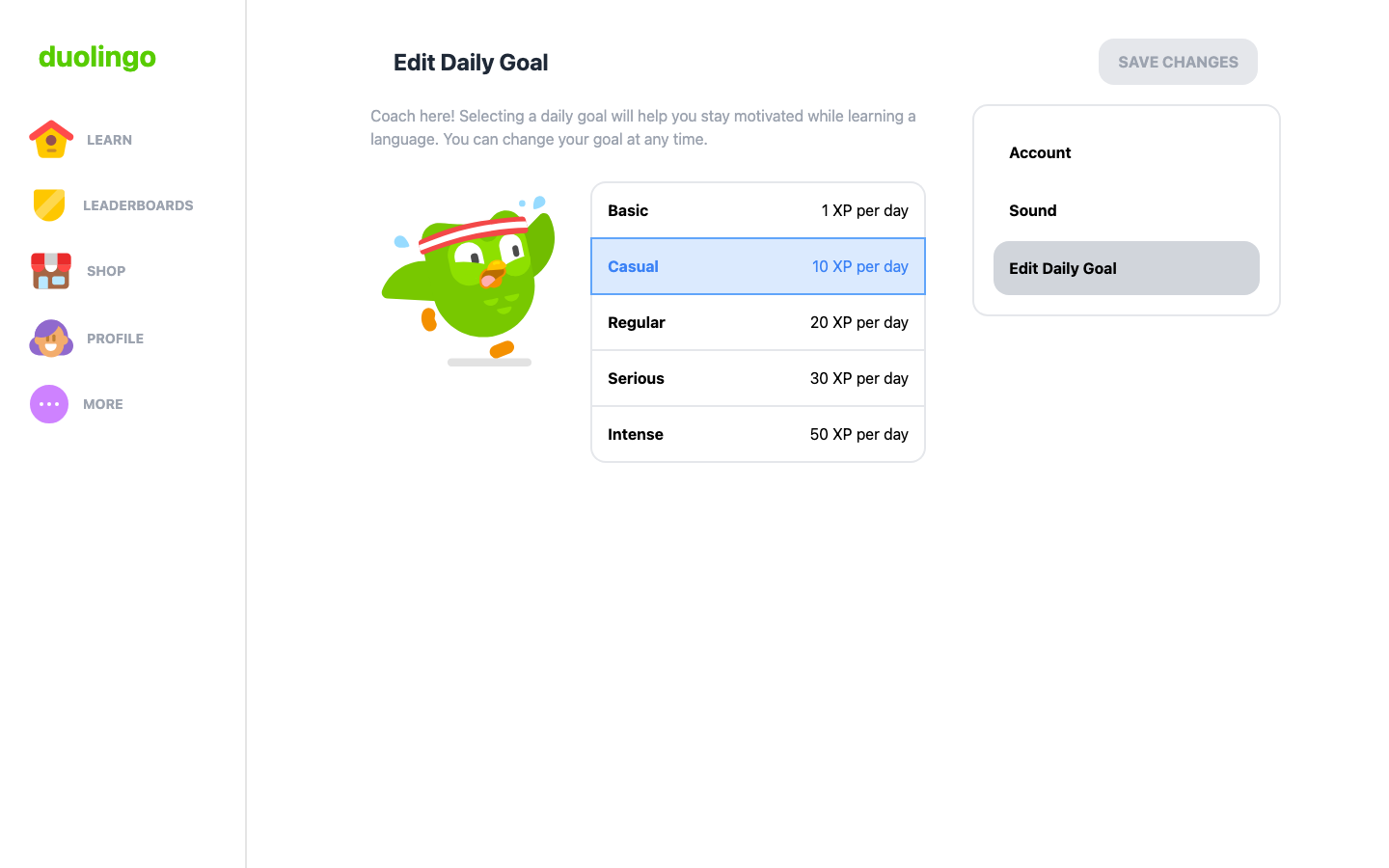}\\[2pt]{\footnotesize\texttt{/settings/coach}}\end{minipage}\hfill
  \begin{minipage}[t]{0.41\linewidth}\centering\includegraphics[width=\linewidth,keepaspectratio]{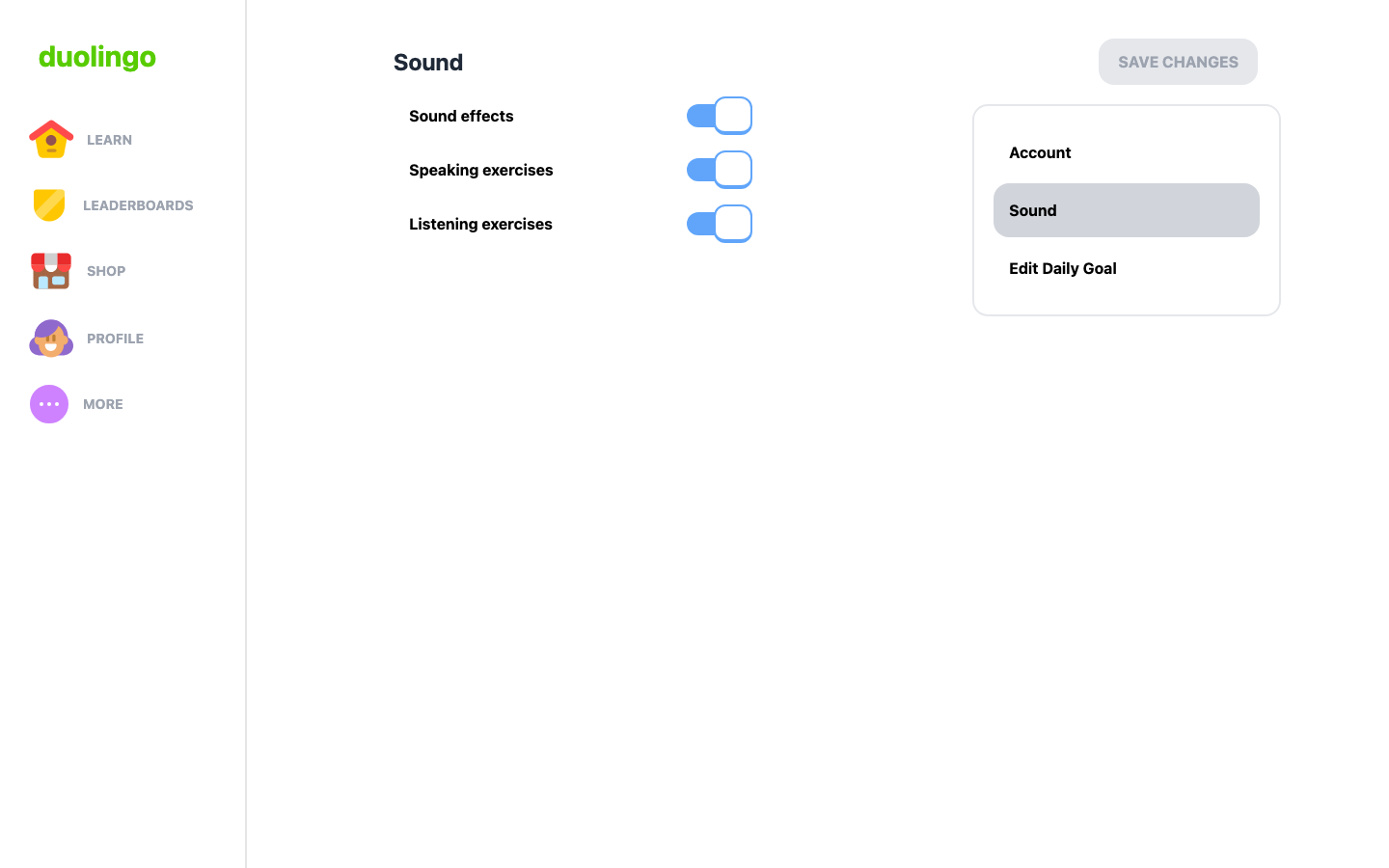}\\[2pt]{\footnotesize\texttt{/settings/sound}}\end{minipage}\par\medskip
\captionof{figure}{\textbf{Full input set for \texttt{61\_react-duolingo} ($10$ screenshots, one per route)} given to every model. The lesson exercise (\texttt{/lesson}) implies an answer-validation + XP-scoring engine; whether each model actually builds it is shown in Fig.~\ref{fig:case_duolingo}.}
\label{fig:case_duolingo_input}
\par}
\medskip

\par\medskip
{\centering
\begin{minipage}[t]{0.33\linewidth}\centering
  \includegraphics[width=\linewidth,keepaspectratio]{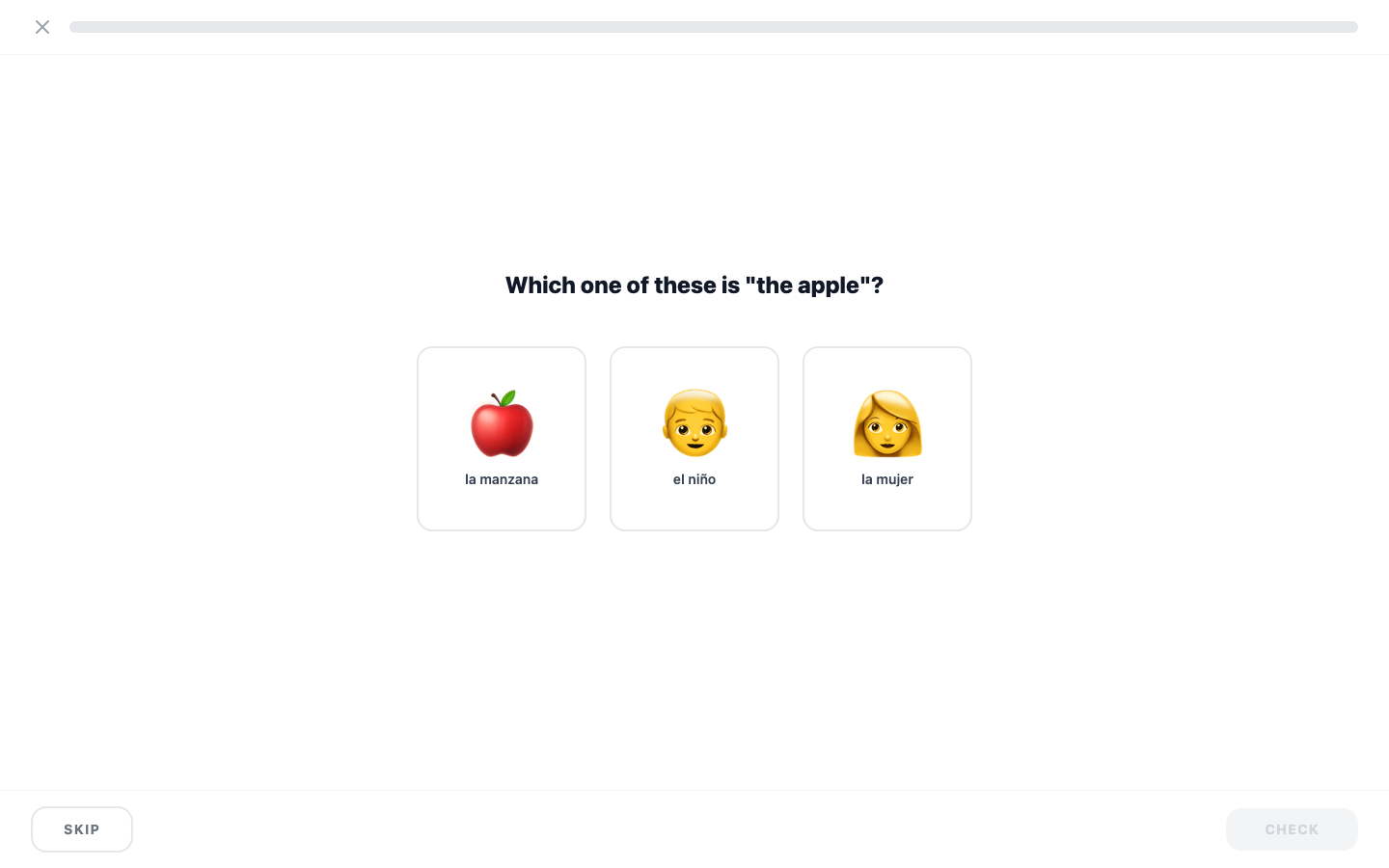}\\[2pt]
  {\footnotesize (a) Claude lesson --- \textbf{before}}
\end{minipage}\hfill
\begin{minipage}[t]{0.33\linewidth}\centering
  \includegraphics[width=\linewidth,keepaspectratio]{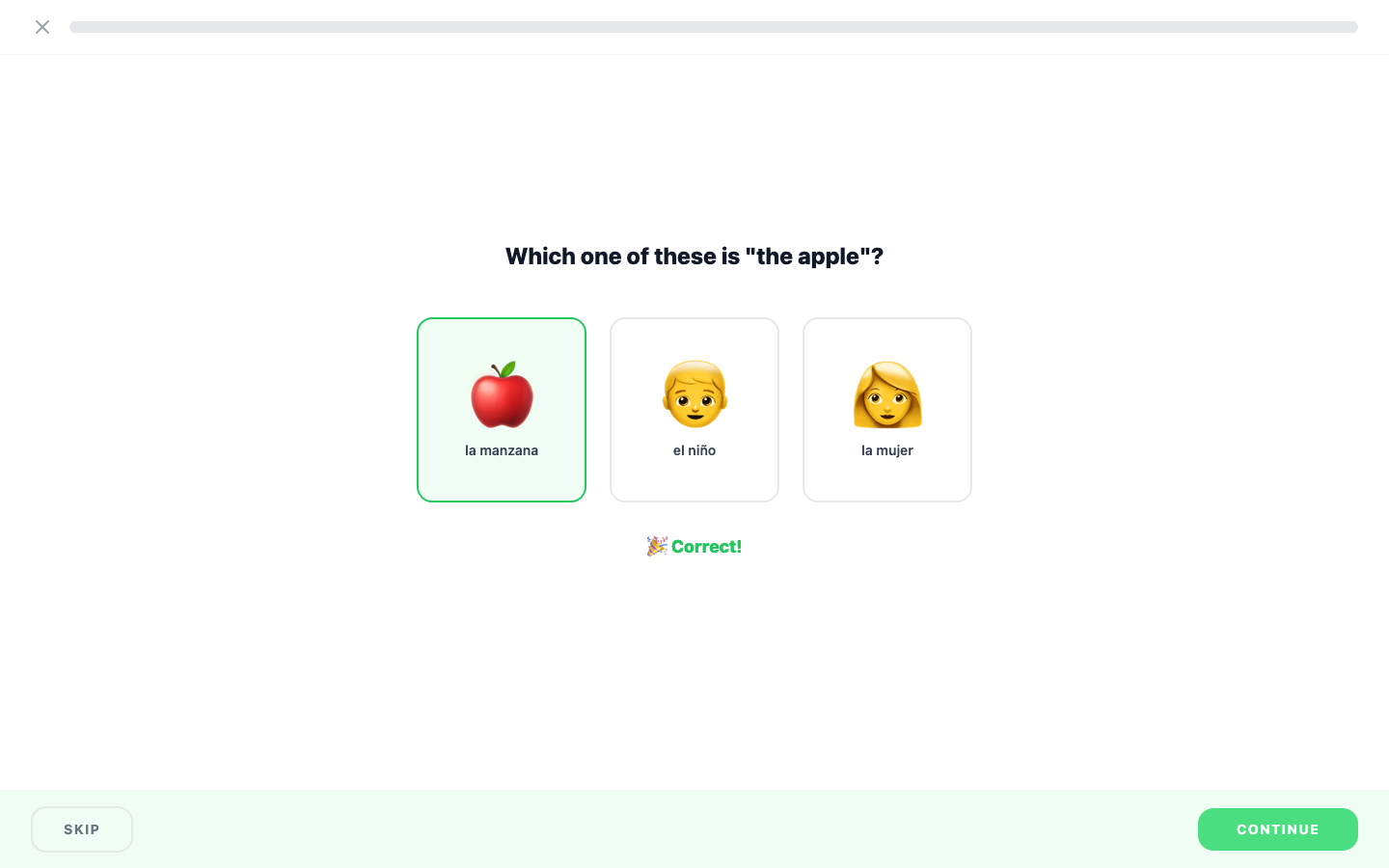}\\[2pt]
  {\footnotesize (b) Claude --- \textbf{after}: \textcolor{blue}{\emph{Correct!} + \textsc{Continue}}}
\end{minipage}\hfill
\begin{minipage}[t]{0.33\linewidth}\centering
  \includegraphics[width=\linewidth,keepaspectratio]{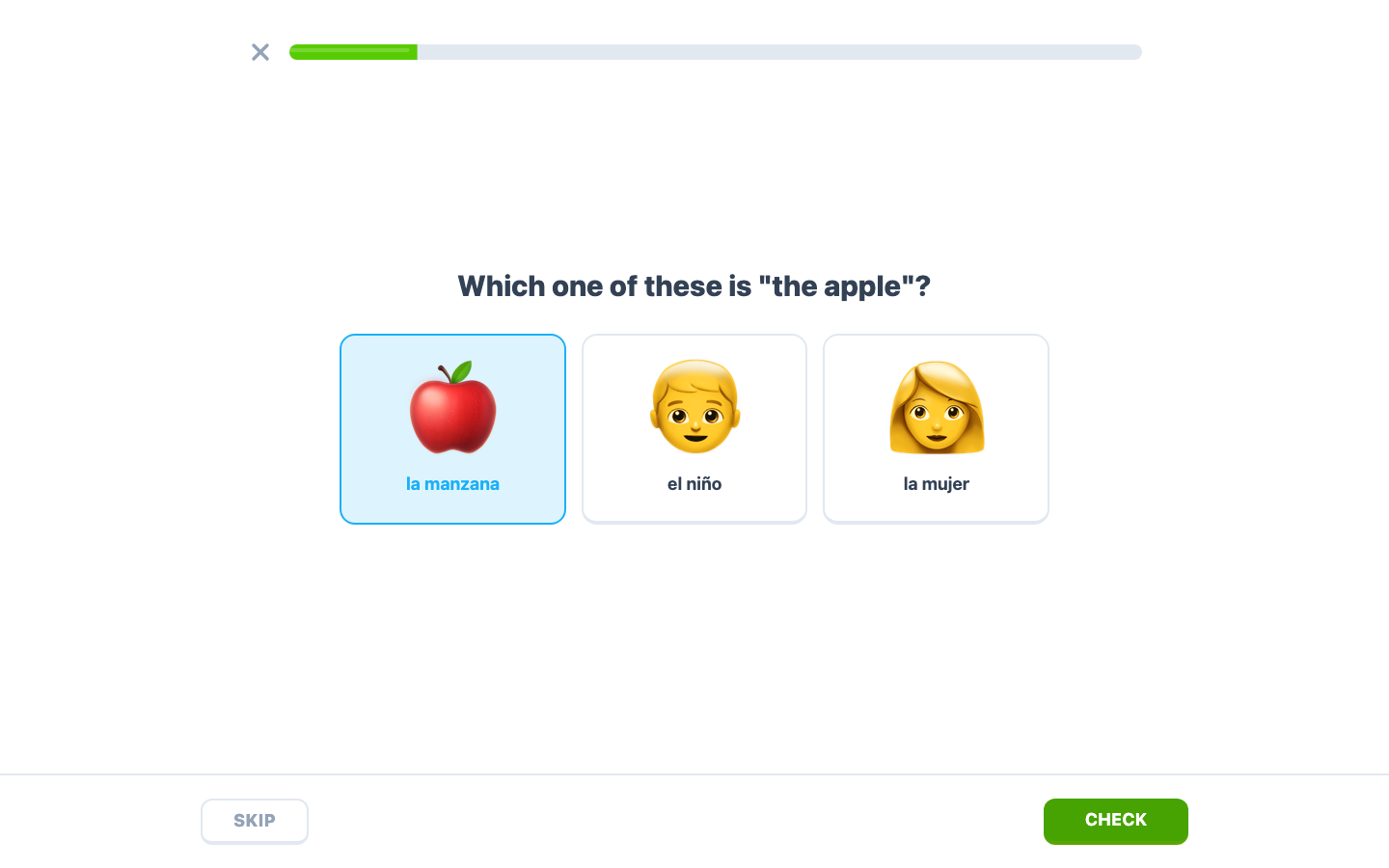}\\[2pt]
  {\footnotesize (c) Gemini --- \textbf{after}: \textcolor{gray}{inert, no feedback}}
\end{minipage}\par\medskip
\begin{minipage}[t]{0.49\linewidth}\centering
  \includegraphics[width=\linewidth,keepaspectratio]{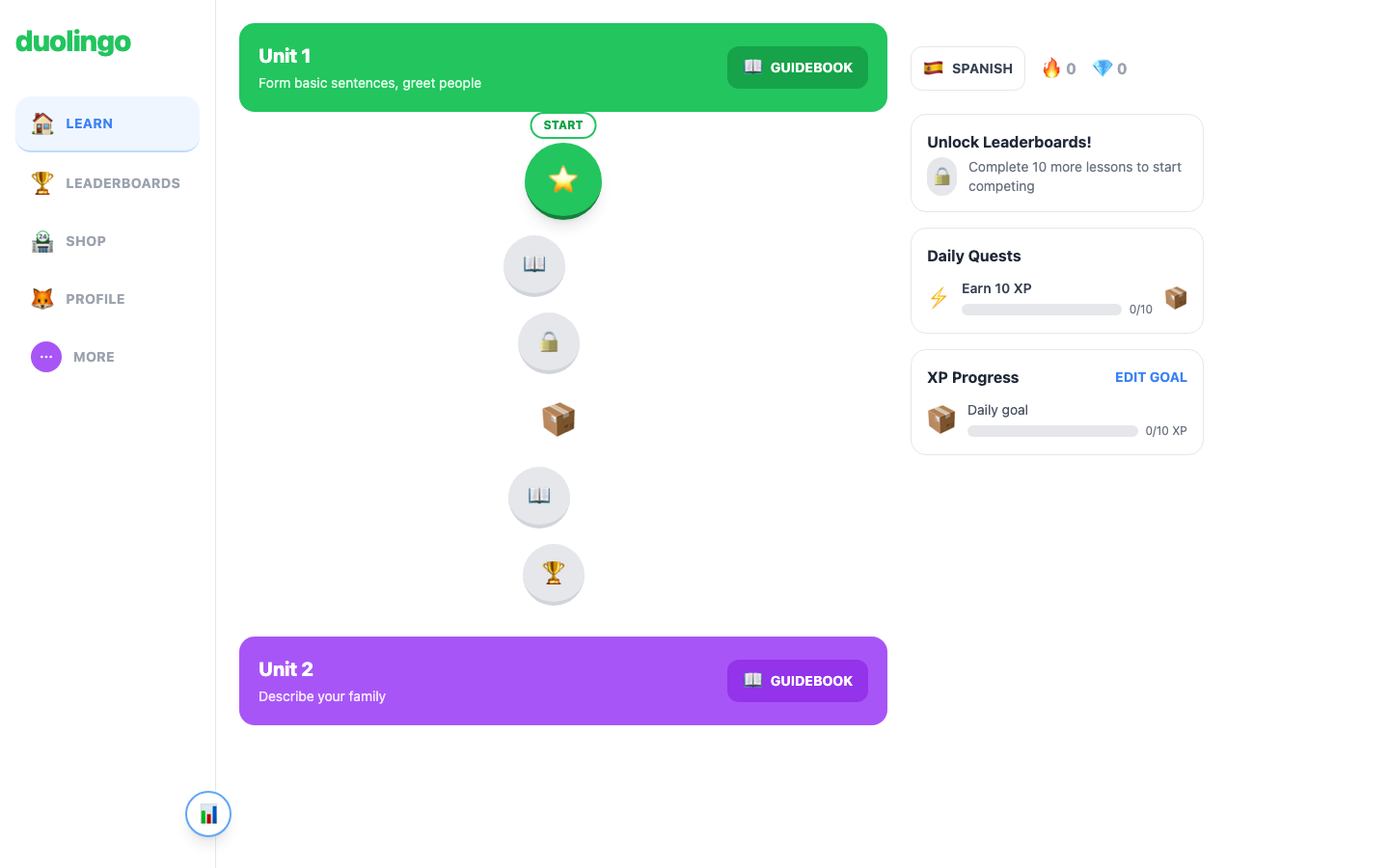}\\[2pt]
  {\footnotesize (d) Claude \texttt{/learn} --- \textbf{before}: Daily Quest \textcolor{gray}{0/10 XP}}
\end{minipage}\hfill
\begin{minipage}[t]{0.49\linewidth}\centering
  \includegraphics[width=\linewidth,keepaspectratio]{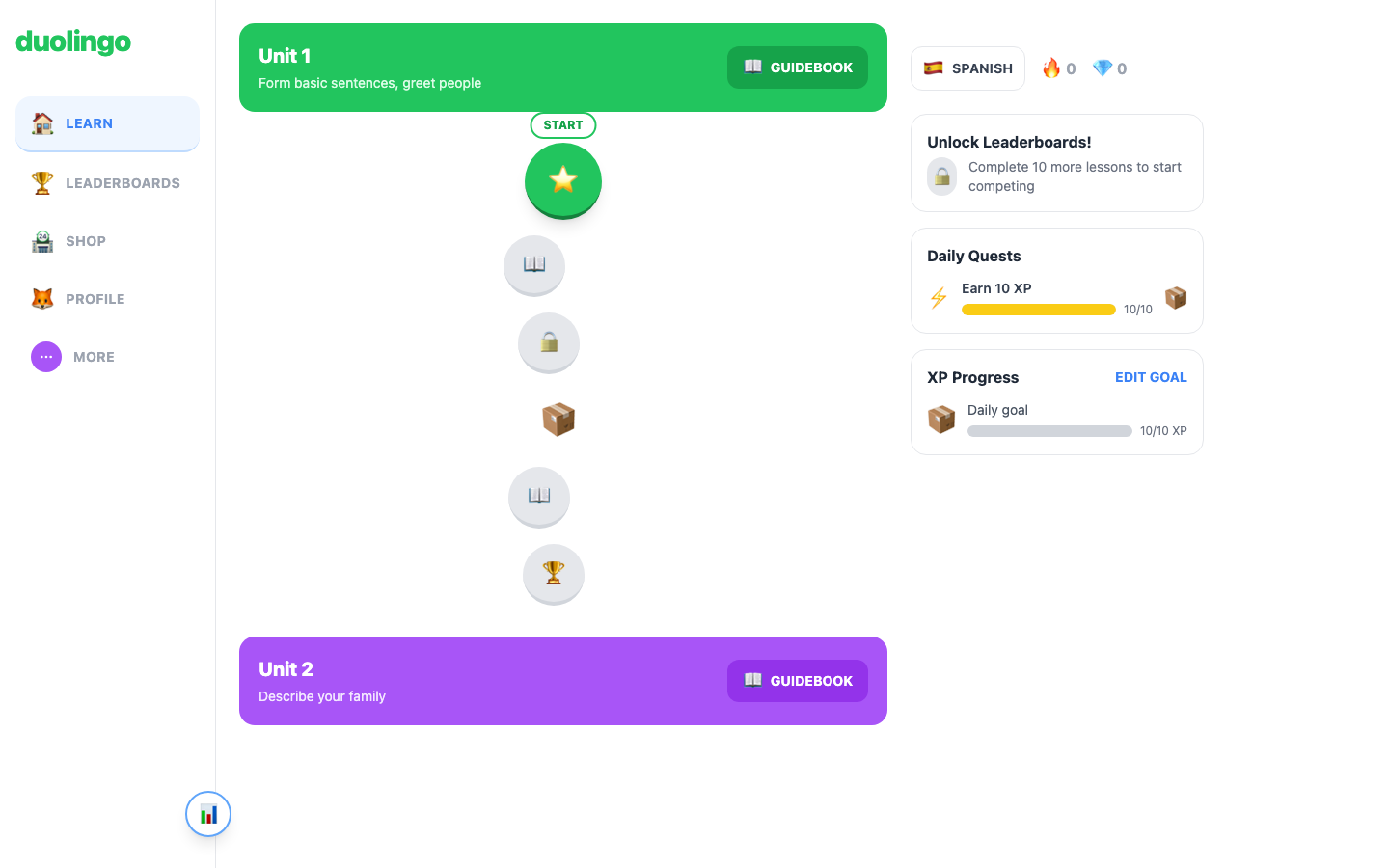}\\[2pt]
  {\footnotesize (e) Claude \texttt{/learn} --- \textbf{after}: Daily Quest \textcolor{blue}{10/10 XP}}
\end{minipage}\par\medskip
\begin{minipage}{0.9\linewidth}
\begin{codecard}{LessonPage.tsx}
const handleCheck = () => {
  if (!selected) return;
  const ok = selected === question.correctId;
  setCorrect(ok); setChecked(true);
  if (ok) incrementLessons();  // +1 lesson, +10 XP
};
\end{codecard}
\captionof{lstlisting}{Claude --- \texttt{LessonPage.tsx} (extract): the answer-validation + XP engine the other five omit.}
\end{minipage}\par\medskip
\captionof{figure}{\textbf{Functional quiz engine on \texttt{61\_react-duolingo} (before/after the same select-answer$\to$\textsc{Check} sequence).} (a,b) Claude validates the answer and shows \emph{Correct!} feedback, advancing to \textsc{Continue}; (c) Gemini's \textsc{Check} has no handler, so the same clicks leave the question unchanged. (d,e) the correct answer also awards XP, filling Claude's daily quest $0/10\to10/10$. Only Claude wires the full engine (Kimi partial; the other four inert) --- an answer-validation + scoring affordance \vfs{} cannot see, since every render is near-identical.}
\label{fig:case_duolingo}
\par}
\medskip

\paragraph{Synthesis across cases.}
The cases illustrate differences in the interaction types implemented by the evaluated models. Kimi~K2.5 implements audio synthesis and cross-session task persistence, while Claude~Sonnet~4.6 implements the demonstrated cart and quiz interactions. Build failures also differ across models: GPT-5.4\textquotesingle s failures are dominated by hallucinated icon exports, whereas GLM-4.6V\textquotesingle s are dominated by scaffold-related instruction violations (Table~\ref{tab:failure_modes}). These observations complement the aggregate scores by showing which behaviors and implementation failures underlie them.

\section{Prompt templates}
\label{app:prompts}

Below are the prompt templates used in the pipeline. Placeholders rendered as \texttt{\{name\}} are populated at run time.

\subsection*{Stage 1 (Plan).}
The Plan prompt is issued in the second user turn after the assistant has emitted the scaffold listing in Turn~1. The same turn carries all $M$ screenshots as image content.

\begin{promptbox}
You are an expert frontend developer. You are given \{M\} screenshots of a web application. Your task is to reproduce the UI as a complete, runnable Vite + React + TypeScript + Tailwind CSS project and implement the interactions implied by the screenshots.

\medskip
The project scaffold has already been set up (you saw it above). src/App.tsx is the root component mounted by main. All source files you generate must be under src/.

\medskip
tsconfig uses "jsx": "react-jsx" so explicit React imports are unnecessary. react-router-dom and lucide-react are available in dependencies.

\medskip
<code\_guidelines> \\
- Use coding best practices. \\
- Use Tailwind CSS utility classes for styling. Match the screenshots as closely as possible. \\
- Use relative imports. Do not add file extensions in imports for .ts/.tsx files. \\
- Files containing JSX must use .tsx extension. \\
</code\_guidelines>

\medskip
Analyze the \{M\} screenshots carefully. Then output a JSON file plan --- a list of all source files you will create, with a one-line description of each file's purpose.

\medskip
Output ONLY valid JSON (no markdown fences, no explanation):

\medskip
\{"extra\_dependencies": \{\}, "plan": [\{"path": "src/App.tsx", "description": "Root component with routing setup"\}, ...]\}

\medskip
Rules: \\
- Include src/App.tsx and all component/page files. All paths start with "src/". \\
- "extra\_dependencies" lists any npm packages beyond the scaffold. \\
- Do NOT include scaffold files. Do NOT generate any code yet --- only the plan.
\end{promptbox}

The four constraints we treat as explicit instruction clauses for the failure-mode attribution in \S\ref{sec:exp_failures} are: (i) all source files under \texttt{src/}; (ii) relative imports; (iii) every non-scaffold package declared in \texttt{extra\_dependencies}; (iv) no regeneration of scaffold files.

\subsection*{Stage 2 (Generate).}
The Generate prompt is issued in the next user turn after the model has returned the plan. Screenshots are not re-sent.

\begin{promptbox}
Now generate the COMPLETE source code for ALL files in the plan below.

\medskip
<project\_plan> \\
\{plan\_summary\} \\
</project\_plan>

\medskip
Output each file using this exact format (no markdown fences, no explanation):

\medskip
--- src/App.tsx --- \\
{[}complete source code for App.tsx{]}

\medskip
--- src/components/Navbar.tsx --- \\
{[}complete source code for Navbar.tsx{]}

\medskip
\ldots and so on for every file in the plan.

\medskip
CRITICAL RULES: \\
- Write COMPLETE file content for every file. Never truncate, abbreviate, or use comments like "// rest of code". \\
- Files containing JSX/TSX syntax (<div>, <Component/>) MUST use .tsx extension. \\
- Use 2 spaces for indentation. \\
- Every file must start with "--- filepath ---" on its own line. \\
- Do NOT output anything other than the file blocks.
\end{promptbox}

If the model's response is truncated (\texttt{finish\_reason}~$\ne$~\texttt{stop}), we send up to three Continue turns: ``Continue writing from exactly where you left off. Do not repeat any content. Continue with the next file or finish the current file.''

\subsection*{Stage 3 (Self-Debug).}
On EXEC failure the same model is re-invoked for up to three rounds. Each round runs three sub-prompts as needed.

\paragraph{(a) File-localization fallback.} When the regex extractor cannot identify which source files to repair from the build error, the model is asked to localize:

\begin{promptbox}
The following build error occurred. Which source file(s) need to be fixed?

\medskip
<error> \\
\{errors\} \\
</error>

\medskip
<project\_files> \\
\{file\_list\} \\
</project\_files>

\medskip
Reply with ONLY the file path(s) that need fixing, one per line. No explanation.
\end{promptbox}

\paragraph{(b) Build-error repair.} For each identified file the model receives the error, the project file listing, and the file's current contents:

\begin{promptbox}
Fix the following build error in \{filepath\}:

\medskip
<error> \\
\{errors\} \\
</error>

\medskip
<project\_files> \\
\{file\_listing\} \\
</project\_files>

\medskip
<current\_code file="\{filepath\}"> \\
\{current\_code\} \\
</current\_code>

\medskip
Output ONLY the fixed complete file content --- no explanation, no markdown fences.
\end{promptbox}

\paragraph{(c) Runtime-error repair.} When the build passes but the home route fails to render, the model receives all candidate files (\texttt{App.tsx}, \texttt{main.tsx}, plus any flagged by error parsing) together for a holistic fix:

\begin{promptbox}
The React app builds successfully but renders a blank page at runtime.

\medskip
<error> \\
\{errors\} \\
</error>

\medskip
\{file\_contents\}

\medskip
For EACH file that needs changes, output: \\
--- filepath --- \\
<complete fixed file content>

\medskip
Output ONLY the fixed files --- no explanation, no markdown fences.
\end{promptbox}

\section{Automated IIS Evaluation}
\label{app:auto_iis}
\paragraph{Evaluator and human validation.}
GPT-6 Luna with medium reasoning effort evaluates generated source code and rendered screenshots without access to the generator identity, assigning coverage, outcome, and scope labels under the interaction rubric. IIS uses the fixed reference inventory and Eq.~\ref{eq:iis}.
We validate the evaluator against human annotations on a 45-application subset: 217 executable artifacts yield 1,519 category-level comparisons; 53 non-executable artifacts receive zero IIS. Coverage agreement is $86.4\%$. On the 868 jointly applicable items, agreement on interaction outcomes (working, partial, or failed) is $81.8\%$ and exact scope agreement is $91.1\%$.
\begin{table}[H]
\centering\small
\caption{Human and GPT-6 Luna IIS on the 45-application validation subset. Differences use unrounded model means.}
\label{tab:auto_human}
\begin{tabular}{lrrr}
\toprule
Model & Human IIS & Auto IIS & Auto $-$ Human \\
\midrule
Claude Sonnet 4.6 & 36.46 & 37.30 & +0.84 \\
Gemini 3.1 Pro Preview & 7.30 & 7.87 & +0.56 \\
Kimi K2.5 & 18.80 & 22.57 & +3.78 \\
GPT-5.4 & 6.76 & 5.20 & -1.56 \\
Qwen3.5-397B-A17B & 12.21 & 13.02 & +0.80 \\
GLM-4.6V & 3.36 & 1.47 & -1.89 \\
\bottomrule
\end{tabular}
\end{table}

Across the six model means, Spearman correlation is $1.00$, Pearson correlation is $0.991$, MAE is $1.57$ IIS points, and mean signed bias is $+0.42$ points. Automated IIS estimates implementation correctness from source and screenshots; the validation labels are based on observed runtime behavior.

\section{Repeated-Generation Evaluation}
\label{app:repeated_evaluation}
To assess generation variability, each application in the 45-app subset is generated independently three times by each of six models, yielding 810 application--model--generation outcomes. GPT-6 Luna (medium) evaluates all 644 executable artifacts using the same rubric and fixed RII; the remaining 166 outcomes receive zero VFS and IIS.
The mean and sample standard deviation in Table~\ref{tab:repeated_runs} are computed from three round-level 45-application model means ($\mathrm{ddof}=1$).
\begin{table}[H]
\centering\small
\caption{Three-generation evaluation on the 45-application subset. Values are the mean $\pm$ sample standard deviation of the three round-level model means. IIS uses GPT-6 Luna (medium), as in Table~\ref{tab:main_results}.}
\label{tab:repeated_runs}
\begin{tabular}{lrrr}
\toprule
Model & EXEC@3 (\%) & VFS & IIS \\
\midrule
Claude Sonnet 4.6 & $98.52 \pm 1.28$ & $69.55 \pm 5.33$ & $36.86 \pm 0.39$ \\
Gemini 3.1 Pro Preview & $98.52 \pm 1.28$ & $71.68 \pm 5.64$ & $8.95 \pm 3.55$ \\
Kimi K2.5 & $76.30 \pm 8.98$ & $54.48 \pm 9.33$ & $18.73 \pm 3.34$ \\
GPT-5.4 & $85.19 \pm 2.57$ & $60.37 \pm 4.04$ & $3.97 \pm 1.68$ \\
Qwen3.5-397B-A17B & $77.78 \pm 0.00$ & $54.13 \pm 1.64$ & $14.64 \pm 1.91$ \\
GLM-4.6V & $40.74 \pm 4.63$ & $24.27 \pm 2.63$ & $2.77 \pm 2.03$ \\
\bottomrule
\end{tabular}
\end{table}

\paragraph{Hierarchical uncertainty.}
We use 20,000 hierarchical bootstrap replicates (seed 20260925). Each replicate samples 45 applications with replacement, preserving application pairing across models, then independently resamples three generations within each selected application--model cell. The 95\% percentile interval for Claude minus Gemini IIS is $[20.95,35.06]$, around a mean difference of $27.91$ points. Claude leads IIS and Gemini leads VFS in every round. This interval captures application composition and generation variability, not evaluator error.
\begin{table}[H]
\centering\small
\caption{Round-level results on the 45-app subset. Within each entry, values are R1 / R2 / R3.}
\label{tab:round_results}
\begin{tabular}{lrrr}
\toprule
Model & EXEC@3 (\%) & VFS & IIS \\
\midrule
Claude Sonnet 4.6 & 100.00 / 97.78 / 97.78 & 75.68 / 67.03 / 65.95 & 37.30 / 36.54 / 36.76 \\
Gemini 3.1 Pro Preview & 100.00 / 97.78 / 97.78 & 78.14 / 67.71 / 69.19 & 7.87 / 6.07 / 12.92 \\
Kimi K2.5 & 86.67 / 71.11 / 71.11 & 65.25 / 48.98 / 49.22 & 22.57 / 17.13 / 16.50 \\
GPT-5.4 & 82.22 / 86.67 / 86.67 & 64.96 / 58.80 / 57.36 & 5.20 / 2.06 / 4.65 \\
Qwen3.5-397B-A17B & 77.78 / 77.78 / 77.78 & 56.02 / 53.21 / 53.16 & 13.02 / 14.17 / 16.75 \\
GLM-4.6V & 35.56 / 44.44 / 42.22 & 21.28 / 25.32 / 26.22 & 1.47 / 1.74 / 5.11 \\
\bottomrule
\end{tabular}
\end{table}

\end{document}